\PassOptionsToPackage{table}{xcolor}
\PassOptionsToPackage{only,llbracket,rrbracket,llparenthesis,rrparenthesis}{stmaryrd}

\documentclass[acmsmall,10pt,screen]{acmart}
\usepackage{suffix}
\usepackage{okgeneralmath}
\usepackage{okcategory}
\usepackage{bbm}
\usepackage{multicol}
\usepackage{mathtools}
\usepackage{mathpartir}
\usepackage{oksemantics}
\usepackage{commath}
\usepackage{listings}
\usepackage[inline]{enumitem}
\usepackage{wrapfig}
\usepackage{diagrams}
\usepackage{framed}
\usepackage{contour}
\usepackage[normalem]{ulem}

\newcommand\insertdiagram[1]{\begin{center}
    \insertdiagram*{#1}
\end{center}}
\WithSuffix\newcommand\insertdiagram*[1]{\begin{tabular}[c]{@{}c@{}}\includegraphics{diagram-#1.mps}\end{tabular}}

\newcommand{\commentout}[1]{}

\newcommand\seclabel[1]{\label{sec:#1}}
\renewcommand\secref[1]{Sec.~\ref{sec:#1}}
\newcommand\subseclabel[1]{\label{subsec:#1}}
\newcommand\subsecref[1]{\S\ref{subsec:#1}}

\newcommand\figlabel[1]{\label{fig:#1}}
\renewcommand\figref[1]{Fig.~\ref{fig:#1}}

\newcommand\AddReferences[2]{
  \expandafter\newcommand\csname#1ref\endcsname[1]{#2~\ref{#1:##1}}
  \expandafter\newcommand\csname#1label\endcsname[1]{\label{#1:##1}}
  \WithSuffix\expandafter\newcommand\csname#1ref\endcsname*[1]{\ref{#1:##1}}
  \WithSuffix\expandafter\newcommand\csname#1label\endcsname+[1]{\hypertarget{#1+:##1}{}\zref@labelbyprops{#1:##1}{oktheoremfreetext}}
  \WithSuffix\expandafter\newcommand\csname#1ref\endcsname+[1]{\hyperlink{#1+:##1}{{{\let\ref\@refstar#2~\zref@extract{#1:##1}{oktheoremfreetext}}}}}
  \WithSuffix\expandafter\newcommand\csname#1ref\endcsname-[1]{\hyperlink{#1+:##1}{{\let\ref\@refstar{\zref@extract{#1:##1}{oktheoremfreetext}}}}}
}

\AddReferences{example}{Ex.}
\AddReferences{lemma}{Lemma}
\AddReferences{theorem}{Thm.}
\AddReferences{corollary}{Cor.}
\AddReferences{definition}{Def.}
\newcommand\defref\definitionref
\newcommand\deflabel\definitionlabel

\newcommand{\citepos}[1]{\citeauthor{#1}'s~\citeyearpar{#1}}
\WithSuffix\newcommand\citepos*[2]{\citeauthor{#1}'s #2~\citeyearpar{#1}}
\WithSuffix\newcommand\citepos-[2]{\citeauthor{#1}#2~\citeyearpar{#1}}
\WithSuffix\newcommand\citepos+[3]{\citeauthor{#1}'s #2~\citeyearpar{#1,#3}}
\makeatletter
\newcommand\ibid[2][\ibid@empty]{
  \citetext{
    \hyper@natlinkstart{#1}
    \textit{ibid.}
    \hyper@natlinkend
    \ifx\ibid@empty#1\ibid@empty
    \else
    , #1
    \fi
  }
}
\makeatother

\newcounter{countitems}
\newcounter{nextitemizecount}
\newcommand{\setupcountitems}{
  \stepcounter{nextitemizecount}
  \setcounter{countitems}{0}
  \preto\item{\stepcounter{countitems}}
}
\makeatletter
\newcommand{\computecountitems}{
  \edef\@currentlabel{\number\c@countitems}
  \label{countitems@\number\numexpr\value{nextitemizecount}-1\relax}
}
\newcommand{\nextitemizecount}{
  \getrefnumber{countitems@\number\c@nextitemizecount}
}
\newcommand{\previtemizecount}{
  \getrefnumber{countitems@\number\numexpr\value{nextitemizecount}-1\relax}
}
\makeatother

\definecolor{shade}{RGB}{223,223,223}
\definecolor{unshade}{RGB}{255,255,255}

\usepackage{tcolorbox}
\newtcbox{\shadebox}{on line,arc=1pt, outer arc=2pt,
  colback=shade,colframe=shade,boxsep=0pt,
  left=1pt,right=1pt,top=2pt,bottom=2pt,
  boxrule=0pt,bottomrule=1pt,toprule=1pt}
\newcommand{\shade}[1]{
        \shadebox{\ensuremath{#1}}
}
\newtcbox{\unshadebox}{on line,arc=1pt, outer arc=2pt,
  colback=unshade,colframe=shade,boxsep=0pt,
  left=1pt,right=1pt,top=2pt,bottom=2pt,
  boxrule=0pt,bottomrule=1pt,toprule=1pt}

\newcommand\syncat[1]{\mspace{-25mu}\synname{#1}}
\newcommand\synname[1]{\qquad\text{#1}}

\newenvironment{syntax}[1][]{
\(
  \rowcolors{100}{white}{white}
  \begin{array}[t]{#1l@{}l@{\,}*2{l@{}}@{\,}l}
}{
\end{array}
\)
}

\newenvironment{sugar}{
  \(
  \begin{array}[t]{@{\bullet\ }l@{\mspace{10mu}}ll}
    \multicolumn{1}{@{\hphantom{\bullet}}l}{\text{Sugar}} & \text{Elaboration}\\
    \hline
}{
  \end{array}
  \)
}

\newenvironment{leansugar}[1][\multicolumn{1}{@{{\bullet\ }}l}]{
  \(
  \begin{array}[t]{@{\bullet\ }l@{\mspace{10mu}}ll}
    #1
}{
  \end{array}
  \)
}

\newcommand\gdefinedby{::=}
\newcommand\gor{\mathrel{\lvert}}
\makeatletter

\newcommand\kind{k}

\newcommand\op[1][1]{
\ifcase #1\or f\or g\or h\or p\or q\else \@ctrerr \fi
}

\newcommand\s[1][1]{
\ifcase #1\or s\or t\or u\or v\or w\else \@ctrerr \fi
}

\newcommand\Alg[1][1]{
\ifcase #1\or A\or B\or C\or D\else \@ctrerr \fi
}

\newcommand\@TyAlph[1]{
\ifcase #1\or \tau\or \sigma\or \rho\else \@ctrerr \fi
}
\newcommand\ty[1][1]{{\@TyAlph{#1}}}

\newcommand\tvar[1][1]{{\@TyVarAlph{#1}}}
\newcommand\@TyVarAlph[1]{
\ifcase #1\or \alpha\or \beta\or \gamma\else \@ctrerr \fi
}

\newcommand\Functor[1][1]{\TypeCons{\@FunAlph{#1}}}
\newcommand\@FunAlph[1]{
\ifcase #1\or F\or G\or H\else \@ctrerr \fi
}

\newcommand\Type[1][1]{{\mathrm{\@TypeAlph{#1}}}}
\newcommand\@TypeAlph[1]{
\ifcase #1\or A\or B\or C\else \@ctrerr \fi
}

\newcommand\var[1][1]{{\@VarAlph{#1}}}
\newcommand\@VarAlph[1]{
\ifcase #1\or x\or y\or z\or u\or v\or w\else \@ctrerr \fi
}

\newcommand\trm[1][1]{{\@TermAlph{#1}}}
\newcommand\@TermAlph[1]{
\ifcase #1\or t\or s\or r\else \@ctrerr \fi
}

\newcommand\val[1][1]{
\ifcase #1\or v\or w\or u\else \@ctrerr \fi
}

\newcommand\ir[1][1]{\monad{\uir[#1]}}

\newcommand\uir[1][1]{\mathop{{}\@IRAlph{#1}}\nolimits}
\newcommand\@IRAlph[1]{
\ifcase #1\or T\or S\or R\else \@ctrerr \fi
}

\newcommand\uIF[1][1]{\mathop{{}\@IFAlph{#1}}\nolimits}
\newcommand\@IFAlph[1]{
\ifcase #1\or F\or G\or H\else \@ctrerr \fi
}

\newcommand\cat[1][1]{\Cat{\@CatAlph#1}}
\newcommand\@CatAlph[1]{
\ifcase #1\or C\or D\or A\or B\or E\else \@ctrerr \fi
}

\newcommand\F[1][1]{\@FunctorAlph#1}
\newcommand\@FunctorAlph[1]{
\ifcase #1\or F\or G\or H\\else \@ctrerr \fi
}

\newcommand\nt[1][1]{\@NatTransAlph#1}
\newcommand\@NatTransAlph[1]{
\ifcase #1\or \alpha\or \beta\or \gamma\else \@ctrerr \fi
}

\newcommand\alg[1][1]{\@AlgAlph#1}
\newcommand\@AlgAlph[1]{
\ifcase #1\or A\or B\or C\else \@ctrerr \fi
}

\newcommand\qmu[1][1]{\underline{\@MuAlph#1}}
\newcommand\@MuAlph[1]{
\ifcase #1\or \mu\or \nu\or \xi\or \zeta\else \@ctrerr \fi
}

\makeatother

\newcommand\Cns{\ell}
\newcommand\Inj[2][\,]{\mathrm{#2}#1}
\WithSuffix\newcommand\Inj*[2][\,]{{#2}#1}
\newcommand\Variant[1]{\{ #1 \}}
\newcommand\vor{\mathrel{\big\lvert}}
\newcommand\Unit{\mathrm{1}}
\WithSuffix\newcommand\t*{*}
\newcommand\rec[2]{\mu#1.#2}
\newcommand\To{\to}
\newcommand\ctx{\Gamma}
\newcommand\env{\gamma}
\newcommand\lctx{\Delta}
\newcommand\Tyctx{\Delta}
\newcommand\kinf{\vdash_{\mathrm k}}
\newcommand\Dkinf[2][]{\Tyctx#1\kinf#2 : \typ}
\newcommand\typ{\mathrm{type}}
\newcommand\context{\mathrm{context}}

\newcommand\TypeCons[1]{\mathop{{}\mathrm{#1}}\nolimits}

\newcommand\List{\TypeCons{List}}
\newcommand\Stoch{\TypeCons{Stoch}}
\newcommand\UntypedL{\Lambda\textnormal{-}\TypeCons{Term}}
\newcommand\boolty{{\TypeCons{bool}}}
\newcommand\tru{\mathop{\Inj[]{True}}}
\newcommand\fls{\mathop{\Inj[]{False}}}

\newcommand\ereals{\overline\reals}
\WithSuffix\newcommand\ereals+{\overline\reals_+}
\WithSuffix\newcommand\reals+{\reals_+}
\newcommand\Real{\underline{\reals}}
\newcommand\lawson{\mathbb{W}}
\newcommand\Borel{\mathcal{B}}
\newcommand\rbox[1][]{\square_{#1}}

\newcommand\elseclause{\,\_}
\newcommand\tInj[3][\,]{#2.#3#1}
\newcommand\tUnit{()}
\newcommand\tpair[2]{(#1, #2)}

\newcommand\troll[2][\,]{#2.\mathrm{roll}#1}
\newcommand\fun[1]{\lambda #1.\,}

\newcommand\vmatch[3][\,]{\mathrm{match}\,#2\,\mathrm{with}#1\{#3\}}
\newcommand\umatch[3][\,]{\mathrm{match}\,#2\,\mathrm{with}#1\tUnit\To#3}
\newcommand\pmatch[5][\,]{\mathrm{match}\,#2\,\mathrm{with}#1\tpair{#3}{#4}\To#5}
\newcommand\rmatch[4][\,]{\mathrm{match}\,#2\,\mathrm{with}#1\roll#3\To#4}
\newcommand\svmatch[2]{\begin{array}[t]{@{}l@{}}\mathrm{match}\,#1\,
\mathrm{with} \\ \{#2\}
\end{array}}
\newcommand\sumatch[2]{\begin{array}[t]{@{}l@{}}\mathrm{match}\,#1\, \mathrm{with}\\ \tUnit\To#2
\end{array}}
\newcommand\spmatch[4]{\begin{array}[t]{@{}l@{}}\mathrm{match}\,#1\, \mathrm{with}\\ \tpair{#2}{#3}\To#4
\end{array}}
\newcommand\srmatch[4][\,]{\begin{array}[t]{@{}l@{}}\mathrm{match}\,#2\,\mathrm{with}\\
#1\roll#3\To#4
\end{array}}

\newcommand\tsample[1][\ ]{\mathrm{sample}#1}

\newcommand\tscore[1][\,]{\mathrm{score}#1}
\newcommand\tfactor[1][\,]{\mathrm{factor}#1}

\newcommand\roll[1][\,]{\mathrm{roll}#1}
\newcommand\unroll[1][\,]{\mathrm{unroll}#1}

\contourlength{0.8pt}
\newcommand{\cnst}[1]{
  {\uline{\phantom{\smash{#1}}}
  \llap{\contour{white}{$#1$}}}}

\newcommand\ifz[4][\,]{\mathrm{match}\,#2\,\mathrm{with}#1\{\cnst{0}\To#3 \vor  \elseclause\To #4\}}
\newcommand\sifz[3]{\begin{array}[t]{@{}l@{}}\mathrm{match}\,#1\,\mathrm{with}\\ \{\cnst{0}\To#2 \vor  \elseclause\To #3\}\end{array}}
\newcommand\ifzun[3]{\mathrm{ifz}\,#1\,\mathrm{then}\,#2\,\mathrm{else}\,#3}

\newcommand\expfun[2]{\lambda #1:#2.}
\newcommand\letin[3]{\mathrm{let}\,#1=#2\,\mathrm{in}\,#3}
\newcommand\handlewith[2]{\mathrm{handle}\,#1\,\mathrm{with}\,#2}
\newcommand\letrecin[3]{\mathrm{letrec}\,#1=#2\,\mathrm{in}\,#3}
\newcommand\expletin[2]{\letin{(#1:#2)}}

\newcommand\randomize{{rand}}

\WithSuffix\newcommand\++{\mathbin{+\kern-.4em+}}

\newcommand\lst[1]{[#1]}
\WithSuffix\newcommand\lst-{\lst{\ }}

\newcommand\tinf{\vdash}
\newcommand\Ginf[3][]{\ctx #1\tinf #2 : #3}
\newcommand\subst[2]{#1{}[#2]}
\newcommand\sfor[2]{#1 \mapsto #2}

\newcommand\gobble[1]{}

\newcommand\Expr[3][\vdash]{\mathrm{Trm}^{#2 #1  #3}}
\newcommand\Val[3][\vdash]{\mathrm{Val}^{#2 #1 #3}}
\newcommand\LExpr[3][\vdash]{\mathrm{TTrm}^{#2 #1 #3}}
\newcommand\LVal[3][\vdash]{\mathrm{TVal}^{#2 #1 #3}}

\newcommand\sem\scottbrackets
\newcommand\psem[1]{\stretchleftright{\llbrace}{\vphantom{(}#1}\rrbrace}
\newcommand\carrier[1]{\left\lvert#1\right\rvert}
\renewcommand\terminal{{1}}

\DeclareMathAlphabet{\mathbbold}{U}{bbold}{m}{n}
\renewcommand\initial{{0}}

\newcommand\uniform[1]{\mathbb{U}_{#1}}
\newcommand\spair[2]{\langle #1, #2\rangle}
\newcommand\sUnit{\langle\rangle}
\newcommand\scopair[2]{[ #1, #2]}
\newcommand\scoUnit{[]}
\newcommand\wqbs{$\omega$qbs}
\newcommand\wchain{$\omega$-chain}
\newcommand\NN{\naturals}
\newcommand\RR{\reals}

\newcommand\lub{\bigvee}
\newcommand\ord[1]{\underline{#1}}
\newcommand{\defeq}{:=}

\newcommand\semroll[1][]{\mathrm{roll}_{#1}}
\newcommand\semunroll[1][]{\mathrm{unroll}_{#1}}

\newcommand\continuum{\mathfrak{c}}
\newcommand\suc[1]{#1^{+}}
\newcommand\IntwCpo[1]{\wCpo(#1)}

\newcommand\DS{\mathfrak{D}}
\newcommand\M{\mathcal{M}}
\newcommand\admon{\mathfrak{M}}
\newcommand\Bor{\mathrm{Borel}}
\newcommand\Sco{\mathrm{Scott}}
\newcommand\BS{\mathrm{BS}}
\newcommand\Fn{\mathrm{Fn}}

\newcommand\pto{\rightharpoonup}
\newcommand\pMap[1]{\Category{p}{#1}}

\newcommand\dmon[1][]{\partial_{#1}}
\newcommand\dfun[1]{\overline{#1}}
\newcommand\dom[1][]{D_{#1}}
\newcommand\Lift[1]{#1_{\bot}}
\newcommand\xpto{\xrightharpoonup}
\newcommand\tot{\mathop{\uparrow}\nolimits}
\newcommand\pleq{\sqsubseteq}

\newcommand\equivalent{\simeq}
\newcommand\E{\Cat{E}_{q}}
\newcommand\wE{\Cat{E}_{\omega q}}
\newcommand\QBS[1]{#1^\sharp}
\newcommand\invQBS[1]{#1^\flat}
\newcommand\wQBS[1]{#1^\sharp}

\newcommand\Sig\Sigma
\newcommand\Pres{\mathcal P}
\newcommand\pospres{\mathrm{pos}}
\newcommand\wcpopres{\boldsymbol{\omega}\mathrm{cpo}}
\newcommand\qbspres{\mathrm{qbs}}
\newcommand\wqbspres{\boldsymbol{\omega}\mathrm{qbs}}
\newcommand\Sorts{\mathcal S}
\newcommand\pSig{\Sig_{\mathrm{p}}}
\newcommand\tSig{\Sig_{\mathrm{t}}}
\newcommand\uSig{\Sig}
\newcommand\Eq{\mathrm{Eq}}
\newcommand\Ops{\mathcal O}
\newcommand\Var{\mathbb V}
\newcommand\arity{\mathop{\mathrm{arity}}\nolimits}
\newcommand\Term{\mathop{\mathrm{Term}}\nolimits}
\newcommand\elem{\mathrm{element}}
\newcommand\ineq{\mathrm{inequation}}
\newcommand\rand{\mathrm{rand}}
\newcommand\newop[2]{\newcommand{#1}{\mathop{\mathrm{#2}}\nolimits}}
\newop\Def{Def}
\newop\source{lower}
\newop\target{upper}
\newop\irrelevance{irrel}
\newop\refl{refl}
\newop\antisym{antisym}
\newop\trans{trans}
\newop\ub{ub}
\newop\least{least}
\newop\Mod{Mod}
\newop\DomainName{Dom}
\newop\ev{ev}
\newop\const{const}
\newop\rearrange{rearrange}
\newop\match{match}
\newop\extensionality{ext}
\newcommand\rlub\bigsqcup
\newcommand\meet\wedge
\newcommand\Meet\bigwedge

\usepackage{extpfeil}
\newextarrow{\xtoto}{{20}{20}{20}{20}}
   {\bigRelbar\bigRelbar{\bigtwoarrowsleft\rightarrow\rightarrow}}
\makeatletter
\newbox\xrat@below
\newbox\xrat@above
\newcommand{\xrightarrowtail}[2][]{
  \setbox\xrat@below=\hbox{\ensuremath{\scriptstyle #1}}
  \setbox\xrat@above=\hbox{\ensuremath{\scriptstyle #2}}
  \pgfmathsetlengthmacro{\xrat@len}{max(\wd\xrat@below,\wd\xrat@above)+.6em}
  \mathrel{\tikz [>->,baseline=-.75ex]
                 \draw (0,0) -- node[below=-2pt] {\box\xrat@below}
                                node[above=-2pt] {\box\xrat@above}
                       (\xrat@len,0) ;}}
\makeatletter
\usepackage{tikz}
\newcommand{\xdasharrow}[2][->]{
\tikz[baseline=-\the\dimexpr\fontdimen22\textfont2\relax]{
\node[anchor=south,font=\scriptsize, inner ysep=1.5pt,outer xsep=2.2pt](x){$ #2$};
\draw[shorten <=3.4pt,shorten >=3.4pt,dashed,#1](x.south west)--(x.south east);
}
}
\makeatother
\newcommand\monad\underline
\newcommand\return{\mathop{\mathrm{return}}\nolimits}

\newcommand{\bind}{\mathrel{\scalebox{0.8}[1]{\(>\!\!>\!=\)}}}
\newcommand{\fbind}{\mathrel{\scalebox{0.8}[1]{\(>\!\!>\)}}}

\newcommand\FromKind[1]{
  \ifthenelse{\equal{#1}{Rep}}
             {}
             {#1\implies}
}

\WithSuffix\newcommand\<-{\mathrel{\leftarrow}}

\newcommand\sscore{\mathop{\mathrm{score}}\nolimits}

\newcommand\constantly\underline

\DeclareSymbolFont{largesymbolsA}{U}{txexa}{m}{n}
\DeclareMathSymbol{\sqintop}{\mathop}{largesymbolsA}{14}
\DeclareMathSymbol{\sqiintop}{\mathop}{largesymbolsA}{80}
\DeclareMathSymbol{\sqiiintop}{\mathop}{largesymbolsA}{82}

\DeclareFontFamily{U}{matha}{\hyphenchar\font45}
\DeclareFontShape{U}{matha}{m}{n}{
        <-6> matha5 <6-7> matha6 <7-8> matha7
        <8-9> matha8 <9-10> matha9
        <10-12> matha10 <12-> matha12
}{}
\DeclareSymbolFont{matha}{U}{matha}{m}{n}
\DeclareMathSymbol{\absc}{3}{matha}{"CE}

\newcommand\Qbs{\Category{Qbs}}
\newcommand{\Meas}{\Category{Meas}}
\newcommand{\Sbs}{\Category{Sbs}}

\newcommand\ssample{\mathrm{sample}}

\newcommand\wcpo{$\omega$cpo}
\newcommand\wcpos{\wcpo{}s}
\newcommand\wCpo{\omega\mathrm{Cpo}}
\newcommand\wQbs{\omega\mathrm{Qbs}}

\DeclareMathDelimiter{\ulcorner}{\mathopen}{okMnSymbolLargeSymbols}{'036}{okMnSymbolLargeSymbols}{'036}
\DeclareMathDelimiter{\urcorner}{\mathclose}{okMnSymbolLargeSymbols}{'043}{okMnSymbolLargeSymbols}{'043}
\DeclareMathDelimiter{\llcorner}{\mathopen}{okMnSymbolLargeSymbols}{'050}{okMnSymbolLargeSymbols}{'050}
\DeclareMathDelimiter{\lrcorner}{\mathclose}{okMnSymbolLargeSymbols}{'055}{okMnSymbolLargeSymbols}{'055}
\DeclareMathDelimiter{\ullcorner}{\mathopen}{okMnSymbolLargeSymbols}{'062}{okMnSymbolLargeSymbols}{'062}
\DeclareMathDelimiter{\ulrcorner}{\mathclose}{okMnSymbolLargeSymbols}{'067}{okMnSymbolLargeSymbols}{'067}
\newcommand\trunc[1]{\left\ulcorner#1\right\urcorner}
\newcommand\floor[2][]{\left\llcorner#2\right\lrcorner_{#1}}

\newcommand\embd[1]{#1^{\mathrm e}}
\newcommand\proj[1]{#1^{\mathrm p}}
\newcommand\ep[1]{#1^{\mathrm{ep}}}
\newcommand\embedding{\hookrightarrow}
\newcommand\pembedding{\lhook\joinrel\rightharpoonup}

\newcommand\epto{\xepto{\hphantom{a}}}
\newcommand\peto{\xpeto{\hphantom{a}}}
\makeatletter
\newbox\xrat@below
\newbox\xrat@above
\newcommand{\xmonomo}[2][]{
  \setbox\xrat@below=\hbox{\ensuremath{\scriptstyle #1\,}}
  \setbox\xrat@above=\hbox{\ensuremath{\scriptstyle #2\,}}
  \pgfmathsetlengthmacro{\xrat@len}{max(\wd\xrat@below,\wd\xrat@above)+.6em}
  \mathrel{\tikz [>-latex,baseline=-.75ex]
                 \draw (0,0) -- node[below=-2pt] {\box\xrat@below}
                                node[above=-2pt] {\box\xrat@above}
                       (\xrat@len,0) ;}}
\makeatother
\newcommand\xepimor\xtwoheadrightarrow

\makeatletter
\def\MTrightharpoonupfill{
  \arrowfill@\relbar\relbar\rightharpoonup}
\def\MTleftharpoondownfill{
  \arrowfill@\leftharpoondown\relbar\relbar}
\def\MTleftharpoonupfill{
  \arrowfill@\leftharpoonup\relbar\relbar}
\def\MTrightharpoondownfill{
  \arrowfill@\relbar\relbar\rightharpoondown}
\newcommand*\xhookrightleftharpoons[2][]{\mathrel{
  \raise.22ex\hbox{
    $\lhook\joinrel\ext@arrow 0359\MTrightharpoonupfill{\phantom{#1}}{#2}$}
  \setbox0=\hbox{
    $\ext@arrow 3095\MTleftharpoondownfill{#1}{\phantom{\lhook\joinrel#2}}$}
  \kern-\wd0 \lower.22ex\box0}}
\newcommand*\xleftrighthookharpoons[2][]{\mathrel{
  \raise.22ex\hbox{
    $\ext@arrow 3095\MTleftharpoonupfill{\phantom{#1\mspace{15mu}}}{#2}$}
  \setbox0=\hbox{
    $\mathrel{\raise-.4837ex\hbox{$\lhook$}}\joinrel\ext@arrow 0359\MTrightharpoondownfill{#1}{\phantom{#2}}$}
  \kern-\wd0 \lower.22ex\box0}}
\makeatother
\newcommand\xepto{\xhookrightleftharpoons}
\newcommand\xpeto{\xleftrighthookharpoons}

\newcommand\Representability{$(\leftadjointto)$}
\newcommand\RepMonotonicity{$(\leftadjointto_{\leq})$}
\newcommand\wCpoEnrichment{$(\cat_{\lub})$}
\newcommand\pMapEnrichment{$(\pMap{}_{\lub})$}
\newcommand\RepContinuity{$(\leftadjointto_{\lub})$}
\newcommand\FullUC{$(\mathrm{fup})$}
\newcommand\Uniformity{$(U)$}
\newcommand\UnitType{$(\terminal)$}
\newcommand\ProductType{$(\times_{\lub})$}
\newcommand\PartialProduct{$(\otimes)$}
\newcommand\PartialUnit{$(\terminal_{\leq})$}
\newcommand\ContinuousPartialProduct{$(\otimes_{\lub})$}
\newcommand\FunType{$(\to_{\leq})$}
\newcommand\wCpoFun{$(\to_{\lub})$}
\newcommand\PartialExponential{$(\pexp_{\lub})$}
\newcommand\Coco{$(CL)$}
\newcommand\pCoco{$(\pMap CL)$}
\newcommand\Coprod{$(+)$}
\newcommand\pCoprod{$(\pMap+_{\lub})$}

\newcommand\ContinuousT{($T_{\lub}$)}

\newcommand\ZeroAdmissible{$(?!)$}
\newcommand\BilimitExpansion{$(BC)$}
\newcommand\apply{\mathrm{eval}}
\makeatletter
\DeclareRobustCommand{\overleftharpoon}{\mathpalette{\overarrow@\leftharpoonfill@}}
\DeclareRobustCommand{\overrightharpoon}{\mathpalette{\overarrow@\rightharpoonfill@}}
\def\leftharpoonfill@{\arrowfill@\leftharpoondown\mn@relbar\mn@relbar}
\def\rightharpoonfill@{\arrowfill@\mn@relbar\mn@relbar\rightharpoonup}

\DeclareMathSymbol{\leftharpoonup}{\mathrel}{okMnSymbolArrows}{'112}
\DeclareMathSymbol{\mn@relbar}{\mathrel}{okMnSymbolArrows}{'320}
\makeatother

\def\pexplen{0mu}
\def\pexpmidlen{4mu}
\newcommand\pexp{
  \mathrel{
    \overrightharpoon{
      {\mspace{-\pexplen}\smash{\mathord{\relbar\mspace{-\pexpmidlen}\relbar}}\vphantom{.}}\mspace{-\pexplen}}
  }
}
\newcommand\shortiff\Leftrightarrow

\usepackage{mathtools}
\newcommand\lebesgue{\lambdaup}
\newcommand\Open{\mathcal{O}}
\newcommand\dirac[1]{\delta_{#1}}
\newcommand\evalto[1][]{\Downarrow_{#1}}

\makeatletter
\newtheorem*{rep@theorem}{\rep@title}
\newcommand{\newreptheorem}[2]{
\newenvironment{rep#1}[1]{
 \def\rep@title{#2 ##1}
 \begin{rep@theorem}\def\@currentlabel{##1}}
 {\end{rep@theorem}}}
\makeatother
\newreptheorem{theorem}{Theorem}
\newreptheorem{corollary}{Corollary}
\newreptheorem{proposition}{Proposition}

\usepackage{booktabs}
\usepackage{subcaption}
\usepackage{tikz,tikz-cd}

\copyrightyear{2019}
\setcopyright{rightsretained}
\acmJournal{PACMPL}
\acmYear{2019}
\acmVolume{3}
\acmNumber{POPL}
\acmArticle{36}
\acmMonth{1}
\acmPrice{}
\acmDOI{10.1145/3290349}

\newcommand\ConferenceArxiv[2]{#2}

\begin{document}

\title{
A Domain Theory for Statistical Probabilistic Programming}
 \author{Matthijs V\'ak\'ar}
\email{mv2745@columbia.edu}
\affiliation{
  \institution{Columbia University}
  \department{Department of Statistics}
  \streetaddress{1255 Amsterdam Avenue}
  \city{New York}
  \postcode{10027}
  \country{USA}
}
\author{Ohad Kammar}
\orcid{0000-0002-2071-0929}
\email{ohad.kammar@cs.ox.ac.uk}
\affiliation{
  \institution{University of Oxford}
  \department{Department of Computer Science}
  \streetaddress{Wolfson Building, Parks Road}
  \city{Oxford}
  \postcode{OX1 3QD}
  \country{UK}
}
\author{Sam Staton}
\email{sam.staton@cs.ox.ac.uk}
\affiliation{
  \institution{University of Oxford}
  \department{Department of Computer Science}
  \streetaddress{Wolfson Building, Parks Road}
  \city{Oxford}
  \postcode{OX1 3QD}
  \country{UK}
}
\newcommand\newshortauthors{
V\'ak\'ar, Kammar, Staton
}
 \authorsaddresses{}
\begin{abstract}
We give an adequate denotational semantics for languages with
recursive higher-order types, continuous probability distributions,
and soft constraints.  These are expressive languages for building
Bayesian models of the kinds used in computational statistics and
machine learning. Among them are untyped languages, similar to Church
and WebPPL, because our semantics allows recursive mixed-variance
datatypes.  Our semantics justifies important program equivalences
including commutativity.

Our new semantic model is based on `quasi-Borel predomains'. These are
a mixture of chain-complete partial orders (cpos) and quasi-Borel
spaces. Quasi-Borel spaces are a recent model of probability theory
that focuses on sets of admissible random elements. Probability is
traditionally treated in cpo models using probabilistic
powerdomains, but these are not known to be commutative on any class
of cpos with higher order functions. By contrast, quasi-Borel
predomains do support both a commutative probabilistic powerdomain and
higher-order functions. As we show, quasi-Borel predomains form both a
model of Fiore's axiomatic domain theory and a model of Kock's
synthetic measure theory.
\end{abstract}


\maketitle

\section{Introduction}

\newtheorem*{theorem*}{Theorem}
\newtheorem*{corollary*}{Corollary}
\newtheorem*{definition*}{Definition}
The idea of statistical probabilistic programming is to use a
programming language to specify statistical models and inference
problems.  It enables rapidly prototyping different models, because:
(1)~the model specification is separated from the technicalities of
the inference/simulation~algorithms; and (2)~software
engineering/programming techniques can be used to manage the
complexity of statistical models.  Here, we focus on the fundamental
programming technique of \emph{recursion}.  We consider both recursive
terms --- looping --- and recursive types, e.g., streams and untyped
languages.

In a traditional programming language, recursion has long been
analyzed using semantic domains based on $\omega$-complete partial
orders. The suitability of this approach is given by the adequacy
theorem, which connects the compositional interpretation in domains
with an operational interpretation.  However, there are long standing
open problems regarding using these domains with probability and
measure. Here we sidestep these problems by introducing a new
notion: \emph{$\omega$-quasi Borel spaces}.  We look at a language
with both statistical constructs and higher-order recursive types and
terms, which is close to languages used in practice for statistical
probabilistic programming, and we show the following adequacy theorem:
\begin{theorem*}[\theoremref*{adequacy}: Adequacy]
If two programs are equal in the \wqbs{} model then they are
contextually equivalent with regard to a Monte-Carlo operational
semantics.
\end{theorem*}

Domain theoretic semantics can verify compositional compiler
optimizations.  As an example, the following reordering transformation
is valid, since it is readily verifiable in the \wqbs es.
\begin{corollary*}[\corollaryref*{comm-sem}: Commutativity]
The following two programs are contextually equivalent:
\[
  \begin{array}{@{}l@{}}
    \letin {\var[1]}{\trm[1]}\\
    \letin {\var[2]}{\trm[2]}
    \trm[3]
  \end{array}
\quad  \approx\quad
  \begin{array}{@{}l@{}}
    \letin {\var[2]}{\trm[2]}\\
    \letin {\var[1]}{\trm[1]}
    \trm[3]
  \end{array}
\]
\end{corollary*}
This property says there is no implicit sequential state in the language.
It is essential
for \citepos*{shan-ramsey:exact-bayesian-inference-by-symbolic-disintegration}{disintegration-based
exact Bayesian inference technique}, implemented in the Hakaru
system~\cite{Hakaru}.
The corollary is related to
Fubini's theorem for reordering integrals: informally, $\int \dif x\int \dif y
\,r(x,y)=\int \dif y \int \dif x \,r(x,y)$.
The important novelty here is that our semantic model extends this
commutativity theorem to higher-order and recursive types, even if
they do not fit easily into traditional measure theory.

\subsection{Introduction to Statistical Probabilistic Programming}
\begin{figure}
\[\begin{array}[t]{@{}ccc@{}}
\begin{array}{@{}l@{}}
\letin{a}{\cnst{\textit{normal-rng}}\,(\cnst{0},\cnst{2})}{}
\\\quad{\tscore[](\cnst{\textit{normal-pdf}}\, (\cnst{1.1}\mid a* \cnst{1},\cnst{0.25} ))};
\\\quad{\tscore[](\cnst{\textit{normal-pdf}}\,(\cnst{1.9}\mid a* \cnst{2}, \cnst{0.25}))};
\\\quad{\tscore[](\cnst{\textit{normal-pdf}}\,(\cnst{2.7}\mid a* \cnst{3}, \cnst{0.25} ))};
\\ \quad
a
\end{array}
&
\begin{array}{@{}r@{}l@{}}
  \text{Prior: }a&{}\sim N(0,2)\\
  \multicolumn2{@{}l@{}}{\text{Observations:}}\\
1.1&{}\sim N(1a,\tfrac14)\\
1.9&{}\sim N(2a,\tfrac14)\\
2.7&{}\sim N(3a,\tfrac14)
\end{array}
&
\hspace{-1cm}
\begin{tabular}{@{}c@{}}
  {\begingroup
  \makeatletter
  \providecommand\color[2][]{
    \GenericError{(gnuplot) \space\space\space\@spaces}{
      Package color not loaded in conjunction with
      terminal option `colourtext'
    }{See the gnuplot documentation for explanation.
    }{Either use 'blacktext' in gnuplot or load the package
      color.sty in LaTeX.}
    \renewcommand\color[2][]{}
  }
  \providecommand\includegraphics[2][]{
    \GenericError{(gnuplot) \space\space\space\@spaces}{
      Package graphicx or graphics not loaded
    }{See the gnuplot documentation for explanation.
    }{The gnuplot epslatex terminal needs graphicx.sty or graphics.sty.}
    \renewcommand\includegraphics[2][]{}
  }
  \providecommand\rotatebox[2]{#2}
  \@ifundefined{ifGPcolor}{
    \newif\ifGPcolor
    \GPcolorfalse
  }{}
  \@ifundefined{ifGPblacktext}{
    \newif\ifGPblacktext
    \GPblacktexttrue
  }{}
  \let\gplgaddtomacro\g@addto@macro
  \gdef\gplbacktext{}
  \gdef\gplfronttext{}
  \makeatother
  \ifGPblacktext
    \def\colorrgb#1{}
    \def\colorgray#1{}
  \else
    \ifGPcolor
      \def\colorrgb#1{\color[rgb]{#1}}
      \def\colorgray#1{\color[gray]{#1}}
      \expandafter\def\csname LTw\endcsname{\color{white}}
      \expandafter\def\csname LTb\endcsname{\color{black}}
      \expandafter\def\csname LTa\endcsname{\color{black}}
      \expandafter\def\csname LT0\endcsname{\color[rgb]{1,0,0}}
      \expandafter\def\csname LT1\endcsname{\color[rgb]{0,1,0}}
      \expandafter\def\csname LT2\endcsname{\color[rgb]{0,0,1}}
      \expandafter\def\csname LT3\endcsname{\color[rgb]{1,0,1}}
      \expandafter\def\csname LT4\endcsname{\color[rgb]{0,1,1}}
      \expandafter\def\csname LT5\endcsname{\color[rgb]{1,1,0}}
      \expandafter\def\csname LT6\endcsname{\color[rgb]{0,0,0}}
      \expandafter\def\csname LT7\endcsname{\color[rgb]{1,0.3,0}}
      \expandafter\def\csname LT8\endcsname{\color[rgb]{0.5,0.5,0.5}}
    \else
      \def\colorrgb#1{\color{black}}
      \def\colorgray#1{\color[gray]{#1}}
      \expandafter\def\csname LTw\endcsname{\color{white}}
      \expandafter\def\csname LTb\endcsname{\color{black}}
      \expandafter\def\csname LTa\endcsname{\color{black}}
      \expandafter\def\csname LT0\endcsname{\color{black}}
      \expandafter\def\csname LT1\endcsname{\color{black}}
      \expandafter\def\csname LT2\endcsname{\color{black}}
      \expandafter\def\csname LT3\endcsname{\color{black}}
      \expandafter\def\csname LT4\endcsname{\color{black}}
      \expandafter\def\csname LT5\endcsname{\color{black}}
      \expandafter\def\csname LT6\endcsname{\color{black}}
      \expandafter\def\csname LT7\endcsname{\color{black}}
      \expandafter\def\csname LT8\endcsname{\color{black}}
    \fi
  \fi
    \setlength{\unitlength}{0.0500bp}
    \ifx\gptboxheight\undefined
      \newlength{\gptboxheight}
      \newlength{\gptboxwidth}
      \newsavebox{\gptboxtext}
    \fi
    \setlength{\fboxrule}{0.5pt}
    \setlength{\fboxsep}{1pt}
\begin{picture}(3168.00,1728.00)
    \gplgaddtomacro\gplbacktext{
      \csname LTb\endcsname
      \put(708,440){\makebox(0,0)[r]{\strut{}$0$}}
      \put(708,611){\makebox(0,0)[r]{\strut{}$0.1$}}
      \put(708,781){\makebox(0,0)[r]{\strut{}$0.2$}}
      \put(708,952){\makebox(0,0)[r]{\strut{}$0.3$}}
      \put(708,1122){\makebox(0,0)[r]{\strut{}$0.4$}}
      \put(708,1293){\makebox(0,0)[r]{\strut{}$0.5$}}
      \put(708,1463){\makebox(0,0)[r]{\strut{}$0.6$}}
      \put(814,330){\makebox(0,0){\strut{}$-2$}}
      \put(1303,330){\makebox(0,0){\strut{}$-1$}}
      \put(1793,330){\makebox(0,0){\strut{}$0$}}
      \put(2282,330){\makebox(0,0){\strut{}$1$}}
      \put(2771,330){\makebox(0,0){\strut{}$2$}}
    }
    \gplgaddtomacro\gplfronttext{
      \csname LTb\endcsname
      \put(176,951){\rotatebox{-270}{\makebox(0,0){\strut{}\begin{tabular}{@{}c@{}}probability\\density\end{tabular}}}}
      \put(1792,154){\makebox(0,0){\strut{}$a$}}
      \csname LTb\endcsname
      \put(1405,1321){\makebox(0,0)[l]{\strut{}prior}}
      \csname LTb\endcsname
      \put(1405,1164){\makebox(0,0)[l]{\strut{}posterior}}
    }
    \gplbacktext
    \put(0,0){\includegraphics{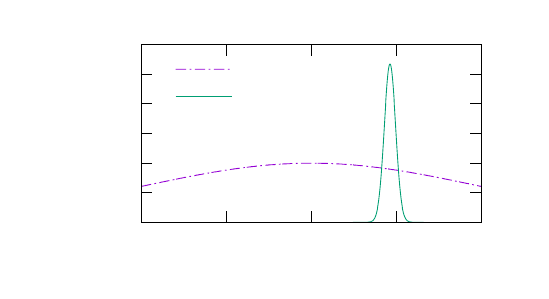}}
    \gplfronttext
  \end{picture}
\endgroup
 }
\end{tabular}
\\
(a)&(b)&(c) \end{array}\]
\caption{Bayesian linear regression as (a)~a first-order probabilistic
  program, (b) an informal specification, and (c) a plot of the prior
  and posterior distributions.  Here
  $N(\mu,\sigma)$ is the normal (Gaussian) distribution with density
  $\textit{normal-pdf}\,(x\mid\mu,\sigma)=
  {(2\pi
    \sigma^2)^{-{\frac 12}}e^{\frac{-(x-\mu)^2}{2\sigma^2}}}$ and
  random number generator
  $\textit{normal-rng}\,(\mu,\sigma)$.
  \figlabel{regression}
}
\end{figure}

We introduce statistical probabilistic programming through a simple
example of a regression problem, in \figref{regression}.  The problem
is: supposing that there is a linear function $x\mapsto ax$ and three
noisy measurements $(1, 1.1)$, $(2, 1.9)$ and $(3, 2.7)$ of it with
postulated noise scale $0.25$, find a posterior distribution on the
slope~$a$.  As indicated, first-order probabilistic programs can be
thought of as a direct translation of a Bayesian statistical problem.
The probabilistic program has an operational reading in terms of
Monte-Carlo simulation: first use a Gaussian sampler/random number
generator, using $\textit{normal-rng}$, to draw from the prior with
mean $0$ and standard deviation $2$; then weight, using $\tscore[]$,
the resulting samples with respect to the three data points according
to the Gaussian likelihood, using $\textit{normal-pdf}$, assuming
noisy measurements with standard deviation $\tfrac14$.  The resulting
program represents the \emph{unnormalised} posterior distribution,
which can then be passed to an inference algorithm to approximate its
normalisation, e.g. through Monte-Carlo sampling.

\paragraph{A case for recursion: higher-order functions over infinite data structures.}
Many probabilistic programming languages also allow other programming
language features, including recursion.  When looking at a whole
closed program of ground type, these extra features pose little
conceptual problem because the entire program will reduce to
a first-order program, albeit a very large or even infinite one.
However, this paper is not concerned with the problem of interpreting whole
closed programs, but rather interpreting individual aspects of a
program in a compositional way.

For example, consider the program in \figref{grw}, which takes
snapshots of a Gaussian random walk at a stream of times ($t$) to produce a
stream of $(t,y)$ co-ordinate pairs.
\begin{figure}
  \[
  \begin{array}[t]{@{}ccc@{}}
    \begin{array}{@{}l}
      \mathrm{reduce}\,\hfill{\texttt{(* $\mathrm{foldl}$ *)}}
      \\
      \quad (
      \lambda((t,y) :: tys,t').\\
      \qquad(t',\cnst{\textit{normal-rng}}(y,\sqrt{{t'-t}}))\\
      \qquad :: (t,y) :: tys)
      \\\quad[(\cnst{0},\cnst{0})]\\\quad
             [\cnst{1},\cnst{2},\cnst{4},\cnst{5.5},\cnst{6},\cnst{6.25}]
    \end{array}
    &
    \begin{array}{@{}r@{\,}r@{(}r@{}l@{)}}
      y_1&\sim N&  0&,\sqrt{{t_1-0}}\\
      y_2&\sim N&y_1&,\sqrt{{t_2-t_1}}\\
      y_3&\sim N&y_2&,\sqrt{{t_3-t_2}}\\
      \vdots \\
      y_6&\sim N&y_5&,\sqrt{{t_6-t_5}}
    \end{array}
    &
    \begin{tabular}{@{}c@{}}
      \begingroup
  \makeatletter
  \providecommand\color[2][]{
    \GenericError{(gnuplot) \space\space\space\@spaces}{
      Package color not loaded in conjunction with
      terminal option `colourtext'
    }{See the gnuplot documentation for explanation.
    }{Either use 'blacktext' in gnuplot or load the package
      color.sty in LaTeX.}
    \renewcommand\color[2][]{}
  }
  \providecommand\includegraphics[2][]{
    \GenericError{(gnuplot) \space\space\space\@spaces}{
      Package graphicx or graphics not loaded
    }{See the gnuplot documentation for explanation.
    }{The gnuplot epslatex terminal needs graphicx.sty or graphics.sty.}
    \renewcommand\includegraphics[2][]{}
  }
  \providecommand\rotatebox[2]{#2}
  \@ifundefined{ifGPcolor}{
    \newif\ifGPcolor
    \GPcolortrue
  }{}
  \@ifundefined{ifGPblacktext}{
    \newif\ifGPblacktext
    \GPblacktexttrue
  }{}
  \let\gplgaddtomacro\g@addto@macro
  \gdef\gplbacktext{}
  \gdef\gplfronttext{}
  \makeatother
  \ifGPblacktext
    \def\colorrgb#1{}
    \def\colorgray#1{}
  \else
    \ifGPcolor
      \def\colorrgb#1{\color[rgb]{#1}}
      \def\colorgray#1{\color[gray]{#1}}
      \expandafter\def\csname LTw\endcsname{\color{white}}
      \expandafter\def\csname LTb\endcsname{\color{black}}
      \expandafter\def\csname LTa\endcsname{\color{black}}
      \expandafter\def\csname LT0\endcsname{\color[rgb]{1,0,0}}
      \expandafter\def\csname LT1\endcsname{\color[rgb]{0,1,0}}
      \expandafter\def\csname LT2\endcsname{\color[rgb]{0,0,1}}
      \expandafter\def\csname LT3\endcsname{\color[rgb]{1,0,1}}
      \expandafter\def\csname LT4\endcsname{\color[rgb]{0,1,1}}
      \expandafter\def\csname LT5\endcsname{\color[rgb]{1,1,0}}
      \expandafter\def\csname LT6\endcsname{\color[rgb]{0,0,0}}
      \expandafter\def\csname LT7\endcsname{\color[rgb]{1,0.3,0}}
      \expandafter\def\csname LT8\endcsname{\color[rgb]{0.5,0.5,0.5}}
    \else
      \def\colorrgb#1{\color{black}}
      \def\colorgray#1{\color[gray]{#1}}
      \expandafter\def\csname LTw\endcsname{\color{white}}
      \expandafter\def\csname LTb\endcsname{\color{black}}
      \expandafter\def\csname LTa\endcsname{\color{black}}
      \expandafter\def\csname LT0\endcsname{\color{black}}
      \expandafter\def\csname LT1\endcsname{\color{black}}
      \expandafter\def\csname LT2\endcsname{\color{black}}
      \expandafter\def\csname LT3\endcsname{\color{black}}
      \expandafter\def\csname LT4\endcsname{\color{black}}
      \expandafter\def\csname LT5\endcsname{\color{black}}
      \expandafter\def\csname LT6\endcsname{\color{black}}
      \expandafter\def\csname LT7\endcsname{\color{black}}
      \expandafter\def\csname LT8\endcsname{\color{black}}
    \fi
  \fi
    \setlength{\unitlength}{0.0500bp}
    \ifx\gptboxheight\undefined
      \newlength{\gptboxheight}
      \newlength{\gptboxwidth}
      \newsavebox{\gptboxtext}
    \fi
    \setlength{\fboxrule}{0.5pt}
    \setlength{\fboxsep}{1pt}
\begin{picture}(3160.00,1720.00)
    \gplgaddtomacro\gplbacktext{
      \csname LTb\endcsname
      \put(357,595){\makebox(0,0)[r]{\strut{}$-6$}}
      \csname LTb\endcsname
      \put(357,751){\makebox(0,0)[r]{\strut{}$-4$}}
      \csname LTb\endcsname
      \put(357,908){\makebox(0,0)[r]{\strut{}$-2$}}
      \csname LTb\endcsname
      \put(357,1064){\makebox(0,0)[r]{\strut{}$0$}}
      \csname LTb\endcsname
      \put(357,1220){\makebox(0,0)[r]{\strut{}$2$}}
      \csname LTb\endcsname
      \put(357,1377){\makebox(0,0)[r]{\strut{}$4$}}
      \csname LTb\endcsname
      \put(357,1533){\makebox(0,0)[r]{\strut{}$6$}}
      \csname LTb\endcsname
      \put(459,409){\makebox(0,0){\strut{}$0$}}
      \csname LTb\endcsname
      \put(826,409){\makebox(0,0){\strut{}$1$}}
      \csname LTb\endcsname
      \put(1192,409){\makebox(0,0){\strut{}$2$}}
      \csname LTb\endcsname
      \put(1559,409){\makebox(0,0){\strut{}$3$}}
      \csname LTb\endcsname
      \put(1926,409){\makebox(0,0){\strut{}$4$}}
      \csname LTb\endcsname
      \put(2293,409){\makebox(0,0){\strut{}$5$}}
      \csname LTb\endcsname
      \put(2659,409){\makebox(0,0){\strut{}$6$}}
    }
    \gplgaddtomacro\gplfronttext{
      \csname LTb\endcsname
      \put(2955,1064){\makebox(0,0)[l]{\strut{}$y$}}
      \csname LTb\endcsname
      \put(1605,130){\makebox(0,0){\strut{}$t$}}
    }
    \gplbacktext
    \put(0,0){\includegraphics{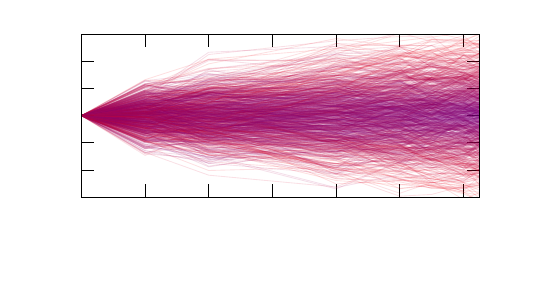}}
    \gplfronttext
  \end{picture}
\endgroup
     \end{tabular}
    \\
    (a)&(b)&(c)
  \end{array}
\]
\caption{A Gaussian random walk as (a)~a higher-order recursive
  program and (b)~an informal specification, with~(c) some
  samples. The function $\mathrm{reduce}$ is sometimes called
  $\mathrm{fold\ left}$.  \figlabel{grw}}
\end{figure}
The meaning of the whole program is clear, and can be reduced to the
first-order statistical model~(b).  But supposing this appears as part
of a bigger model, we would like to understand each part separately.
What is the mathematical meaning of $\mathrm{reduce}$ here?  It takes
a parameterized random operation, an initial value, and a stream, and
produces a random stream. By providing mathematical objects that
represent these recursive concepts, we can understand
$\mathrm{reduce}$ as a first-class construct and reason about it.

\paragraph{A case for recursion: untyped programs.}
Rather than distinguish between ground types and higher-order
recursive types, an alternative approach is to combine the full \emph{untyped} lambda calculus
with the constructions of the simple statistical programming language
(\figref{untyped}).  Many probabilistic programming languages
take this approach (
Church~\cite{goodman_uai_2008},
Anglican~\cite{wood-aistats-2014},
WebPPL~\cite{Goodman2014},
Venture~\cite{Mansinghka-venture14}
).
Recall that we can express untyped calculi using recursive
types, using a single recursive type like
$\Lambda=(\terminal \vor \Lambda
* \Lambda \vor \cdots \vor \Lambda\to \Lambda)$.
Thus a language with recursive types can be thought of as
generalising this untyped situation.

\subsection{Summary of Semantics in $\omega$-Quasi Borel Spaces}
\label{sec:summary-of-wqbs}
The usual method for interpreting a programming language is as follows:
\begin{itemize}
\item types denote spaces
(in an untyped
  language, there is just one universal space);
\item closed programs denote points in a space; and
\item program phrases denote functions
  assigning a point to every valuation of their free variables.
\end{itemize}
Probabilistic programs, on the other hand, vary this by saying
\begin{itemize}
\item closed programs denote \emph{measures} (or distributions) on a space; and
\item program phrases denote kernels between spaces.
\end{itemize}
Recursive probabilistic programming has a tension between what a space
is and what it is used for:
\begin{itemize}
\item In traditional domain theory, a type denotes a
  topological space or cpo in which continuity and convergence model
  recursion using fixed points.
\item For first-order probabilistic programs, a type denotes
  a topological or measurable space whose purpose is to support
  measures and expectations.
\end{itemize}
It is tempting to use a single topological  (or cpo) structure to interpret
both the probability and recursion.
By contrast with existing mainstream approaches \cite{jones1989probabilistic}, however,
we choose to keep both structures separate but compatible, through the following
observation.
In probability theory it is widely acknowledged that topological and
measurable structure are only a precursor to the notion of random
element.  Recall that a random element in a set $P$ is a function
$
\alpha:\Omega\to P
$
where $\Omega$ is
some space of random seeds,
e.g. $\Omega=\RR$ with a Lebesgue measure.
When $P=\RR$, one would typically ask that $\alpha$ be measurable
so that there is an expected (average, mean) value of $\alpha$.
But random elements are relevant beyond their expectation.  A measure
on $P$ can be understood as a random element modulo being `equal in
distribution', but it is helpful to also keep the distinction between
random element and measure.
This leads us to the following definition.
\begin{definition*}[\ref{def:wqbs}]
An \emph{\wqbs} comprises a set $P$ together with the following structure:
\begin{itemize}
\item a partial order $\leq$ on $P$ such that limits of
  $\omega$-chains exist: to model recursion;
\item a set $M$ of functions $\alpha : \RR\to P$:  these are thought of as
   \emph{random elements}, to interpret probability;
\end{itemize}
all subject to some compatibility conditions.
\end{definition*}

For example:
\begin{enumerate}
\item Let $P$ be the set of subsets of $\NN$. This is a space to
  interpret deterministic programs with a natural number argument:
  a subset $X \subset \NN$ represents the program that returns for
  each member in $X$.  The order $\leq$ is the inclusion order, so
  program $X$ is below $Y$ if $X$ diverges whenever $Y$ diverges.  The
  random elements $M$ are the functions $\alpha\colon \RR\to P$ such
  that $\inv\alpha[\{S~|~n\in S\}]$ is (Borel) measurable for all
  $n\in \NN$.
\item Let $\reals$ be the set of real \emph{values}. When thought
  of as values, we use the discrete order. The random elements $M$ are
  the measurable functions $\reals\to\reals$.
\item Let $\lawson = [0, \infty]$ be the non-negative extended
  reals. Its elements stand for computation \emph{weights} with the
  linear order. The random elements $M$ are the measurable functions
  $\reals\to[0,\infty]$.
\end{enumerate}

The semantics in \wqbs es supports higher-order functions.
Moreover, we can interpret recursive types by using the recipe of
\citepos*{fiore1994axiomatisation}{axiomatic domain theory}.
That is to say:
\begin{theorem*}[\corollaryref{wqbs-structure}, \subsecref{bilimit-compact}]
The category $\wQbs$ has products, sums, function spaces, and a
bilimit compact expansion (sufficient structure to interpret recursive types).
\end{theorem*}

\subsection{A Probabilistic Powerdomain}
With higher-order functions and recursive types dealt with, the
remaining ingredient is measures (and probabilistic programs that
generate measures) as first-class constructions.
To this end, for every \wqbs{} $P$ we will associate an
\wqbs{} $T(P)$ of measures on $P$.
Following \citet{Moggi89}, we turn $T$ into a monad encapsulating the
probabilistic aspects of the programming language.

Recall that an \wqbs{} $P$ comes with a set $M$ of
random elements, viz.~functions $\alpha\colon \RR\to P$. Because a
statistical probabilistic program naturally describes an unnormalized
posterior measure, we consider the basic space $\RR$ with the full Lebesgue
measure, which has infinite total measure.
To consider finite measures we consider partial random elements, which
are given by pairs $(\alpha,D)$ where $\alpha\in M$ and $D\subseteq \RR$ is
Borel.
Given any partial random element $(\alpha,D)$ and any morphism
$f\colon P\to \lawson$ ($\lawson=[0,\infty]$), the composite $f\compose \alpha$ is measurable,
so we can find an expectation for $f$:
\newcommand{\Expect}{\mathbb E}
\begin{equation}\label{eqn:expop}
\Expect_{(\alpha,D)}[f]:=\int_D f (\alpha(x))\ \dif x.\end{equation} We
say that
$(\alpha,D)$ and $(\alpha',D')$
are \emph{equivalent} when they give the same expectation operator:
\begin{equation}\label{eqn:equalexp}
\text{for all $f\colon P\to \lawson$, \quad $\int_D f
  (\alpha(x))\ \dif x=\int_{D'} f (\alpha'(x))\ \dif
  x\text.$}\end{equation}
\begin{definition*}
A measure on an \wqbs\ $P$ is an equivalence class of a partial
random element $(\alpha, D)$, modulo the equivalence
relation~\eqref{eqn:equalexp}.
Equivalently, a measure is a morphism $\lawson^P\to
\lawson$ of the form $\Expect_{(\alpha,D)}$ for some partial
random element~$(\alpha,D)$.
\end{definition*}

Here we run into a technical problem: the set of all measures has a
natural (pointwise) partial order structure but this set might not
be closed under
suprema of $\omega$-chains.
On the other hand, we know that
$J(P) := \lawson^{(\lawson^P)}$ is always closed, because
$J$ is the continuation monad which makes sense in any category with
function spaces.
Thus we take the closure $T(P)$ of the set of measures in $J(P)$ as our
space of measures. In other words, $T(P)\subseteq J(P)$ contains those
expectation operators that arise as iterated suprema of \wchain{}s on
$P$.

Our approach follows existing continuation-passing-style
techniques~\citep[e.g.][]{keimel2009predicate,kiselyov2009embedded,olmedo2016reasoning}.
CPS semantics is analogous to working with the full continuation monad
$J$ or a fragment of it. This fragment must be chosen carefully, or
else the commutativity property fails in the model. Indeed $J$ also
has constants such as $\mathrm{exit}_r=\lambda k.r\in J(P)$ violating
the commutativity equation: put $t_1=\mathrm{exit}_1$,
$t_2=\mathrm{exit}_2$. As we have the commutativity property, these
constants lie outside our monad $T(P)$, hence are not definable in the
language.

\paragraph{Aside on the Jung-Tix problem.}
A long standing problem in traditional domain theory is to find a
category of Scott domains that is closed under function spaces and a
commutative probabilistic powerdomain \cite{JUNG199870}.  This remains
an open problem. We side-step this problem by using \wqbs es instead
of Scott domains. They inherit many of the properties and intuitions
of $\omega$-cpos, are closed under function spaces, and support a
commutative probabilistic powerdomain.
We summarize further work in this direction in \secref{conclusion}.

\subsection{Summary}
We have provided a domain theory for recursion in
statistical probabilistic programming.
The main contributions of this work are the following
novel constructions:
\begin{enumerate}
\item a Cartesian closed category of (pre)-domains~(\secref{wqbs}), that admits the solution of recursive domain equations (\secref{axiomatic-domain-theory}),
\item a commutative probabilistic power-domain~(\secref{monad});
\item an adequate denotational model  (\secref{semantics})
for probabilistic
  programming with recursive types (\secref{calculus}),
  and in consequence
  \begin{itemize}
  \item an adequate denotational model for a higher-order
    language with sampling from continuous distributions, term
    recursion and soft constraints (\subsecref{spcf});
  \item an adequate denotational model for untyped
    probabilistic programming (\subsecref{idealised-church}).
  \end{itemize}
\end{enumerate}

\newcommand{\hide}[1]{}\hide{
The rest of this paper proceeds as follows.
\secref{calculus}
presents the syntax of our recursively typed, simply typed, and
untyped calculi.
\secref{wqbs}
presents our category of pre-domains $\wQbs$ and its basic
properties.
\secref{monad} constructs the
probabilistic powerdomain and establishes its
properties.
\secref{axiomatic-domain-theory} outlines the axiomatic
development of $\wQbs$'s domain theory.
\secref{semantics}
gives semantics to our calculi: operational and denotational semantics
related via an adequacy result using a relational semantics.
\secref{characterising-wqbs} presents equivalent
characterisations of $\wQbs$ and establishes more advanced categorical
properties.
 \secref{conclusion} discusses related work and concludes.
}

\section{Calculi for statistical probabilistic programming}
\seclabel{calculus} We consider three call-by-value calculi for
statistical probabilistic programming. The main calculus, Statistical
FPC (SFPC) is a statistical variant of
\citepos*{fiore1994axiomatisation}{Fixed-Point Calculus (FPC)}.  SFPC
has sum, product, function, and recursive types, as well as a ground
type $\Real$ of real numbers, constants $\cnst{f}$ for all measurable
functions $f:\reals^n\to\reals$, a construct $\ifz{-}{-}{-}$ for
testing real numbers for zero, a construct $\tsample[]$ for drawing a
random number using the uniform (Lebesgue) measure $\uniform{[0,1]}$
on $[0,1]$ and a construct $\tscore[]$ for reweighting program traces
(to implement soft constraints).  We express the other two calculi as
fragments of SFPC. The first, Idealised Church, is an untyped
$\lambda$-calculus for statistical probabilistic programming based on
the Church modelling language \cite{Wingate2011}. The second, the
Call-by-Value Statistical PCF (CBV SPCF), is a call-by-value variant
of \citepos{plotkin:pcf} and \citepos{scott:pcf} simply typed
$\lambda$-calculus with higher-order term recursion, extended with
statistical primitives.

\subsection{Preliminaries: Borel Measurability}
We need the following fundamentals of measure theory. The \emph{Borel
  subsets} of the real line $\reals$ are given inductively by taking
every interval $[a,b]$ to be a Borel subset, and closing under
complements and countable unions. More generally, for every natural
number $n$, the Borel subsets of $\reals^n$ are given inductively by taking
every $n$-dimensional box $[a_1, b_1]\times \cdots \times [a_n, b_n]$
to be a Borel subset and closing under complements and countable
unions. A (partial) function $f : \reals^n \pto \reals$ is
\emph{Borel-measurable} when its inverse image maps every Borel subset
$B \subset \reals$ to a Borel subset $\inv f[B] \subset \reals^n$. The
set of Borel-measurable functions contains, for example, all the
elementary functions.

A \emph{measure} $\mu$ on $\reals^n$ is an assignment of
possibly-infinite, non-negative real values $\mu(B)$ to every Borel
subset $B \subset \reals^n$, assigning $0$ to the empty set
$\mu(\emptyset) = 0$, linear on disjoint unions $\mu(B_1 \uplus B_2) =
\mu(B_1) + \mu(B_2)$, and continuous with respect to countably
increasing sequences: if for every $n$, $B_n \subset B_{n+1}$, then
$\mu(\Union_{n=0}^\infty B_n) = \lim_{n \to \infty} \mu(B_n)$. The
\emph{Lebesgue measure} $\lebesgue$ is the unique measure on $\reals$
assigning to each interval its length $\lebesgue([a, b]) = b-a$. A
\emph{probability measure} on $\reals^n$ is a measure $\mu$ whose
\emph{total measure} $\mu(\reals^n)$ is $1$. The \emph{uniform
  probability measure} $\uniform{[a,b]}$ on an interval $[a,b]$
assigns to each Borel set $B$ the relative Lebesgue measure it
occupies in the interval $[a,b]$: $\uniform{[a,b]}(B) \definedby
\tfrac1{b-a}\lebesgue(B \intersect [a,b])$. Measurable functions $f :
\reals^n \to \reals^m$ let us transport measures $\mu$ on $\reals^n$
to their \emph{push-forward} measure $f_*\mu$ on $\reals^m$, by
setting $f_*\mu(B) \definedby \mu(\inv f[B])$.

We think of whole statistical probabilistic programs of real type as a
formalism for describing measures. To work compositionally, we need
the following analogous concept for program fragments, i.e., terms
with unbound variables. A \emph{probability kernel} $k$ from
$\reals^n$ to $\reals$, written as $k : \reals^n \leadsto \reals$ is a
function assigning to every $\vec x \in \reals^n$ a probability
measure $k(\vec x, -)$ on $\reals$, such that, for every Borel set $B$, the
function $k(-, B) : \reals^n \to \reals$ is measurable. We will use
the following key result about probability kernels, the
\emph{randomisation lemma}, which says that a straightforward random
number generator $\uniform{[0,1]}$ suffices to implement any of them.
\begin{lemma}[{\cite[Lemma~3.22]{kallenberg2006foundations}}]
\lemmalabel{randomize}
For every probability kernel $k : \reals^n \leadsto\reals$, there is a
measurable function $\randomize_k:\reals^{n+1}\to\reals$, such that
$\randomize_k(x_1,\ldots,x_n,-)_*\uniform{[0,1]}=k(x_1,\ldots,x_n)$.
\end{lemma}

\subsection{SFPC: Bayesian Statistical Modelling with Recursive Types}
\subsubsection*{Syntax.}
As recursive types contain type-variables, we use a kind system to
ensure types are well-formed.  \figref{kinds}~(top left) presents the
kinds of our calculus, and \figref{types}~(bottom left) presents the
types of SFPC. We include type variables, taken from a countable set
ranged over by $\tvar, \tvar[2], \tvar[3]$.  We include simple types:
unit, product, function, and variant types.  Variant types use
constructor labels taken from a countable set ranged over by $\Cns,
\Cns_1, \Cns_2, \ldots$.  In our abstract syntax, variant types are
\emph{dictionaries}, partial functions with a finite domain, from the
set of constructor labels to the set of types.  The recursive type
former $\rec{\tvar}{\ty}$ binds $\tvar$ in $\ty$.  In our abstract
syntax, term variable contexts $\ctx$ are dictionaries from the
countable set of variables, ranged over by $\var, \var[2], \var[3],
\ldots$, to the set of types.

\begin{figure}
  \begin{tabular}[t]{@{}ccc@{}}
  \begin{syntax}
    \kind && \gdefinedby{}& \syncat{kinds}\\
    &&\typ & \synname{type}\\
    &\gor& \context & \synname{context}    \\
    \ty&,&\ty[2],\ty[3] \gdefinedby            &\syncat{types}   \\
    &    & \tvar     &\synname{variable}\\
    &\gor & \Real     &\synname{reals}\\
    &\gor& \Unit      &  \synname{unit}   \\
    &\gor& \ty \t* \ty[2]  &  \synname{product}      \\
    &\gor& \ty \To \ty[2]                 &  \synname{function}     \\
    &\gor& \begin{array}[t]{@{}r@{\,}l@{}}\Variant{&
      \Inj*{\Cns_1}{\ty_1}
      \\\vor& \ldots
      \\[2pt]\vor&
      \Inj*{\Cns_n}{\ty_n}
      }
    \end{array}
      &  \synname{variant}          \\
    &\gor& \rec{\tvar}{\ty}                        &  \mathrlap{\synname{iso-recursive}}\\
  \end{syntax}
  &
  \begin{syntax}
    \trm&,& \trm[2], \trm[3] \gdefinedby & \syncat{terms}     \\
    &    & \var                   &  \synname{term variable} \\
    &\gor& \cnst{f}(\trm_1,\ldots,\trm_n)  &  \synname{primitive}     \\
    &\gor& \mathrlap{\sifz{\trm}{\trm[2]}{\trm[3]}} & \synname{conditional}\\
    &\gor& \tsample & \synname{sampling}\\
    &\gor& \tscore{\trm} & \synname{conditioning}\\
    &&&\syncat{constructors:}\\
    &\gor& \tUnit                          &  \synname{unit}    \\
    &\gor& \tpair \trm {\trm[2]}          &  \synname{pairing} \\
    &\gor&\tInj\ty\Cns\trm               & \synname{variant}        \\
    &\gor& \troll\ty(\trm)                     & \synname{iso-recursive} \\
  \end{syntax}
  &
  {\begin{syntax}
      &&&\syncat{function:}\\
        &\gor& \expfun \var\ty \trm              &  \synname{abstraction} \\
        &\gor& \trm\ \trm[2]                  &  \synname{application}\\
          &&&\syncat{pattern matching:}\\
      &\gor& \mathrlap{
        \sumatch
            {\trm}
            {\trm[2]}}
      & \synname{unit} \\
      &\gor& \spmatch
        {\trm}
        {\var[1]}{\var[2]}
        {\trm[2]}
        & \synname{product} \\
        &\gor& \svmatch {\trm  }
        {
          \Inj*{\Cns_1}{\var_1}\To{\trm[2]_1}
          \vor\cdots  \\\,\vor\!
          \Inj*{\Cns_n}{\var_n}\To{\trm[2]_n}
        }  \hspace*{-17pt}         & \synname{variant}\\
        &\gor&
        \srmatch
            {\trm}
            \var{\trm[2]}
            & \synname{recursive} \\
  \end{syntax}}
  \\
  \multicolumn{3}{@{}c@{}}{
    \(
    \ctx \definedby{} x_1 : \ty_1, \ldots, x_n : \ty_n
  \syncat{variable contexts}
    \)
  }
\end{tabular}
   \caption{SFPC kinds and types (left) and terms (right)}
  \figlabel{kinds}\figlabel{types}\figlabel{terms}
\end{figure}

We desugar stand-alone labels in a variant type $\Variant{\cdots \vor
  \Inj\Cns \vor \cdots}$ to the unit type $\Variant{\cdots \vor
  \Inj\Cns\Unit \vor \cdots}$. We also desugar top-level-like
recursive type declarations $\ty \definedby \subst{\ty[2]}{\sfor
  \tvar\ty}$ to $\ty \definedby \rec\tvar\ty[2]$.

\begin{example}\examplelabel{types}
  The type of booleans is \(
  \boolty \definedby \Variant{\tru \vor \fls}
  \).
  The type of natural numbers is \(
  \naturals
  \definedby
                  \Variant{\Inj[]{Zero}
                  \vor     \Inj{Succ}\naturals}
  \)
  desugaring to \(
  \naturals
  \definedby \rec \tvar
                  \Variant{\Inj[]{Zero}
                  \vor     \Inj{Succ}\tvar}
  \).
  The type of $\ty$-lists is
  \(
  \List\ty
  \definedby
  \Variant{
    \Inj[]{Nil}
    \vor
    \Inj{Cons}{\ty\t* \List\ty}
  }
  \), desugaring to
  \(
  \List\ty
  \definedby
  \rec{\tvar}\Variant{
    \Inj{Nil}\Unit
    \vor
    \Inj{Cons}{\ty\t* \tvar}
  }
  \).
  The type of (infinite) stochastic processes in $\ty$ is
  \(
  \Stoch\ty
  \definedby
  \Unit\To(\ty\t* \Stoch\ty)
  \), desugaring to
  \(
  \Stoch\ty
  \definedby
  \rec{\tvar}
  \Unit\To(\ty\t* \tvar)
  \).
  The type of untyped $\lambda$-terms with values in $\ty$ is
  \(
  \UntypedL\ty
  \definedby
  \Variant{
    \Inj{Val}{\ty}
    \vor
    \Inj{Fun}{(\UntypedL\ty\To \UntypedL\ty)}
  }
  \), desugaring to
  \(
  \UntypedL\ty
  \definedby
  \rec{\tvar}\Variant{
    \Inj{Val}{\ty}
    \vor
    \Inj{Fun}{(\tvar\To \tvar)}
  }
  \).
\end{example}

\figref{terms}~(right) presents the terms of SFPC. Value variables are
taken from a countable set ranged over by $x$, $y$, $z$, $\ldots$. We
include primitive function constants $\cnst f(t_1, \ldots, t_n)$ for
every measurable (partial) function $f : \reals^n \pto \reals$. The
conditional construct tests whether its argument of type $\Real$
evaluates to $0$. We include an effect $\tsample[]$ for sampling a real
number uniformly from the interval $[0,1]$, for defining the prior
distribution of the program. We also include an effect $\tscore r$ for
reweighting the posterior distribution by the non-negative factor
$\abs r \in [0, \infty)$. We include standard constructors and
  pattern-matching constructs for the simple types, and standard
  function abstraction and application constructs. Finally, we include
  the standard \emph{iso-recursive} constructors and pattern matching,
  which require an explicit rolling and unrolling of the recursive
  definition such as $\troll\naturals{(\Inj[]{Zero}\tUnit)}$.  Variant
  and iso-recursive constructor terms as well as function abstraction
  annotate their binding occurrences with the appropriate closed type
  $\ty$ to ensure unique typing.

\begin{figure}
  \begin{tabular}{cc}
  {\begin{sugar}
  \_                            & \text{fresh variable}                          \\
  \expletin\var\ty\trm{\trm[2]} & (\expfun \var\ty\trm[2])\,\trm                 \\
  \letrecin{(\var:\ty[3])}{\trm}{\trm[2]} & (\expfun{\var}{\ty[3]}{\trm[2]})
  (\rec{\var:\ty[3]}\trm)\\
  \trm;\trm[2] & \umatch{\trm}{\trm[2]}\\
  \unroll\trm                       & \rmatch\trm{\var}{\var} \\
\end{sugar}
}&{
   \begin{leansugar}[]
     \multicolumn{2}{c}{~}\\
     \rec{\var:\ty}\trm &
     \begin{array}[t]{@{}l@{}}
       \expletin {body}{\ty[2]\To\ty)}{
       \expfun{\var[2]}{\ty[2]}
       \\\mspace{10mu}
       \expletin {\var}\ty{\expfun{\var[3]}{\ty_1}\unroll{\var[2]}\, \var[2]}
       \trm\\
       }{
       body\, (\troll{\ty[2]}{body})
       }
     \end{array}
\\
\multicolumn2{@{\hphantom\bullet}l@{}}{
  \text{where $\ty \defeq \ty_1\To\ty_2$ and $\ty[2] \defeq \rec\tvar\tvar\To\ty$}
}
\\
  \bot_{\ty[3]} & \rec{\var:\ty[3]}{\var}
\end{leansugar}
}
\end{tabular}
   \caption{Term-level syntactic sugar}\figlabel{sugar}
\end{figure}

To aid readability, we use the standard syntactic sugar of
\figref{sugar}, e.g. $\letin\var \trm{\trm[2]}$ for $(\fun
\var\trm)\trm[2]$, $\rec{\var:\ty\To\ty[2]}{\trm}$ for the usual encoding of
term level recursion using type level
recursion~\cite{fiore:thesis,abadi1996syntactic}.  When dealing with a
recursive variant type $\ty[2] = \rec\tvar\Variant{\ldots \vor
  \Inj\ell \ty \vor \ldots}$, we write $\Inj\ell \trm$ for the more
cumbersome constructor $\troll{\ty[2]}(\tInj {\Variant{\ldots \vor
    \Inj\ell \ty \vor \ldots}}\ell \trm)$ as long as $\ty[2]$ is clear
from the context.  We can encode constructs $\cnst{k}$ in our calculus
for drawing from arbitrary probability kernels $k:\reals^n\leadsto
\reals$. Using the Randomisation \lemmaref{randomize}, first find the
appropriate measurable function $\randomize_k : \reals^{n+1} \to
\reals$ and express $\cnst{k}(\trm_1,\ldots,\trm_n)$ by
$\cnst{\randomize_k}(\trm_1,\ldots,\trm_n,\tsample[])$.

\begin{example}\examplelabel{terms}
  For $\Stoch \Real =\rec{\tvar} \Unit\To(\Real\t* \tvar)$,
  define $ \mathrm{draw} \definedby \expfun{\var}{\Stoch
    \Real}{\unroll(\var)\, \tUnit} $, which draws a value from the
  process and moves the process to the corresponding new state.  As an
  example stochastic process, writing
  $normal\_rng:\reals\times\reals\leadsto \reals$ for a Gaussian
  probability kernel taking the mean and standard deviation as
  arguments, we define an example Gaussian random walk
  $\mathrm{RW}\,\cnst{\sigma}\, \cnst{\mu}$ with initial position
  $\mu$ and standard deviation $\sigma$ by setting:
  \[
  \mathrm{RW} =
  \expfun{\var}{\Real}{ \rec{\var[2]:\Real\to
      \Stoch\Real}{\expfun{\var[3]}{\Real}{\troll{\Stoch \Real}(
        \expfun{\_}{\Unit}{\tpair{\var[3]}{\var[2](\cnst{normal\_rng}(\var[3],\var))}})}}}
  \]
\end{example}

\subsubsection*{Kind and Type Systems}
\begin{figure}
  \[
\begin{array}{llll}
  \inferrule{
      ~
  }{
    \Dkinf \tvar
  }(\tvar \in \Tyctx)

&
  \inferrule{
    ~
  }{
    \Dkinf{\Real}
  }
&
  \inferrule{
    ~
  }{
    \Dkinf\Unit
  }
&
  \inferrule{
    \Dkinf{\ty}
    \\
    \Dkinf{\ty[2]}
  }{
    \Dkinf{\ty\t*\ty[2]}
  }

  \end{array}
  \]
  \[
  \begin{array}{llll}
    \inferrule{
  	\mbox{$
	\begin{array}{l}
    \Dkinf \ty
    \\
    \Dkinf{\ty[2]}
	\end{array}$}
  }{
    \Dkinf{\ty \To \ty[2]}
  }
&
      \inferrule{
	\mbox{$
	\begin{array}{l}
    \text{for all $1 \leq i \leq n$: }\\
    \Dkinf {\ty_i}
	\end{array}$
	}
  }{
    \Dkinf{\Variant{
                \Inj*{\Cns_1}{\ty_1}
                \vor \ldots \vor
                \Inj*{\Cns_n}{\ty_n}
              }}
  }
  &
  \inferrule{
    \Dkinf[, \tvar]{\ty}
  }{
    \Dkinf{\rec \tvar \ty}
  }
&
  \inferrule{
  	\mbox{$
	\begin{array}{l}
    \text{for all $(x : \ty) \in \ctx$: }\\
    \kinf \ty : \typ
	\end{array}$}
  }{
    \kinf{\ctx} : \context
  }
\end{array}
\]
   \caption{SFPC kind system}\figlabel{kind system}
\end{figure}

To ensure the well-formedness of types, which involve type variables,
we use the kind system presented in \figref{kind system}. Each
kinding judgement $\Dkinf\ty$ asserts that a given type $\ty$ is
well-formed in the \emph{type variable context} $\Tyctx$, which is a
finite set of type variables.  The kinding judgements are
standard. All type variables must be bound by the enclosing context,
or by a recursive type binder.  \emph{Variable contexts} $\ctx$ must
assign closed types.  We treat $\alpha$-conversion and capture
avoiding substitution of variables as usual, possessing the
standard structural properties.

\figref{type system} presents the resulting type system, including the
derivable typing judgement for the sugar $\rec{\var:\ty\To\ty[2]}{\trm}$ for
term recursion.  Each typing judgement $\Ginf\trm\ty$ asserts that the
term $\trm$ is well-typed with the well-formed closed type $\kinf \ty
: \typ$ in the variable context $\kinf \ctx : \context$.  The rules
are standard. By design, every term has at most one type in a given
context.

\begin{example}
  The types from \exampleref{types} are well-kinded:
    $ \boolty, \naturals, \List\Real, \Stoch\Real, \UntypedL\Real~:~\typ$.
  The terms from \exampleref{terms} are well-typed:
    $\mathrm{draw} : \Stoch\Real\To(\Real\t* \Stoch\Real)$,
	$\mathrm{RW}:\Real\To\Real\To \Stoch\Real$.
\end{example}

\begin{figure}[tb]
  \[
\begin{array}{lll}
  \inferrule{
    (\var : \ty) \in \ctx
  }{
    \Ginf \var\ty
  }
&
  \inferrule{
    {f:\reals^n\pto\reals}
    \quad
    {
      \text{for all $1 \leq i \leq n$: }
      \Ginf{\trm_i}{\Real}
    }
    }{
      \Ginf{\cnst{f}(\trm_1,\ldots,\trm_n)}{\Real}
    }
&
  \inferrule{
      \Ginf{\trm}{\Real}\\
	  \Ginf{\trm[2]}{\ty}\\
	  \Ginf{\trm[3]}{\ty}
    }{
      \Ginf{\ifz{\trm}{\trm[2]}{\trm[3]}}{\ty}
    }
\end{array}
\]
\[
\begin{array}{lllll}
  \inferrule{
      ~
  }{
      \Ginf{\tsample[]}{\Real}
  }
&
  \inferrule{
      \Ginf{\trm}{\Real}
  }{
      \Ginf{\tscore{\trm}}{\Unit}
  }
&
  \inferrule{
    ~
  }{
    \Ginf\tUnit\Unit
  }
  &
    \inferrule{
    \Ginf {\trm[1]}{\ty[1]}
    \quad
    \Ginf {\trm[2]}{\ty[2]}
  }{
    \Ginf{\tpair{\trm[1]}{\trm[2]}}{\ty[1] \t* \ty[2]}
  }
\end{array}
\]
\[
\begin{array}{lll}
  \inferrule{
    \Ginf\trm{\ty_i}
  }{
    \Ginf{\tInj{\ty}
               {\Cns_i}\trm
         }{\ty}
    }
    (
    \ty = \Variant{
                \Inj*{\Cns_1}{\ty_1}
                \vor \ldots \vor
                \Inj*{\Cns_n}{\ty_n}
      }
    )
    &
  \inferrule{
    \Ginf{\trm}{\subst{\ty[2]}{\sfor{\tvar}{\ty}}}
  }{
    \Ginf{\troll\ty(\trm)}{\ty}
  }(\ty = \rec\tvar\ty[2])
  &
  \inferrule{
    \Ginf[, \var : \ty]{\trm}{\ty[2]}
  }{
    \Ginf{\expfun \var\ty\trm}{\ty\To\ty[2]}
  }
\end{array}
\]
\[
\begin{array}{lll}
  \inferrule{
    \Ginf{\trm}{\ty[2]\To\ty}
    \\
    \Ginf{\trm[2]}{\ty[2]}
  }{
    \Ginf{\trm\, \trm[2]}{\ty}
  }
&
  \inferrule{
    \Ginf{\trm}{\ty[2]\t*\ty[3]}
    \\
    \Ginf[,{\var[1] : \ty[2], \var[2] : \ty[3]}]{\trm[2]}\ty
  }{
    \Ginf{\pmatch
           \trm
           {\var[1]}{\var[2]}
           {\trm[2]}}{\ty}
  }

  \inferrule{
    \Ginf{\trm}{\Unit}
    \\
    \Ginf{\trm[2]}\ty
  }{
    \Ginf{\umatch
           \trm
           {\trm[2]}}{\ty}
  }
\end{array}
\]
\[
\begin{array}{ll}
  \inferrule{
    \Ginf\trm{\Variant{
                \Inj*{\Cns_1}{\ty_1}
                \vor \ldots \vor
                \Inj*{\Cns_n}{\ty_n}}}
    \\
    \text{for each $1 \leq i \leq n$: }
    \Ginf[, \var_i : \ty_i]{\trm[2]_i}{\ty}
  }{
    \Ginf{\vmatch \trm
                {\begin{array}[t]{@{}l@{\,}l@{}l@{}}
                    \Inj*{\Cns_1}{\var_1}\To{\trm[2]_1}
                    \vor\cdots
                    \vor\Inj*{\Cns_n}{\var_n}&\To{\trm[2]_n}
    }}
    \ty
                  \end{array}
  }
\end{array}
\]
\[
\begin{array}{lll}
  \inferrule{
    \Ginf{\trm}{\rec\tvar\ty[2]}
    \\
    \Ginf[,{\var : \subst{\ty[2]}{\sfor\tvar{\rec\tvar\ty[2]}}}]{\trm[2]}\ty
  }{
    \Ginf{\rmatch
           \trm
           {\var}
           {\trm[2]}}{\ty}
  }
  &\shade{
				    \inferrule{
    \Ginf[, \var : {\ty\To \ty[2]}] {\trm}{\ty\To\ty[2]}
  }{
    \Ginf {\rec{\var:\ty\To\ty[2]}{\trm}}{\ty\To\ty[2]}
  }}
  \end{array}
\]
   \caption{SFPC type system, including (in gray) the derivable typing rule for
  term recursion}\figlabel{type system}
\end{figure}

\subsection{Idealised Church: Untyped Statistical Modelling}
\subseclabel{idealised-church}
\begin{figure}
\begin{tabular}{@{}ccc@{}}
  {
    \begin{syntax}
      \trm&,& \trm[2], \trm[3] \gdefinedby & \syncat{terms}                          \\
      &    & \var                   &  \synname{variable} \\
      &\gor& \cnst{k}(\trm_1,\ldots,\trm_n)  &  \synname{kernel}     \\
      &\gor& \ifzun{\trm}{\trm[2]}{\trm[3]} & \synname{conditional}\\
    \end{syntax}
  }&
  {
    \begin{syntax}\\
      &\gor& \tfactor{\trm} & \synname{conditioning}\\
      &\gor& \lambda \var. \trm              &  \synname{abstraction} \\
      &\gor& \trm\ \trm[2]                  &  \synname{application}
    \end{syntax}
  }&{
    \begin{syntax}
      \multicolumn3{@{}l@{}}{\val}
      & \syncat{values}           \\
      &\gdefinedby{}& \var                   &  \synname{variable} \\
      &\gor& \cnst{r}  &  \synname{real number}     \\
      &\gor& \lambda \var. \trm              &  \synname{abstraction}
    \end{syntax}
  }
\end{tabular}
 \caption{Idealised Church, terms and values}\figlabel{untyped}
\end{figure}
In applied probabilistic programming systems, there has been
considerable interest in using an \emph{untyped} $\lambda$-calculus as
the basis for probabilistic programming.  For instance, the Church
modelling language~\cite{Wingate2011}, in idealised form, is an
untyped call-by-value $\lambda$-calculus over the real numbers with
constructs $\cnst{k}(\var_1,\ldots,\var_n)$ for drawing from
probability kernels $k:\reals^n\leadsto \reals$ (including all real
numbers, measurable functions like a Gaussian density
$\cnst{{normal\_pdf}}(y\mid \mu,\sigma)$ and proper kernels like a
Gaussian random number generator $\cnst{{normal\_rng}}(\mu,\sigma)$),
and a construct $\tfactor[]$ for reweighting program traces, to enforce
soft constraints \cite{BorgstromLGS-corr15}.  \figref{untyped}
presents the syntax of Idealised Church, and we desugar
$\letin\var\trm{\trm[2]}$ to mean $(\lambda x. \trm[2])\, \trm$.

\begin{figure}
\[
\begin{array}[t]{@{}ll@{}}
\begin{array}[t]{@{}l@{}}
  \hline
  \text{Values}\;\; \val\\
  \hline
\begin{array}{@{}*3{l@{{}\mapsto{}}l}@{}}
\var  & \var&
\cnst{r}  & \Inj{Val}\cnst{r}&
\lambda \var.\trm  & \Inj{Fun}(\expfun{\var}{\ty}{\trm^\dagger})
 \end{array}
\\
\hline
\text{Terms}\;\; \trm,\trm[2],\trm[3]\\
\hline
\begin{array}{@{}l@{{}\mapsto{}}l@{}}
\val  & \val^\dagger\\
\trm\ \trm[2]  &
(\var[4] := \trm^\dagger;)\bullet \var[4]\ \trm[2]^\dagger
\\
\cnst{k}(\trm_1,\ldots, \trm_n) &
\begin{array}[t]{@{}l@{}}
(;\var_1 := \trm_1^\dagger, \ldots, \var_n := \trm_n^\dagger)\bullet\\
  \Inj{Val}\cnst{k}(\var_1,\ldots,\var_n)
\end{array}
\\
\ifzun{ \trm}{\trm[3]}{\trm[2]} &
\begin{array}[t]{@{}l@{}}
(;\var_1 := \trm^\dagger)\bullet
  \ifz[\\\quad]{\var}{\trm[3]^\dagger}{\trm[2]^\dagger}
\end{array}
\\
\tfactor(\trm)  &
(;\var := \trm)\bullet
\tscore \var; \Inj{Val} \var
\end{array}
\end{array}
&
\begin{array}[t]{@{}l@{}}
  \hline
    \text{Auxiliary sequential unpacking }\\
  (\var[4]_1 := \trm_1, \ldots \var[4]_n := \trm_n; \var_1 := \trm[2]_1,
\ldots, \var_m := \trm[2]_m) \bullet \trm[3]
  \\
  \hline
\begin{array}[t]{@{}l@{{}\mapsto{}}l@{}}
(;) \bullet \trm[3] & \trm[3]
\\
(; \var_1 := \trm[2]_1, \ldots)\bullet \trm[3]
&
\begin{array}[t]{@{}l@{}}
  \vmatch{\unroll \trm[2]_1}{\\
    \begin{array}[t]{@{}l@{}l@{}}
      \hphantom{\vor{}}
      \Inj{Val}{\var_1}&\To(; \ldots)\bullet\trm[3]
      \\\vor
      \Inj{Fun}{\_}&\To\bot_{\ty[2]}
      }
    \end{array}
\end{array}
\\
(\var[4]_1 := \trm_1, \ldots; \ldots)\bullet \trm[3]
&
\begin{array}[t]{@{}l@{}}
  \vmatch{\unroll \trm_1}{\\
    \begin{array}[t]{@{}l@{}l@{}}
    \hphantom{\vor{}}
    \Inj{Fun}{\var[4]_1}&\To(\ldots; \ldots)\bullet\trm[3]
    \\\vor
    \Inj{Val}{\_}&\To\bot_{\ty[2]}
  }
    \end{array}
\end{array}
\end{array}
\end{array}
\end{array}
\]
 \caption{A faithful translation $(-)^\dagger$ of Idealised Church into
  SFPC using the type $\ty[2] := \UntypedL\Real$.}

\figlabel{church-translation}
\end{figure}

Idealised Church arises as a sublanguage of SFPC.  We encode Idealised
Church terms as SFPC terms of the type $\UntypedL\Real \definedby
\Variant{ \Inj{Val}{\ty} \vor \Inj{Fun}{(\UntypedL\Real\To
    \UntypedL\Real)} }$ from \exampleref{types}, using the translation
$(-)^\dagger$ in \figref{church-translation}. The translation uses an
auxiliary SFPC construct $(\ldots;\ldots) \bullet \trm[3]$ for
sequentially evaluating its arguments $\var := \trm$, unpacking each
term $\trm$, ensuring it is either a function or a real value, and
binding its unpacked value to $\var$ in $\trm[3]$. This translation is faithful~(\lemmaref{PCFadequacy}).

\subsection{CBV SPCF: simply typed recursive modelling}\subseclabel{spcf}
We consider a simply typed sublanguage of SFPC, a call-by-value (CBV)
probabilistic variant of \citeauthor{plotkin:pcf} and
\citeauthor{scott:pcf}'s PCF~\citeyearpar{plotkin:pcf,scott:pcf}.
The types and terms are given by the following grammars:
\[
\begin{array}{lllr@{}l}
  \ty,\ty[2],\ty[3] &\gdefinedby \Real \gor \ty \To \ty[2]\qquad\qquad\qquad &
  \trm, \trm[2], \trm[3]  &\gdefinedby \var
  \gor{} &\cnst{f}(\trm_1,\ldots,\trm_n)
  \gor \ifz{\trm}{\trm[2]}{\trm[3]} \\
&&&\gor{}& \tsample[]
  \gor \tfactor{(\trm)}
  \gor \expfun \var\ty \trm
  \gor \trm\ \trm[2]
  \gor \rec{\var:\ty\To\ty[2]}{\trm},
  \end{array}
\]
SPCF is a fragment of SFPC: term recursion
$\rec{\var:\ty\To\ty[2]}{\trm}$ is interpreted as in \figref{sugar} and conditioning
$\tfactor{(\trm)}$ as
$\letin{(\var:\Real)}\trm\tscore\var; \var$.
SPCF derives its kind and type systems from SFPC.
 \section{Quasi-Borel Pre-domains}\seclabel{wqbs}
\label{wqbs}
Previous works on quasi-Borel spaces (qbses) give a denotational
semantics for higher-order probabilistic languages with a range of
types, but crucially excludes higher-order term recursion and
recursive types.  To do that, we further equip a qbs with a compatible
\wcpo{} structure.  We call this new semantic structure a
\emph{quasi-Borel pre-domain} or an \emph{\wqbs}.

\subsection{Preliminaries}
\subsubsection*{Category theory}
We assume familiarity with categories $\cat$, $\cat[2]$, functors
$\F, \F[2] : \cat \to \cat[2]$, natural transformations
$\nt, \nt[2] : \F \to \F[2]$, and their theory of (co)limits
and adjunctions.
We write:
\begin{itemize}
\item unary, binary, and $I$-ary products
as $\terminal$, $X_1\times X_2$, and $\prod_{i\in I}X_i$, writing
$\projection_i$ for the projections and
$\sUnit$, $\spair{x_1}{x_2}$, and $\seq[i\in I]{x_i}$ for the tupling maps;
\item unary, binary, and $I$-ary coproducts
as $\initial$, $X_1 + X_2$, and $\sum_{i\in I}X_i$, writing
$\injection_i$ for the injections and
$\scoUnit$, $\scopair{x_1}{x_2}$, and $\coseq[i\in I]{x_i}$ for the cotupling
maps;
\item exponentials as $X^Y$, writing $\Lambda$ for the currying
 maps.
\end{itemize}
 \subsubsection*{Domain theory}
We recall some basic domain theory. Let $\omega=\set{0 \leq
1 \leq \ldots}$ be the ordinary linear order on the naturals. An \emph\wchain{}
in a poset $P = \pair{\carrier P}{\leq}$ is a monotone function $a_-
: \omega \to P$. A poset $P$ is an \emph{\wcpo}
when every \wchain{} $\seq[n \in \NN]{a_n}$ has a least upper bound
(lub) $\lub_{n \in \NN}a_n$ in $P$.

\begin{example}\examplelabel{simple wcpos}
  Each set $X$ equipped with the discrete partial order forms an \wcpo{}
  $\pair X=$. E.g., the discrete \wcpo{} $\reals$ over the
  real line.  The non-negative extended reals equipped with the
  ordinary order, $\lawson \defeq \pair{[0,\infty]}\leq$, is an
  \wcpo{}. The Borel subsets of $\reals^n$ ordered by inclusion form
  an \wcpo{} $\Borel_n$.
\end{example}

For every pair of \wcpo s $P$ and $Q$, a \emph{Scott-continuous}
function $f : P \to Q$ is a monotone function $f : \carrier P \to
\carrier Q$ such that for every \wchain{} $a_-$, we have: $f(\lub_n
a_n) = \lub_n f(a_n)$.  A Scott-continuous function $f : P \to Q$ is a
\emph{full mono} when, for every $a,b \in \carrier P$, we have $f(a)
\leq f(b) \implies a \leq b$. Recall that the category $\wCpo$ of
\wcpo s and Scott-continuous functions is Cartesian closed: products
are taken componentwise and the exponential $Q^P$ has carrier
$\wCpo(P,Q)$ and order $f\leq_{Q^P} g$ iff $\forall p\in
|P|.f(p)\leq_Q g(p)$. A \emph{domain} is an \wcpo{} with a least
element $\bot$. A \emph{strict} function between domains is a
Scott-continuous function that preserves their least elements.

\begin{example}\examplelabel{lebesgue integral}
A measure on $\reals^n$ is a strict continuous function $\mu : \Borel_n \to
\lawson$ that is linear on disjoint subsets.  The measurable functions
$\Borel(\reals, [0, \infty])$ ordered pointwise fully include into the
\wcpo{} $\lawson^{\reals}$. The \emph{integral} is the unique
Scott-continuous function $\int \mu : \Borel(\reals, [0,
  \infty]) \to \lawson$ satisfying, for all measurable partitions
$\reals = \sum_{n \in \naturals} U_n$ and weights $\seq{w_n}$ in
$\lawson$: $\int\mu\coseq[n \in \naturals]{\lambda r : U_n.w_n} = \sum_{n
  \in \naturals}x_n\cdot\mu U_n$ .
\end{example}

An \emph{$\wCpo$-(enriched) category} $\cat$ consists of a
locally-small category $\ord\cat$ together with an assignment of an
\wcpo{} $\cat(A, B)$ to every $A, B \in \Obj{\smash{\ord\cat}}$ whose carrier is
the set $\ord\cat(A, B)$ such that composition is Scott-continuous. An
\emph{$\wCpo$-functor}, a.k.a.~a \emph{locally-continuous functor},
$F : \cat \to \cat[2]$ between two $\wCpo$-categories is an ordinary
functor $\ord F : \ord\cat \to \ord{\cat[2]}$ between the underlying
ordinary categories, such that every morphism map
$\ord F_{A, B} : \cat(A, B) \to \cat[2](FA, FB)$ is Scott-continuous.

\begin{example}
  Every locally-small category is an $\wCpo$-category whose
  hom-\wcpos{} are discrete. The category $\wCpo$ itself is an
  $\wCpo$-category. If $\cat$ is an $\wCpo$-category, its categorical
  dual $\opposite\cat$ is an $\wCpo$-category. The category of
  locally-continuous functors $\opposite\cat \to \wCpo$, with the
  order on natural transformations $\alpha : F \to G$ given
  componentwise, is an $\wCpo$-category when $\cat$ is.
\end{example}
 \subsubsection*{Measure theory}
A measurable space $X = \pair{\carrier X}{\Sigma_X}$ consists of a
carrier set $\carrier X$ and a set of subsets
$\Sigma_X \subset \Powerset X$, called its \emph{$\sigma$-algebra},
containing the empty set, and closed
under complements and countable unions, thus axiomatising the
measurable subsets of $\reals^n$. A measurable function $f : X \to Y$
is a function $f : \carrier X \to \carrier Y$ whose inverse image maps
measurable subsets to measurable subsets. Thus every $n$-dimensional
Borel set, together with its Borel subsets, forms a measurable
space. The measurable spaces that are measurably isomorphic to a Borel
set are called \emph{standard Borel spaces}. A fundamental result in
descriptive set theory is that every standard Borel space is
measurably isomorphic to $\set{i \in \naturals \suchthat i < n}$ for
some $n = 0, 1, \ldots, \omega$, or to $\reals$ \cite{kechris2012classical}.
We write $\Meas$ and
$\Sbs$ for the categories of measurable and standard Borel spaces and
measurable functions between them.

\subsubsection*{Quasi-Borel spaces}\subseclabel{qbs}
\newcommand\baseR{\reals} As a Cartesian closed alternative to measure
theory, \citet{HeunenKSY-lics17} introduced the category $\Qbs$ of
\emph{quasi-Borel spaces} (qbses). Measure theory axiomatises
measurable subsets of a space $X$, and deriving the \emph{random
  elements}: measurable functions $\alpha : \baseR \to X$, for pushing
measures forward. Qbses axiomatise random elements directly.

A \emph{quasi-Borel space (qbs)} $X = \pair{\carrier X}{M_X}$ consists
of a carrier set $|X|$ and a set of functions $M_X\subseteq
|X|^{\baseR}$, called the \emph{random elements}, such that $(i)$~all
the constant functions are in $M_X$, $(ii)$~$M_X$ is closed under
precomposition with measurable functions on $\baseR$, and $(iii)$
if $\baseR=\Union_{n\in \naturals} U_n$, where $U_n$ are
pairwise-disjoint and Borel measurable, and $\alpha_n\in M_X$ for all
$n$, then the countable case-splitting
$\coseq[n\in\naturals]{\alpha_n|_{U_n}}$ is in $M_X$.
A \emph{morphism} $f\colon X\to Y$ is a structure-preserving function
$f : |X|\to|Y|$, i.e.~if $\alpha\in M_X$ then $(f \compose \alpha)\in
M_Y$.  Morphisms compose as functions, and we have a category $\Qbs$.

\begin{example}\examplelabel{borel spaces are qbs}
  We turn the $n$-dimensional space $\reals^n$ into a qbs by taking
  the random elements to be the measurable functions
  $M_{\reals^n} \definedby \Meas(\baseR, \reals^n) \isomorphic
  \prod_{i=1}^n \Meas(\baseR, \reals)$, i.e., $n$-tuples of correlated
  random variables. We also turn every set $X$ into a qbs by taking
  the random elements to be measurably piece-wise constant functions,
  i.e., the \emph{step functions}.
\end{example}

Both these examples are special cases of a more abstract
situation. Every measurable space $X$ can be turned into a qbs by
setting $M_X \defeq \Meas(\baseR, X)$.  This defines a functor
$M_{-}:\Meas\to\Qbs$. It has a left adjoint
$\Sigma_{-}:\Qbs\to\Meas$ which equips a qbs $X$ with the largest
$\sigma$-algebra such that all random elements $\alpha\in M_X$ are
measurable.  This adjunction restricts to an adjoint embedding of the
category of standard Borel spaces $\Sbs$ as a full subcategory of
$\Qbs$.  This embedding makes $\Qbs$ a conservative extension of the
well-behaved standard Borel spaces.

The category of qbses possesses substantial pleasant categorical
properties: it has all (co)limits and is Cartesian closed, and so can
interpret simple types, quotients, and refinements. In fact, $\Qbs$ is
a Grothendieck quasi-topos, and so can interpret an expressive
internal logic. The conservativity of the embedding $\Sbs \embed \Qbs$
means that interpreting closed programs of ground type and reasoning
about them in $\Qbs$ have standard measure-theoretic counterparts. The
benefit comes from doing so \emph{compositionally}: program fragments
that are higher-order functions have a compositional interpretation
and reasoning principles in $\Qbs$, but not in $\Sbs$ nor in $\Meas$.

\subsection{Definition and Some Simple Examples}
The difficulty inherent in combining domain and measure theory stems
from the following considerations~\cite{JUNG199870}. Each (pre-)domain
induces a topological space whose open subsets are the
\emph{Scott-open} subsets (see \exampleref{scott open subsets}
and \exampleref{scott open}), from which
one generates a measurable space structure by closing over countable
unions and complements. Both the domain-theoretic structure and the
induced measure-theoretic structure possess a cartesian product
construction. However, without further assumptions, the two product
structures may differ. To get a commutative probabilistic powerdomain,
one requires conditions that ensure these two structures agree while
maintaining, e.g., cartesian closure. The search for such a category
is known as the \emph{Jung-Tix problem}. We circumvent the Jung-Tix
problem, without solving it, by keeping the two structures, the
domain-theoretic and the measurable, separate but compatible. Doing so
also allows us to replace the measure-theoretic structure, which is
usually incompatible with higher-order structure, with a quasi-Borel
space structure. The result is the following definition:

\begin{definition}\label{def:wqbs}
  An \emph{\wqbs} $P$ consists of a triple $P = \triple{\carrier P}{M_P}{\leq_P}$
  where:
  $\pair{\carrier P}{M_P}$ is a qbs;
  $\pair{{\carrier{P}}}{\leq_P}$ is an \wcpo{} over
    $\carrier P$; and
  $M_P$ is closed under pointwise sups of \wchain s w.r.t.~the
  pointwise order.
  A \emph{morphism} between \wqbs es $f : P \to Q$ is a
  Scott-continuous function between their underlying \wcpo s that is
  also a $\Qbs$-morphism between their underlying qbses.
  We denote the category of \wqbs es and their morphisms by $\wQbs$.
\end{definition}

\begin{example}[Real Values]\examplelabel{real values}
We have the \wqbs{}
$\reals=(\reals,\Meas(\reals,\reals),=_{\reals})$
with the discrete order $=_{\reals}$.
This pre-domain represents a space of \emph{values} returned by probabilistic computations.
\end{example}
\begin{example}[Real Weights]\examplelabel{real weights}
Contrast this pre-domain with $\lawson = ([0,\infty],\Meas(\reals,{[0,\infty]}),
\leq_{[0,\infty]}~)$ with the linear order.
This pre-domain represents a space of \emph{weights}  of computation traces.
Compare this space to the Sierpi\'nski space
$\{0,1\}_{\leq}$, the full subspace $\{0,1\}$ in $\lawson$.
\end{example}

The category $\wQbs$ is an $\wCpo$-category,
with each homset in $\wQbs$ ordered \emph{pointwise} by
setting, for every pair of morphisms
$f, g \in \wQbs(X, Y)$:
$
  f \leq g$ when $\forall x \in \carrier X. f(x) \leq_Y g(x)
$.

\subsection{Interpreting Simple Types and Partiality}\subseclabel{simple types}
We turn every qbs into the \emph{discrete} \wqbs{} over it by taking
the discrete \wcpo{} structure, i.e., equality as an order. This
construction is the left adjoint to the evident forgetful functor
$\carrier- : \wQbs \to \Qbs$. Similarly, we turn every \wcpo{} into
the \emph{free} \wqbs{} over it whose random elements are lubs of step
functions. This construction is the left adjoint to the evident
forgetful functor $\carrier- : \wQbs \to \wCpo$.

\begin{example}
  The discrete \wqbs{} on the qbs
  structure of the real line is the pre-domain $\reals$ of real values.
  The free \wqbs{} on the \wcpo{} of weights (\exampleref{simple
    wcpos}) is the pre-domain $\lawson$ of real weights
  (\exampleref{real weights}).
\end{example}

These adjoints equip $\wQbs$ with well-behaved limits and coproducts:

\begin{lemma}
\lemmalabel{forget-preserves-lim}
The forgetful functors $\carrier- : \wQbs \to \wCpo$,
$\carrier- : \wQbs \to \Qbs$ preserve limits~and coproducts.
This uniquely determines the limits and coproducts of $\wQbs$,
which exist for  small diagrams.
\end{lemma}
In \secref{characterising-wqbs} we will see that $\wQbs$
also has quotients, but their construction is
more subtle.

\begin{figure}
  \(
  \begin{array}[t]{@{}l@{\ }l@{\ \ }l@{\ \ }l@{}}
    &\textbf{Carrier} &\textbf{Random elements} &\textbf{Partial order}
    \\
    \terminal
    &
    \{r\mapsto \sUnit\}
    &
    \{\sUnit\}
    &
    =_{\terminal}
    \\
    P\times Q
    &
    |P|\times |Q|
    &
    \{\spair{\alpha}{\beta}\mid \alpha\in M_P, \beta\in M_Q\}
    &
    \spair{p}{q}\leq \spair{p'}{q'}{:{}} p\leq_P p',  q\leq_Q q'
    \\
    \sum\limits_{i=1}^n P_i
    &
    \sum\limits_{i=1}^n |P_i|
    &
    \set{
      \coseq[i=1][n]{\injection_i\compose \alpha_i}\suchthat
      \begin{array}{@{}l@{}}
        \seq[1\leq i \leq n]{A_i}\in \Borel^n,\\
        \reals = \biguplus_{i=1}^n A_i \text{ partition}
        ,\\
        \forall 1\leq i\leq n. \alpha_i\in M_{P_i}
      \end{array}
    }
    &
    \spair{j}{p} \leq \spair{k}{q}{:{}} j=k,
    p\leq_{P_j} q
    \\
    Q^P
    &
    \wQbs(P,Q)
    &
    \set{
      \Lambda(\alpha):\reals\to \carrier{Q^P}\suchthat \alpha\in \wQbs(\reals\times P,Q)
    }
    &
    f\leq_{Q^P} g{:{}} \forall p \in |P|.f(p)\leq_Q g(p)
    \\
    \Lift P
    &
    \carrier P + \set{\bot}
    &
    \set{
      [B.\tot\compose \alpha, B^{\complement}.\bot] \suchthat B \in \Borel, \alpha \in M_P
    }
    &
    \bot \leq_{\Lift P} x, (\tot a \leq_{\Lift P} \tot b \iff a \leq_P b)
  \end{array}
  \)
  \caption{The simply-typed structure of $\wQbs$}
  \figlabel{bi-ccc}
\end{figure}

The category $\wQbs$ is bi-Cartesian closed, with the concrete
structure given in \figref{bi-ccc}. The structural maps, such as
tupling, projections, and so forth, are given as for sets.  The figure
also depicts a locally-continuous \emph{lifting} monad\footnote{See \subsecref{monad prelims} for the definition of monads.}
$\triple{\Lift-}{\return}{\bind}$, for interpreting partiality, where:
\begin{gather*}
  \tot x \defeq \injection_1
  x
  \quad
  \bot \defeq\injection_2\bot
  \quad
    \return x \defeq \tot x
    \quad
    \parent{(\tot x) \bind f} \defeq f(x)
    \quad
    \parent{\bot \bind f} \defeq \bot.
  \end{gather*}
This monad jointly lifts the partiality monad $P \mapsto \Lift P$ over
$\wCpo$ and the exception monad $X \mapsto X + \terminal$ over
$\Qbs$.
\begin{corollary}\corollarylabel{wqbs-structure}
  The functor $\carrier- : \wQbs \to \Set$ preserves the
  bi-Cartesian and partiality structures.
\end{corollary}

So simple types, when interpreted in $\wQbs$, retain
their natural set-theoretic interpretations.
 \section{A Commutative Statistical Powerdomain}\seclabel{monad}
\label{monad}
Our powerdomain construction combines two classical ideas in
probability theory. The first idea is Schwartz's treatment of
distributions as expectation operators. We construct the powerdomain
monad $T$ as a submonad of the continuation monad
$J:=\lawson^{\lawson^-}$, where $\lawson$ is the space of weights over
$[0,\infty]$ from \exampleref{real weights}. The second idea is the
Randomisation \lemmaref{randomize}: kernels $X \to TY$ should arise by
pushing forward the Lebesgue measure along a partial function $X\times
\reals \pto Y$, i.e., a morphism $X \to (\Lift Y)^{\reals}$.  Each
element of $(\Lift Y)^{\reals}$ induces a \emph{randomisable}
expectation operator via the Lebesgue integral. Combining the two
ideas, we take $T$ to be the smallest full submonad of the
Schwartz distribution monad that contains all the randomisable
distributions.

\hypertarget{sample and score}{
Sampling and conditioning} have natural interpretations as expectation
operators:
\[
\ssample:\terminal  \to  J\reals,
\ssample\sUnit[w] \definedby \int_{[0,1]}w(r)\dif r;
\qquad
\sscore:\reals \to J\terminal,
\sscore r[w] \definedby \abs r \cdot w(\sUnit)
\]
We will see that $T$ is also the smallest full submonad of $J$ which is
closed under $\ssample$ and $\sscore$.

\subsection{Preliminaries}\subseclabel{monad prelims}

\paragraph{Monads.}A \emph{strong monad structure} $\monad T$ over a
Cartesian closed category $\cat$ is a triple $\triple T\return\bind$
consisting of an assignment of an object $TX$ and a morphism
$\return_X : X \to TX$ for every object $X$, and an assignment of a
morphism $\bind_{X,Y} : TX \times (TY)^X \to TY$. A \emph{strong
  monad} is a strong monad structure $\monad T$ satisfying the monad
laws below, expressed in the Cartesian closed internal language:
\[
  \parent{(\return x) \bind f} = f(x) \qquad \parent{a \bind \return} = a
  \qquad \parent{(a \bind f) \bind g} = \parent{a \bind \lambda x. f(x) \bind g}
\]
Every monad yields an endofunctor $T$ on $\cat$-morphisms:
$T(f):=\id\bind^T(\return^T\compose f)$.  The \emph{Kleisli} category
$\cat_{\ir}$ consists of the same objects as $\cat$, but morphisms are
given by $\cat_{\ir}(X, Y) \defeq \cat(X, \uir Y)$.  A strong monad
$\ir$ is \emph{commutative} when, for every pair of objects $X,Y$:
\[
a : \uir X,b : \uir Y\vdash
  a \bind \lambda x. b \bind \lambda y. \return (x,y)
  =
  b \bind \lambda y. a \bind \lambda x. \return (x,y)
\]

\paragraph{Factorisation systems.}
We use the following concepts to factorise our powerdomain as a
submonad of the Schwartz distribution monad.  Recall that a \emph{orthogonal
  factorisation system} on a category $\cat$ is a pair
$\pair{\mathcal{E}}{\M}$ consisting of two classes of morphisms of
$\cat$ such that:
  \begin{itemize}
  \item
    Both $\mathcal{E}$ and $\M$ are closed under composition, and contain
    all isomorphisms.
  \item
    Every morphism $f : X \to Y$ in $\cat$ factors into $f = m \compose e$
    for some $m \in \M$ and $e \in \mathcal{E}$.
  \item Functoriality: for each situation as on the
    left, there is a unique $h : A\to A'$ as on the right:
    \centering
      \begin{tikzcd}[row sep=scriptsize]
        X \arrow[r, "e \in \mathcal E\ \ " above, two heads]
        \arrow[d, "f" left] &
        A \arrow[r, "m \in \mathcal M" above, tail]
        \arrow[d, "=", phantom]
        &
        Y
        \arrow[d, "g" right]
        \\
        X' \arrow[r, "e' \in \mathcal E\ \ " below, two heads] &
        A' \arrow[r, "m' \in \mathcal M" below, tail]
        &
        Y'
      \end{tikzcd}
      $\implies$
      \begin{tikzcd}[row sep=scriptsize]
        X \arrow[r, "e" above, two heads]
        \arrow[d, "f" left]
        \arrow[rd, "=", phantom] &
        A \arrow[r, "m" above, tail]
        \arrow[d, "{h}" , dashed]
        &
        Y
        \arrow[d, "g" right]
        \\
        X' \arrow[r, "e'" below, two heads] &
        A' \arrow[r, "m'" below, tail]
        &
        Y'
        \arrow[ul, "=", phantom] &
      \end{tikzcd}
\end{itemize}
 
\subsubsection*{S-finite measures and kernels}
Let $X$ be a measurable space. Define measures and kernels by direct
analogy with their definition on $\reals^n$.  A measure $\mu$ is
\emph{finite} when $\mu(X) < \infty$, and a kernel $k : X \leadsto Y$
is \emph{finite} when there is some bound $B \in [0, \infty)$ such
  that for all $x \in \carrier X$, $k(x, Y) < B$.  We compare measures and
  kernels pointwise, and both collections are
  \wcpo{}s. Measures and kernels are closed under countable pointwise
  sums given as the lubs of the finite partial sums. A measure $\mu$
  is \emph{s-finite} when it is a lub of finite measures, equivalently
  $\mu = \sum_{n \in \naturals} \mu_n$ for some countable sequence of
  finite measures, and similarly a kernel $k$ is \emph{s-finite} when
  it is a lub of finite kernels, equivalently $k = \sum_{n \in
    \naturals} k_n$ for some countable sequence of finite
  kernels~\cite{sfinite}. The Randomisation \lemmaref{randomize}
  generalises to s-finite kernels~\cite[Theorem~15]{vakarongsfinite}:
  for every s-finite kernel $k : X \leadsto Y$, with $Y$ standard
  Borel, there is a partial measurable function $f : X\times \reals
  \pto Y$ such that, for every $x \in \carrier X$, $k(x, -) = f(x,
  -)_*\lebesgue$. Recall the $2$-dimensional Lebesgue measure
  $\lebesgue\otimes\lebesgue$, which assigns to each rectangle $[a_1,
    b_1]\times [a_2, b_2]$ its area $(b_1 - a_1)\times(b_2 -
  a_2)$. Applying the Randomisation Lemma to
  $\lebesgue\otimes\lebesgue$ yields the \emph{transfer principle}:
  there is a measurable $\phi : \reals \pto \reals\times\reals$ such
  that $\phi_*\lebesgue = \lebesgue\otimes\lebesgue$.

\subsection{Randomisable Expectation Operators}
A \emph{randomisation} of an \wqbs{} $X$ is a partial $\wQbs$-morphism
$\alpha:\reals\pto X$, equivalently a total $\wQbs$ morphism
$\reals\to\Lift{X}$. Thanks to the Cartesian closure, we have an
\wqbs{} of randomisations $RX \defeq (\Lift X)^{\reals}$.  A
randomisation $\alpha \in RX$ represents an intensional description of
a measure on $X$ by pushing forward the Lebesgue measure $\lebesgue$.
The undefined part of $\alpha$ shaves some of the measure leaving us
with the restriction of $\lebesgue$ to $\Dom\alpha \defeq
\inv\alpha\coseq{x \in \carrier {\Lift X} \suchthat x \neq \bot}$. By
construction, $\Dom\alpha$ is a Borel set, and for every weighting
function $w : X \to \lawson$, the composition $w \compose \alpha :
\Dom\alpha \to X \to \lawson$ is an \wqbs{} morphism. Underlying this
composition is a qbs morphism $w \compose \alpha : \Dom\alpha \to
\lawson$. Because $\Qbs$ is a conservative extension of $\Sbs$, this
morphism is a Borel measurable function. Thus, every
randomisation induces an expectation operator. Moreover, this assignment is
an \wqbs{} morphism:
\[
\Expect: RX \to JX;\qquad\qquad
 \Expect_{\alpha} [w] \defeq \int_{\Dom\alpha}
 \lebesgue(\dif r)w(\alpha(r))
 \]

A \emph{randomisable} expectation operator $\mu \in JX$ is one where
$\mu = \Expect_\alpha$ for some randomisation $\alpha \in RX$.  Let
$|SX|$ be the set of randomisable operators
$\Expect_{\left[\Lift{X}^{\reals}\right]}\subseteq |JX|$. Similarly,
consider the randomisable random operators $M_{SX} \defeq
\Expect\compose[M_{RX}]$.  Let $|TX|$ be the \wchain-lub-closure of
$|SX| \subset \carrier{JX}$, and $M_{TX}$ be the closure of $M_{SX}
\subset M_{JX}$ under (pointwise) lub of \wchain s. The \wqbs{}
$TX$ is the smallest \wqbs{} that is a full sub-\wcpo{} of $JX$ and
containing the randomisable random operators.

\begin{example}
  When $X$ is a standard Borel space with the discrete order, each
  randomisable operator defines an s-finite measure, and each randomisable
  random operator defines an s-finite kernel. By the Randomisation Lemma
  for s-finite kernels, conversely, every s-finite measure/kernel arises from a
  randomisable (random) operator. Moreover,
  each s-finite measure/kernel is a lub of
  finite measures/kernels. So $TX$ is the smallest sub-\wcpo{} of $JX$
  containing the finite measures as elements and kernels as random
  elements, and consists of the s-finite measures and kernels.
\end{example}

The randomisation functor $R : \wQbs \to \wQbs$ has the following
monad structure. Using the transfer principle, fix any measurable
$\phi : \reals \xpto{} \reals\times\reals$ satisfying
$\phi_*\lebesgue = \lebesgue\otimes\lebesgue$, and define:
\[
  \return x \definedby \coseq{[0,1].\tot x, [0,1]^{\complement}.\bot}
\qquad
  \parent{\alpha\bind f} : \reals \xpto{\phi} \reals\times \reals \xpto{\reals\times \alpha}
  \reals \times X \xto{\reals\times f} \reals\times (\Lift Y)^{\reals} \xpto{\apply} Y
\]
The unit is a randomisation of the Dirac distribution, shaving from
Lebesgue all but a probability distribution concentrated on $x$. The
monadic bind splits the source of randomness, using the transfer
principle, into two independent sources of randomness, one for
$\alpha$ and one for the kernel $f$. The expectation morphism $\Expect : RX \to JX$
preserves this monad structure. The monad structure $R$ does \emph{not} satisfy the monad laws. While $RX$ depends on the choice of
$\phi$, $TX$ is independent of the choice.

\subsection{Factorising Monad Structure Morphisms}
To show that the monad structure of $J$ restricts to $T$, we rely on
a general theory, recently developed by
\citet{mcdermott-kammar:factorisations}.  The full monos between
\wcpo{}s form the $\M$-class of an orthogonal factorisation system on
$\wCpo$, where the $\mathcal E$-class consists of the \emph{dense}
epis: Scott-continuous functions $e : P \to Q$ whose image is dense,
i.e., the closure of $e[P]$ is
$Q$. \citet{kammar-plotkin:effect-systems} and
\citet{mcdermott-kammar:factorisations} use this factorisation system
to decompose a locally-continuous monad over $\wCpo$ into appropriate
sub-monads indexed by the sub-collection of effect operation
subsets. We use this construction to carve a sub-monad for
sampling and conditioning.

A \emph{full mono} between \wqbs{}es is a full mono between them as
\wcpo{}s, i.e., an order reflecting \wqbs{} morphism. A \emph{densely
  strong epi} $e : X \epimor Y$ is an \wqbs{} morphism that maps the
random elements $M_X$ into a Scott dense subset of $M_Y$ w.r.t. the
pointwise order.
\begin{lemma}
Densely strong epis and full monos form an orthogonal factorisation system
on $\wQbs$.
Moreover, the densely strong epis are closed under countable products,
exponentiation with standard Borel spaces, and the lifting monad $\Lift{(-)}$.
\end{lemma}
Therefore, the densely strong epis are closed under the randomisation
functor $R$. We can now directly apply
\citepos*{mcdermott-kammar:factorisations}{construction} to turn $TX$
into a canonical monad:
\begin{theorem}\theoremlabel{powerdomain strong monad}
  The unit and bind of $\monad J$ restrict to $T$. The (densely strong
  epi, full mono)-factorisation of the expectation operator $\Expect :
  RX \epimor TX \monomo JX$ preserves these monad structures.
\end{theorem}

\subsection{Sampling and Conditioning}
In the \hyperlink{sample and score}{introduction} to this section, we
defined sampling and conditioning as expectation operators. Both arise
as expectation operators for the following randomisations:
$$
\begin{array}{@{}l@{}l@{\qquad}l@{}l@{}}
  \ssample &: \terminal \to R\reals
  &
  \sscore &: \reals \to R\terminal
  \\
\ssample &{}\definedby [[0,1].\tot,[0,1]^{\complement}.\bot];
&
\sscore(r) &{}\definedby [[0,\abs r].\tot\sUnit,[0,\abs r]^{\complement}.\bot]
\end{array}
$$ Post-composing with $\Expect : R \to T$, we have analogous
operations for $T$.  Let $FX$ be the \emph{free monad} over $\wQbs$
with Kleisli arrows $\ssample : \terminal \to F\reals$ and $\sscore :
\reals \to F\terminal$. It exists because, for example, $\wQbs$ is
locally presentable (see \ConferenceArxiv{Appx.~\ref{characterising-wqbs}}{\secref{characterising-wqbs}}). The monad
morphism $m_J : \monad F \to \monad J$ preserving $\ssample$ and
$\sscore$ given by initiality of $F$ factors through the full
inclusion of $T$ in $J$ as $m_J : F \xto{m_T} T \xto{\Expect} J$,
where $m_T : F \to T$ is the unique monad morphism given by
initiality.  The randomisable operators are \emph{fully definable} by
$\ssample$ and $\sscore$, lubs, and the monad operations in the
following sense:
\begin{lemma}
  The unique monad morphism from the free monad preserving sampling
  and conditioning is a component-wise densely strong epi $m_T : F
  \epimor T$.
\end{lemma}
\begin{proof}[Proof (sketch)]
  Define the Lebesgue integral $\Expect_{\tot}(w) \definedby
  \int_{\reals}\lebesgue(\dif r)w(r)$ by rescaling the normal
  distribution. Let $\textrm{normal-rng} : [0,1] \to \reals$ be a
  randomisation of the gaussian with mean $0$ and standard deviation
  $1$, and let $\textrm{normal-pdf} : \reals \to \reals$ be the probability
  density function of this gaussian. Set $\xi \in F\reals$ as on the
  left, and calculate as on the right:
  \[
  \begin{array}{@{}l@{}}
  \xi \defeq \begin{array}[t]{@{}l@{}}
    \ssample \bind \lambda s. \\
    \letin r{\textrm{normal-rng}(s)}\\
    \sscore\tfrac1{\textrm{normal-pdf}\ r};
    \return r
  \end{array}
  \end{array}
  \quad
  (m_T\xi)(w) = \int_{[0,1]}\lebesgue(\dif
  s)\tfrac{w(\textrm{normal-rng}(s))}{\textrm{normal-pdf} \compose
  \textrm{normal-rng}\ s}
  =
  \int_{\reals}\lebesgue(\dif r) w(r)
\]
For each non-empty \wqbs{} $X$, choose $x_0 \in \carrier X$, and
consider a randomisable random operator
$\alpha = \Expect \compose \Lambda f$, for some uncurried
$f : \reals \times \reals \pto X$. Set
$\zeta(s) \definedby \xi \bind \lambda r. \handlewith{f(s, r)}x_0$
where the auxiliary function \( \handlewith -{x_0} : \Lift X \to FX; \)
is given by: \( \handlewith {\tot x}{x_0} \definedby \return x \) and
\( \handlewith {\bot}{x_0} \definedby \sscore0; \return x_0 \).
Then:
$m_T \compose \zeta(s)(w) =
\int_{\reals}\lebesgue(\dif r)f(s, r) = \alpha(s, w)$.
\end{proof}
Using the functoriality of the factorisation system, we deduce:
\begin{proposition}
The monad
$T$ is the minimal full submonad of $J$ that contains $\ssample$ and $\sscore$.
\end{proposition}

\subsection{Synthetic Measure Theory}

Synthetic mathematics identifies structure and axioms from which we
can recover the main concepts and results of specific mathematical
theories, and transport them to new settings.
\citet{kock:commutative-monads-as-a-theory-of-distributions}
demonstrates that some measure theory can be reproduced for any
suitable monad on a suitable category.
\citet{Scibior:2017:DVH:3177123.3158148} impose this categorical
structure on $\Qbs$ and use it to verify implementations of Bayesian
inference algorithms. Our statistical powerdomain $T$ is a very
well-behaved monad.  On the full subcategory $\Qbs\subseteq \wQbs$, it
restricts to the distribution monad given by
\citet{Scibior:2017:DVH:3177123.3158148}.  Like there, $T$ makes
$\wQbs$ into a model of synthetic measure theory, enabling the
interpretation of a statistical calculus.  We hope to replicate their
proofs for the more expressive SFPC in the future, and only state that
$\wQbs$ has this structure.

Recall the central definition to synthetic measure theory.  A
\emph{measure category} $\pair\cat{\ir}$ consists of a
Cartesian closed category $\cat$ with countable limits and coproducts;
and a commutative monad $\ir$ over $\cat$ such that the morphisms
$\sUnit_{T\initial} : T\initial \to \terminal$ and
$\bind\coseq[i]{\seq[j]{\delta_{i, j}}} : T\sum_{i \in \naturals}X_i
\to \prod_{j \in \naturals}T X_j$ are invertible, where
$\delta_{i,i}=\return^T_{X_i}$ and $\delta_{i\neq j} = x\mapsto
(\sUnit_{T\initial}^{-1}\sUnit) \bind (\return^T\compose\scoUnit)$.
The intuition is that elements in $T\sum_i X_i$, thought of as a
measure on a countable coproduct spaces are in one-to-one
correspondence with tuples of measures on the component
spaces. Surprisingly, this short definition guarantees that
$T\terminal$ is a countably-additive semi-ring whose elements resemble
\emph{scalars}, and measures have a countably-additive structure and
scalar multiplication operations. We then also have a morphism
$\mathrm{total} \defeq T\sUnit : TX \to T\terminal$, which we think of as
assigning to each measure the scalar consisting of its \emph{total
  measure}.

\begin{theorem}\theoremlabel{commutative}
  The statistical powerdomain $T$ equips $\wQbs$ with a measure
  category structure. The countable semiring of scalars is given by
  weights $T\terminal\cong \lawson$ with addition and multiplication.
  In particular, elements of $TX$ are linear and $T$ is a commutative
  monad.
\end{theorem}

Recall that we obtain a \emph{probabilistic} powerdomain as a
full submonad $\monad P$ of $\monad T$ as the equalizer of
$1,\mathrm{total}:TX\to T\terminal$, and a
\emph{sub}probabilistic powerdomain similarly.
We make no use of these
additional powerdomains in this work, except to note that there is a
continuous normalization function $TX\to (PX)_\bot+\{\top\}$,
where $\top$ is maximal, which returns $\bot$ or $\top$ if the overall measure is
$0$ or $\infty$ respectively, and a normalized probability measure otherwise.
In probabilistic programming, this normalization operation is usually used at the top level
to extract a posterior distribution.

\subsection{Valuations on Borel-Scott Open Subsets}\subseclabel{valuations}

We compare our approach to traditional notions of probabilistic
powerdomains using the following concepts. Let $X$ be a set. A
\emph{lub-lattice of $X$-subsets} is a family of subsets of $X$ that
is closed under finite unions and intersections, and countable unions
of \wchain{}s $B_0 \subset B_1 \subset \ldots$ w.r.t.~inclusion.

\begin{example}[Scott open subsets]\examplelabel{scott open subsets}
  Let $P$ be an \wcpo{}. A subset $U \subset \carrier P$ is
  \emph{Scott open} when it is
  \begin{itemize}
    \item upwards closed: for all $x \in  U$, and $y \in \carrier
    Q$, if $x \leq y$ then $y \in U$;
  \item $\omega$-inaccessible: for every $\omega$-lub $\lub_{n \in
    \NN}y_n \in U$, there is some $n \in \NN$ for which $y_n \in U$.
  \end{itemize}
  The Scott open subsets form a lub-lattice of subsets.
\end{example}

\begin{example}[Borel subsets]\examplelabel{borel subsets}
  Let $X$ be a qbs. A subset $B \subset \carrier X$ is \emph{Borel}
  when, for every random element $\alpha$, the subset $\inv\alpha[B]
  \subset \reals$ is Borel. The Borel subsets form a lub-lattice of
  subsets.
\end{example}

\begin{example}[Borel-Scott open subsets]
  Let $X$ be a qbs. A subset $D \subset \carrier X$ is
  \emph{Borel-Scott open} when it is Borel w.r.t.~the underlying qbs
  and a Scott open w.r.t.~the underlying \wcpo{}. The Borel-Scott open
  subsets form a lub-lattice of subsets.
\end{example}

The measurable functions into a Borel space are characterised as
approximated by step functions (see \exampleref{borel spaces are
  qbs}). The same is true for \wqbs{}es. Let $X$ be a \wqbs{}, and $B$
a Borel-Scott open subset of $X$. The \emph{characteristic function}
of $B$ is given by $[-\in B] \defeq [B.1, B^\complement.0] : X \to
\set{0,1} \subset \lawson$. A morphism $f : X \to \lawson$ is a
\emph{step function} when it is a finite weighted sum of
characteristic functions.

\begin{lemma}[Approximation by Simple Functions]
Let $f:X\to \lawson$ be a \wqbs-morphism.  Then, there is an
\wchain{} $f_n:X\to \lawson$ of step functions such that
$\lub_{n\in\NN}f_n=f$.
\end{lemma}
\begin{proof}[Proof (Sketch)]
Define
\(
f_n:=\sum_{1\leq i\leq 4^n} \tfrac{[-\in f^{-1}[(\tfrac{i}{2^n},\infty]]]}{2^{n}}
\).
\end{proof}

Let $\Open = \pair{\carrier{\smash\Open}}{\subset}$ be a lub-lattice
over a set $X$ ordered w.r.t.~inclusion. An
\emph{$\Open$-valuation}~\cite{lawson:valuations-on-continuous-lattices}
is a strict Scott-continuous function $v : \Open \to \lawson$
satisfying the binary \emph{inclusion-exclusion principle}: $v(A
\union B) + v(A \intersect B) = v(A) + v(B)$ for all $A, B \in
\carrier\Open$. Every lub-lattice of subsets on $X$ induces a
measurable space structure on $X$ by closing under complements and
countable unions. Every valuation induces a unique measure on this
measurable space. A valuation $v$ over $X$ is \emph{finite} when $v(X)
< \infty$, and \emph{s-finite} when it is the pointwise lub of finite
valuations. The s-finite valuations form an \wchain{}-closed subset of
the valuations.

The linearity and Scott continuity of expectation operators in $TX$
now gives us the following.
\begin{corollary}\corollarylabel{determining operators by char}
Expectation operators are determined by their value on characteristic
functions.
\end{corollary}
This corollary tells us that elements of $TX$ can be thought of as certain
(s-finite) valuations on the lub-lattice of Borel-Scott open subsets.
If $P$ is an \wcpo{}, the valuations over the free \wqbs{} generated
by $P$ (see \subsecref{simple types}) coincide with traditional
valuations on the Scott opens of $P$. When $X$ is a qbs with the
discrete order, restricting the expectation operators in $TX$ to
characteristic functions yields s-finite measures on the underlying measurable space $\Sigma_X$, and
by \corollaryref{determining operators by char}, each such s-finite
measure uniquely determines the operator. Similarly, when $X$ and $Y$
are qbses with the discrete order, each Kleisli arrow $X \to TY$
determines at most one s-finite kernel $\Sigma_X \leadsto \Sigma_Y$ by
restricting to characteristic functions.  For an sbs $X$, $TX$
consists of all s-finite measures on $X$ and $M_{TX}$ of all s-finite
kernels $\reals \leadsto X$, by the randomisation lemma for s-finite
kernels \cite{vakarongsfinite}.
 \section{Axiomatic Domain Theory}\seclabel{axiomatic-domain-theory}
\label{axiomatic-domain-theory}
Domain theory develops order-theoretic techniques for solving
recursive domain equations. The theorems guaranteeing such solutions
exist are technically involved.
\citepos+{fiore1994axiomatisation}{\emph{axiomatic domain
    theory}}{fiore:thesis} axiomatises categorical structure
sufficient for solving such equations. They aggregate axioms of
different strengths dealing with the same domain-theoretic aspects of
the category at hand; the strength of one axiom can compensate for the
weakness of another.  This theory allows us to treat recursive domain
equations in $\wQbs$ methodically.

This section is technical, and the main result is that $\wQbs$ has an
expansion to a category $\pMap\BS$ of \wqbs{}es and \emph{partial
  maps} between them, supporting the solution of recursive domain
equations. The type-formers of SFPC, that will denote locally
continuous mixed-variance functors over $\wQbs$, then have a locally
continuous extension to $\pMap\BS$, allowing us to use the solutions
of recursive domain equations as denotations of recursive types.

\subsection{Axiomatic Structure}
We begin by isolating the structure \citeauthor{fiore:thesis}
postulates as it applies to $\wQbs$. For the full account, see
\citeauthor{fiore:thesis}'s thesis. A \emph{($\Pos$)-domain structure}
$\DS$ is a pair $\seq{\cat_{\DS}, \admon_{\DS}}$ consisting of a
$\Pos$-enriched category $\cat_{\DS}$, and a locally small
full-on-objects subcategory $\admon_{\DS}$ consisting solely of
monomorphisms such that for every $m : D \monomo Y$ in $\admon_{\DS}$ and $f : X
\to Y$ in $\cat_{\DS}$:
\begin{itemize}
\item the pullback of $m$ along $f$ exists in $\cat_{\DS}$; and
\item in every pullback diagram, the pulled back morphism $f^*m: f^* D\monomo X$ is in $\admon_{\DS}$.
\end{itemize}
We call $\cat_{\DS}$ the category of \emph{total maps}
and the monos in $\admon_{\DS}$ \emph{admissible}.

\begin{example}[Scott open monos]\examplelabel{scott open}
  A \emph{Scott open mono} $m : P \to Q$ between \wcpo{}s is a
  Scott-continuous function that is a full mono such that $m[P]
  \subset Q$ is Scott open (see \exampleref{scott open
    subsets}). Equivalently, $m$ is mono and $m$-images of Scott open
  subsets are Scott open. Taking the Scott open monos as admissible
  monos yields a domain structure $\Sco$ over $\wCpo$.
\end{example}

To define partial maps for $\Qbs$ and $\wQbs$, we use the following
two examples. First, define a \emph{strong mono} $m : X \monomo Y$ between
qbses to be an injective function $m : \carrier X \monomo \carrier Y$
such that $m \compose [M_X] = M_Y \intersect |m[X]|^{|\reals|}$. The
strong monos in $\Qbs$ coincide with the regular monos, and so are closed under
pullbacks. They are also closed under composition and include all
isomorphisms.

\begin{example}[Borel open monos]
  Let $X$, $Y$ be qbses. A \emph{Borel open} mono $m : X \monomo Y$ is
  a strong mono whose image $m[X] \subset Y$ is Borel (see
  \exampleref{borel subsets}). Equivalently, a strong mono such that
  $m$-images of Borel subsets are Borel.
  The Borel open monos are closed under pullbacks, composition, and
  contain all isos.  By the completeness of $\Qbs$ we have all
  pullbacks, so taking the Borel open monos as admissible monos yields a
  domain structure $\Bor$ over $\Qbs$.
\end{example}
\begin{example}[Borel-Scott open monos]
  Taking the $\wQbs$-morphisms that are both Borel open and Scott open
  yields $\BS$, our domain structure of interest over $\wQbs$. It is a
  domain structure as by \lemmaref{forget-preserves-lim} the pullbacks
  in $\wQbs$ are computed using the pullbacks of $\wCpo$ and $\Qbs$.
\end{example}
To incorporate effects, we extend \citepos{fiore:thesis} account in
the following special case. When the \emph{representability} axiom
\Representability{}, which we define below,
holds, there is a strong monad over $\cat$ called~the
\emph{lifting} monad $\Lift-$. We define an \emph{effectful} domain
structure to be a triple $\triple{\DS}Tm$ consisting of~a
representable domain structure $\DS$ with finite products; a
strong monad $T$ over the category of~total maps; and a
strong monad morphism $m : \Lift- \to T$, thought of as \emph{encoding}
partiality using $T$'s~effects.

Our effectful domain structure consists of $\wQbs$, together with the
Borel-Scott open monos; the probabilistic power-domain of
\secref{monad}; and, as the resulting lifting monad is the partiality
monad of \subsecref{simple types}, the monad morphism $m$ is the
function mapping $\bot$ to the zero measure and every other element
$x$ to the Dirac measure $\delta_x$.

\subsection{Axioms and Derived Structure}
We develop the domain theory following \citeauthor{fiore:thesis}'s
development and describe the axioms it validates,
summarised in \figref{axioms}. While doing so, we recall the structure
\citeauthor{fiore:thesis} derives from these axioms.

\begin{figure}
  \begin{tabular}[t]{@{}c@{}}
  \begin{tabular}[t]{@{}ll@{}}
    \textbf{Structure} &  \textbf{Axioms}\\
    \begin{tabular}[t]{@{}l@{\ }l@{}}
        $\cat_{\DS}$ & total map category\\
                    & $\wQbs$           \\
        $f\leq g$   &$\Pos$-enrichment  \\
                    & pointwise order   \\
        $\admon_{\DS}$   & admissible monos  \\
                    & Borel-Scott opens \\
        $T$         & monad for effects \\
                    & power-domain      \\
        $m$ & partiality encoding\\
                    & $m : \Lift-\to T$, $\bot \mapsto \constantly 0$\\
    \end{tabular}
    &
    \begin{tabular}[t]{@{}l@{}}
      \begin{tabular}[t]{@{}l@{\ }l@{}}
    \Representability&every object has a partial \\&map classifier
    $\tot_X : X \to \Lift X$\\
    \FullUC & every admissible mono is full\\
            & and upper-closed\\
    \RepMonotonicity&$\floor - $ is locally monotone\\
    \wCpoEnrichment&$\cat_{\DS}$ is $\wCpo$-enriched\\
    \Uniformity &$\omega$-colimits behave uniformly\\
                &(\lemmaref{uniformity})\\
        \UnitType & $\cat_{\DS}$ has a terminal object
      \end{tabular}
    \end{tabular}
      \begin{tabular}[t]{@{}l@{\ }l@{}}
        \FunType  & $\cat_{\DS}$ has locally monotone\\
                  &exponentials\\
        \Coprod   &locally continuous total\\
        &coproducts\\
    \ZeroAdmissible&
    $\initial \to \terminal$ is admissible\\
        \ProductType & $\cat_{\DS}$ has a locally \\
                     & continuous products\\
        \Coco &$\cat_{\DS}$ is cocomplete\\
        \ContinuousT & $T$ is locally continuous\\
      \end{tabular}

  \end{tabular}
  \\
\end{tabular}
\begin{tabular}{@{}l@{}}
\textbf{Derived axioms/structure}\\
\begin{tabular}{@{}l@{}ll@{}}
  \begin{tabular}[t]{@{}l@{\ }l@{}}
    $\pMap\DS$ &partial map category\\
    $\Lift-$   &partiality monad    \\
    \RepContinuity & the adjunction $J\leftadjointto L$\\
    &is locally continuous\\
    \pMapEnrichment&$\pMap\DS$ is $\wCpo$-enriched\\
    \PartialUnit   &$\pMap\DS$ has a partial terminal\\
  \end{tabular}
  &
  \begin{tabular}[t]{@{}l@{\ }l@{}}
    \PartialProduct&$\pMap\DS$ has partial products\\
    \ContinuousPartialProduct&$(\otimes)$ is locally continuous\\
    \wCpoFun&$\cat_{\DS}$ has locally continuous\\
    & exponentials\\
  \PartialExponential&$\pMap\DS$ has locally continuous\\
                     & partial exponentials\\
  \end{tabular}
  &
  \begin{tabular}[t]{@{}l@{\ }l@{}}
  \pCoco &$\pMap\DS$ is cocomplete\\
  \pCoprod  &$\pMap\DS$ has locally continu-\\
            &ous partial coproducts\\
  \BilimitExpansion&$J : \cat \embed \pMap\DS$ is a bilimit\\
  &compact expansion
  \end{tabular}
\end{tabular}
\end{tabular}
   \caption{The axiomatic domain theory of $\wQbs$ and its probabilistic power-domain}
  \figlabel{axioms}
\end{figure}

\subsubsection*{Partial maps}
Each domain structure $\DS = \pair\cat\admon$ constructs a
\emph{category $\pMap\DS$ of partial maps}. Let $X$, $Y$ be
$\cat$-objects. A \emph{partial map description} $u:X\pto Y$ from $X$ to $Y$, is a pair $u = \pair{\dmon[u]}{\dfun u}$ consisting of an
admissible mono $\dmon[u] : \dom[u] \monomo X$ and a $\cat$-morphism
$\dfun u : \dom[u] \to Y$. Two descriptions $u$ and $v$ are
\emph{equivalent}, $u \equiv v$, when there is an isomorphism $i :
\dom[u] \xto{\isomorphic} \dom[v]$ satisfying: $\dfun{v}\compose i =
\dfun{u}$ and $\dmon[v]\compose i = \dmon[u]$.

A \emph{partial map} $u : X \pto Y$ is the equivalence class of a
description, bearing in mind that $\admon$ is locally small. E.g., for
$\BS$, choose inclusions as the canonical representatives $\dmon [u] :
\dom[u] \subset X$, which uniquely determine the description. Partial
maps form a category $\pMap\DS$, with identities given by $[\id,
  \id]$, and the composition $v \compose u : X \xpto{u} Y \xpto{v} Z$
via pullback (see \figref{partial-composition}, left).  We have an
identity-on-objects, faithful functor $J : \cat \to \pMap\DS$ mapping
each total map $f : X \to Y$ to $[\id, f] : X \pto Y$.

\begin{figure}
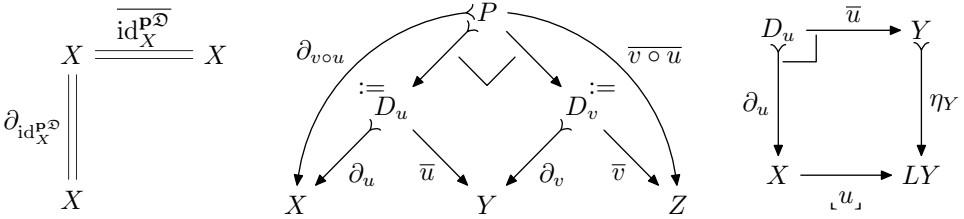

  \begin{tabular}{@{}c@{}}
  \begin{tabular}{@{}c@{\qquad}c@{\qquad}c@{}}
      \insertdiagram*{63}
      &
      \insertdiagram*{32}
      &
      \insertdiagram*{29}
  \end{tabular}
  \end{tabular}
  \caption{Partial identities and composition (left) and
    characteristic maps (right).}
  \figlabel{partial-composition}\figlabel{characteristic-maps}
\end{figure}

\subsubsection*{Representability}
Given a domain structure, a \emph{classifier of partial maps} is a
collection of admissible monos $\seq[X \in \cat]{\tot_X : X \monomo
  LX}$, indexed by the objects of $\cat$, such that for every $Y$, and
every partial map $f : X \pto Y$, there is a unique (total) map
$\floor f : X \to LY$ such that \figref{characteristic-maps} (right)
is a pullback square. We call $\floor f$ the \emph{total
  representation} of $f$.  Each classifier $\tot$ of partial maps
induces a right adjoint $J \leftadjointto L : \pMap\DS \to \cat$
with $\tot$ as unit. We give two well-known and two novel examples.

\begin{example}[partiality in $\Set$]
The collection $\seq[X \in \Set]{\injection_1 : X \to X + \set\bot}$
classifies the partial maps of $\Fn$, the domain structure over $\Set$
with injections as admissible monos. The representation of a partial function
maps the elements outside the domain to $\bot$. The induced
adjunction is isomorphic to the Kleisli resolution of the partiality
monad $(-) + \set\bot$.
\end{example}
\begin{example}[partiality in $\wCpo$]
Let $\Lift X$ be the lifting of the \wcpo{} $X$ by adjoining a new
bottom element. The collection $\seq[X \in \wCpo]{\tot : X \to
  \Lift X}$ classifies the $\Sco$-partial maps with $\floor-$ as in
$\Fn$.  The induced adjunction is isomorphic to the Kleisli resolution
of the lifting monad $\Lift{(-)}$.
\end{example}
\begin{example}[partiality in $\Qbs$]
  The collection $\seq[X \in \Qbs]{\injection_1 : X \to X +
    \set{\bot}}$ classifies the $\Bor$-partial maps with $\floor-$ as
  in $\Fn$. The induced adjunction is isomorphic to the Kleisli
  resolution of the partiality monad $(-) + \set\bot$.
\end{example}

\begin{example}[partiality in $\wQbs$]
  The collection $\seq[X \in \Qbs]{\tot : X \to \Lift X}$ classifies
  the $\BS$-partial maps with $\floor-$ as in $\Fn$. The induced
  adjunction is the Kleisli resolution of the monad $\Lift{(-)}$ of
  \subsecref{simple types}.
\end{example}

\subsubsection*{Enrichment}
The order enrichment of $\cat_\BS = \wQbs$ is an $\wCpo$-enrichment. Moreover, the
partial map category inherits a potential $\Pos$-enrichment: for
$u,v:X\pto Y$, write $u\pleq v$ to mean that there is some $i :
\dom[u] \to \dom[v]$ such that $ \dmon[u] = \dmon[v]\compose i$ and $\dfun{u}
\leq \dfun{v}\compose i$. The isomorphism $\pMap\DS \isomorphic
\wQbs_{\Lift-}$ respects $\pleq$ and $\leq$, i.e., $u \pleq v:X\pto Y$ iff
$\floor u \leq \floor v$ as morphisms $X\to \Lift Y$ in $\wQbs$, and
so $\pMap\DS$ is $\wCpo$-enriched. We can also deduce this fact from
more fundamental axioms. First, every admissible mono is full and its
image is upper-closed. As a consequence of \citeauthor{fiore:thesis}'s
Prop.~4.2.4, the order $\pleq$ is a partial order and $\Pos$-enriches
$\pMap\DS$. Denote the inverse map to representation by
$\trunc-:\cat_{\BS}(X, \Lift Y) \to \pMap\BS(X, Y)$. Because $\trunc-$
is monotone, the adjunction $J \leftadjointto L : \pMap\DS$ is locally
monotone~\ibid[Prop.~4.5.4]{fiore:thesis}, and, as a consequence
$\pMap\DS$ is $\wCpo$-enriched~\ibid[Prop.~4.5.3]{fiore:thesis}.

\subsubsection*{Uniformity}
\ConferenceArxiv
    {As $\wQbs$ is cocomplete (\corollaryref{cocomplete} in
      Appx.~\ref{sec:characterising-wqbs}), it}
    {While by \corollaryref{cocomplete} $\wQbs$}
has local $\omega$-lubs, but these lubs can behave
pathologically~\ibid[Sec.~4.3.2]{fiore:thesis}. The following
 \emph{uniformity} axiom avoids such pathologies.
\begin{lemma}[Uniformity]\lemmalabel{uniformity}
  Let $\seq[n \in \NN]{i_{n+1} : D_n \monomo D_{n+1}}$ be an \wchain{}
  of Borel-Scott open monos.
  \begin{itemize}
  \item Every colimiting cocone $\pair{D}{\seq[n \in \NN]{\mu_n : D_n \to D}}$
  consists of Borel-Scott opens.
  \item The mediating morphism into any other cocone of Borel-Scott opens is
  also Borel-Scott open.
  \end{itemize}
\end{lemma}

\subsubsection*{Unit type}
As $\wQbs$ has a terminal object $\terminal$, the partial maps have
$\terminal$ as a \emph{partial} terminal object: every hom-poset
$\pMap\BS(A, \terminal)$ has a unique maximal morphism~\ibid[p.~84]{fiore:thesis}.

\subsubsection*{Product types}
As $\wQbs$ has binary products $X\times Y$, we can extend the binary product
functor $(\times) : \cat_{\DS} \times \cat_{\DS} \to \cat_{\DS}$ to a
\emph{partial-product} functor $\otimes : \pMap\DS\times\pMap\DS \to
\pMap\DS$~\ibid[Prop.~5.1.1]{fiore:thesis}. Because $\wQbs$ has
exponentials, the functor $-\times X$ preserves colimits.  As a consequence,
as products are locally continuous, and as axioms \Uniformity,
\pMapEnrichment{} hold, $(\otimes)$ is locally
continuous~\ibid[Prop.~5.1.3]{fiore:thesis}.

\subsubsection*{Variant types}
\ConferenceArxiv
    {Because $\wQbs$ is cocomplete, using}
    {By \corollaryref{cocomplete}, $\wQbs$ is cocomplete. Using}
     axioms
\Representability,\UnitType{} we deduce that $\pMap\DS$ is
cocomplete~\ibid[Theorem~5.3.14]{fiore:thesis}, and
colimiting cocones of total diagrams comprise of total maps and are colimiting in
$\cat_{\DS}$~\ibid[Prop.~5.2.4]{fiore:thesis}. As the coproducts of
$\wQbs$ are locally continuous, so are the coproducts in
$\pMap\BS$~\ibid[Prop.~5.3.13]{fiore:thesis}. As a consequence, the
locally continuous finite-coproduct functor $\sum_{i \in I}: \wQbs^I
\to \wQbs$ extends to a locally continuous functor $\coprod_{i \in I}
: \pMap\DS^I \to \pMap\DS$ on partial maps~\ibid[remark following
  Cor.~5.3.10]{fiore:thesis}.

\subsubsection*{Effectful function types}
The following development is new, as \citeauthor{fiore:thesis} only
considered the partiality effect. As we now saw, when the
representability axioms \Representability{},\RepMonotonicity{}, and
the enrichment axioms \pMapEnrichment{},\ContinuousPartialProduct{}
hold, we have a locally continuous strong lifting monad $\Lift-$ over
$\cat$. Recall the additional structure given in an effectful domain
structure $\triple{\DS}Tm$, namely the locally monotone strong monad
$T$ over $\cat_{\DS}$, and the strong monad morphism $m : \Lift- \to
T$. We further assume that $T$ is locally continuous, which we call
axiom \ContinuousT{}. \theoremref{powerdomain strong monad} validates
it for the statistical powerdomain.

We have an identity-on-objects locally continuous functor $\floor[T]-
: \pMap\DS \to \cat_T$, from the category of partial maps into the
Kleisli category for $T$, given for every partial map $u : X \pto Y$
by: \[
\floor[T]u : X \xto{\floor u} \Lift Y \xto{m} TY
\]
When axiom \FunType{} holds, the composite functor $(\odot X) :
\floor[T]{J-\otimes X} : \cat \to \cat_T$ has a right adjoint:
\[
(X\pexp_T-) : \cat_T \to \cat
\quad
X \pexp_T Y := (TY)^X
\quad
X \pexp_T v := \lambda X. \parent{(TY_1)^X\times X \xto{\apply} TY_1 \xto{\bind v} TY_2}
\]
for every $v : Y_1 \to TY_2$. Axiom \FunType{} implies the exponential
adjunction is locally continuous, and as a consequence, $(\pexp_T)$ is
locally continuous. We use it to extend Kleisli exponentiation
$(-\Rightarrow T-) : \opposite\cat \times \cat \to \cat$ to a locally
continuous functor $(\pexp_T) : \opposite{\pMap\DS}\times \pMap\DS \to
\pMap\DS$ by setting, for every $u : X_2 \pto X_1$ and $v : Y_1 \pto{v} Y_2$:
\[
u \pexp_T v := J\lambda X. \parent{(TY_1)^{X_1}\times X_2 \xto{\floor[T]{\id \otimes u}}T((TY_1)^{X_1}\times X_1)\xto{\bind\apply} TY_1 \xto{\bind \floor[T]v} TY_2}
\]

\subsubsection*{Recursive types}
\seclabel{bilim-cpt}
To solve recursive domain equations, we synthesise axiomatic domain
theory with \citepos*{levy2012call}{more modern account}. Recall that
an embedding-projection-pair (ep-pair) $u : A\epto B$ in an
$\wCpo$-enriched category $\cat[4]$ is a pair consisting of a
$\cat[4]$-morphism $\embd u:A\to B$, the \emph{embedding}, and a
$\cat[4]$-morphism $\proj u:B\to A$, the \emph{projection}, such that
$e\compose p \leq \id$ and $p\compose e = \id$. An \emph{embedding} $u
: A \embedding B$ is the embedding part of some ep-pair $A \epto
B$. Every embedding $u : A \pembedding B$ in a partial map category with axiom
\pMapEnrichment{} is a total map $u : A \to
B$~\cite[Prop.~5.4.2]{fiore:thesis}.

An \emph{\wchain{} of ep-pairs} $\pair{\seq[n \in
    \naturals]{A_n}}{\seq[n \in \naturals]{a_n}}$ in $\cat[4]$
consists of a countable sequence of objects $A_n$ and a countable
sequence of ep-pairs $a_n : A_n \epto A_{n+1}$. A \emph{bilimit}
$\pair Dd$ of such an \wchain{} consists of an object $D$ and a
countable sequence of ep-pairs $\smash{d_n : A_n \epto D}$ such that,
for all $n \in \naturals$, $\smash{d_{n+1} \compose a_n = d_n}$, and
$\smash{\lub_{n \in \naturals} \embd d_n \compose \proj d_n =
  \id[D]}$. The celebrated \emph{limit-colimit
  coincidence}~\cite{smyth-plotkin:rde} states that the bilimit
structure is equivalent to a colimit structure $\pair{D}{\embd d}$ for
$\pair*{\seq{A_n}}{\seq{\embd a_n}}$, in which case $\proj d_n$ are
uniquely determined, and similarly equivalent to a limit structure
$\pair*{D}{\proj d}$ for $\pair*{\seq{A_n}}{\seq{\proj a_n}}$, in
which case $\embd d_n$ are uniquely determined. As we saw, the partial
map category $\pMap\BS$ has all colimits (derived axiom \pCoco), and
so $\pMap\BS$ has bilimits of \wchain{}s of ep-pairs.

A \emph{zero object} is an object that is both initial and
terminal. An \emph{ep-zero object} in an $\wCpo$-category is a zero
object such that every morphism into it is an embedding and every
morphism out of it is a projection. We say that axiom
\ZeroAdmissible{} holds in a domain structure $\DS$ in which $\cat$
has both an initial object $\initial$ and terminal object $\terminal$,
when the unique morphism $\initial \to \terminal$ is an admissible
mono. The Borel-Scott domain structure satisfies
\ZeroAdmissible{}. When axioms
\Coprod,\UnitType,\ProductType,\FunType,\Coco,\ZeroAdmissible{},\Representability,
and \RepMonotonicity{} hold, $\pMap\DS$ has $\initial$, the initial
object of $\cat$, as an ep-zero object. So in $\pMap\BS$ the empty
\wqbs{} is an ep-zero.

A \emph{bilimit compact category} is an $\wCpo$-category $\cat[4]$
with an ep-zero and ep-pair \wchain{} bilimits. So $\pMap\BS$ is
bilimit compact. When $\cat[3], \cat[4]$ are bilimit compact, every
locally continuous, mixed-variance functor $F :
\opposite{\cat[3]}\times\opposite{\cat[4]}\times{\cat[3]}\times\cat[4]
\to \cat[4]$ has a parameterised solution to the recursive equation
$\semroll : F(A, X, A', X) \xto{\isomorphic} X$, for every $A$, $A'$ in
$\cat[3]$, qua the bilimit $\pair{\mu B.F(A,B,A',B)}{d_{F, A, A'}}$
of
\[
\initial \epto F(A, \initial, A'\!, \initial)
\xepto{\ep F(A, \peto, A'\!, \epto)} F(A, F(A, \initial, A'\!, \initial), A'\!, F(A, \initial, A'\!, \initial)) \epto \cdots
\epto F^n(A, A') \epto \cdots
\]
The solution is minimal in the sense of \citet{pitts1996relational},
and we denote the inverse to $\semroll$ by $\semunroll$.  The
assignments $\mu B. F(A, B, A', B)$ extend to a mixed-variance functor
$\mu B. F(-, B, -, B) : \opposite{\cat[3]}\times\cat[3] \to \cat[4]$
by $\mu B. F(f, B, g, B) := \lub_{n} \embd d_n \compose F^n(f,
g)\compose \proj d_n$.

Finally, a \emph{bilimit compact expansion} $J : \cat \embed \cat[4]$
is a triple consisting of an $\wCpo$-category $\cat$, a bilimit
compact category $\cat[4]$; and an identity-on-objects, locally
continuous, order reflecting functor $J : \cat \to \cat[4]$ such that,
for every ep-pair \wchain{}s $\pair Aa$, $\pair Bb$ in $\cat[4]$,
their bilimits $\pair Dd$, $\pair Ee$, and countable collection of
$\cat$-morphisms $\seq[n \in \naturals]{\alpha_n : A_n \to B_n}$ such
that for all $n$:
\[
J\alpha_{n+1}\compose \embd a_n = \embd b_n
\compose J\alpha_n, \qquad J\alpha_{n}\compose \proj a_n = \proj b_n
\compose J\alpha_{n+1}
\]
(i.e., $J\alpha : \pair A{\embd a} \to \pair B{\embd b}$ and $J\alpha
: \pair A{\proj a} \to \pair B{\proj b}$ are natural transformations),
there is a $\cat$-morphism $f : A \to B$ such that $Jf = \lub_{n}
\embd e_n \compose \alpha_n \compose \proj d_n$. The motivation for
this definition: given two bilimit compact expansions $I :
\cat[2] \embed \cat[3]$, $J : \cat \embed \cat[4]$, and two
locally continuous functors \[ F : \opposite{\cat[2]}\times \opposite\cat \times \cat[2]
\times \cat \to \cat \quad G : \opposite{\cat[3]}\times
\opposite{\cat[4]} \times \cat[3] \times \cat[4] \to \cat[4]
\quad\text{s.t.}\quad G\compose \opposite I\times \opposite J \times I
\times J = J \compose F
\]
the functor $\mu B. G(-, B, -, B) : \opposite{\cat[3]}\times \cat[3]
\to \cat[4]$ restricts to $\mu B. G(\opposite J-, B, J-, B) :
\opposite{\cat[2]}\times \cat[2] \to \cat$. Bilimit compact expansions
are closed under products and opposites~\cite{levy2012call}.

Returning to axiomatic domain theory, when axioms
\Representability,\RepMonotonicity,\UnitType,\ProductType,\FunType,\Coco,
and \ZeroAdmissible{} hold, the embedding $J : \cat \embed \pMap\DS$
is a bilimit compact expansion of total maps to partial maps. To show
the third condition in the definition of expansion, note that
embeddings in $\pMap\DS$ are total, and we can use the limit-colimit
coincidence to reflect the two bilimits along $J$ to the colimits in
$\cat$, and find a mediating morphism $f : D \to E$. This morphism
maps to the mediating morphism in $\pMap\DS$, which is precisely
$\lub_{n} \embd e_n \compose \alpha_n \compose \proj d_n$.

\subsection{Summary: Semantics of Types}\subseclabel{bilimit-compact}
Using axiomatic domain theory, we constructed an expansion $J : \wQbs
\embed \pMap\BS$ to a suitable category of partial maps.  We use it to
interpret SFPC's type-system. We define two interpretations for
type-variable contexts $\Tyctx$ as $\wCpo$-categories, and two locally
continuous interpretations of well-kinded types $\Dkinf{\ty}$: a total
map interpretation $\sem-$ and its extension $\psem-$ to partial maps,
i.e.~$J\compose \sem- = \psem-\compose\opposite J\times J$:
\[
\sem\Tyctx \defeq \prod_{\alpha \in \Tyctx}\wQbs
\quad
\psem\Tyctx \defeq \prod_{\alpha \in \Tyctx}\pMap\BS
\qquad
 \sem\ty : \opposite{ \sem\Tyctx}\times  \sem \Tyctx \to \wQbs
\quad
\psem\ty : \opposite{\psem\Tyctx}\times \psem \Tyctx \to \pMap\BS
\]
\[
\begin{array}{@{}c@{}}
\begin{array}{@{}*2{c}@{}}
\begin{array}{@{}*2{l@{{}\defeq{}}l}@{}}
    \sem\alpha & \projection_\alpha&
   \psem\alpha & \projection_\alpha\\
    \sem\Real  & \reals&
    \psem\Real  & \reals\\
    \sem\Unit  & \terminal&
   \psem\Unit  & \terminal\\
\end{array}
&
\begin{array}{@{}*2{l@{{}\defeq{}}l}@{}}
    \sem{\ty[1]\t*\ty[2]}  &  \sem{\ty[1]} \times \sem{\ty[2]}&
   \psem{\ty[1]\t*\ty[2]}  & \psem{\ty[1]}\otimes\psem{\ty[2]}\\
    \sem{\ty[1]\To\ty[2]}&  \sem{\ty[1]}  \Rightarrow T  \sem{\ty[2]} &
   \psem{\ty[1]\To\ty[2]}& \psem{\ty[1]}\pexp_T \psem{\ty[2]}\\
\end{array}
\end{array}
\\
\begin{array}{@{}*2{c}@{}}
  \begin{array}{@{}*2{l@{{}\defeq{}}l}@{}}
     \sem{\begin{array}{@{}l@{}}
       \Variant{
                \Inj{\Cns_1}{\ty_1}\\
                \,\vor \ldots \\[1pt]
                \,\vor \Inj{\Cns_n}{\ty_n}}
     \end{array}
   }
                            & \sum   \limits_{i = 1}^n  \sem{\ty_i}&
   \psem{\begin{array}{@{}l@{}}
       \Variant{
                \Inj{\Cns_1}{\ty_1}\\
                \,\vor \ldots \\[1pt]
                \,\vor \Inj{\Cns_n}{\ty_n}}
     \end{array}
   }
                            & \coprod\limits_{i = 1}^n \psem{\ty_i}\\

  \end{array}
  &
    \begin{array}{@{}*2{l@{{}\defeq{}}l}@{}}
    \sem{\mu\alpha.\ty} & \mu\alpha. & \psem{\mu\alpha.\ty} &\mu\alpha. \\
    \multicolumn{2}{l}{\psem\ty(J-, \alpha, J-, \alpha)}&
    \multicolumn{2}{l}{\mspace{10mu}\psem\ty( -, \alpha,  -, \alpha)}\\
  \end{array}
  \end{array}
\end{array}
\]
 \section{Semantics for SFPC}\seclabel{semantics}
\label{semantics}

Operational semantics for statistical probabilistic calculi such as
SFPC require a mathematically technical development. The difficulty
comes from dealing with continuous distributions over programs that
manipulate continuous data. We follow the recipe previously employed by
\citet{staton2016semantics}
(others have used broadly similar methods, e.g.~\citet{BorgstromLGS-corr15}).
Each program phrase involving the real numbers $r_1, \ldots, r_n$ can
be represented as an open term $t$ with free variables $x_1, \ldots,
x_n$ each used linearly and in sequence, together with the tuple
$\seq{r_1, \ldots, r_n} \in \reals^n$. This separation lets us equip
the abstract syntax with a measurable space structure, and define a
big-step operational semantics as kernels between these syntactic
spaces~\cite{staton2016semantics}.

\subsection{Measure Theoretic Preliminaries}
\paragraph{Constructing spaces}
Every set is the carrier of a \emph{discrete} measurable space, in
which all subsets are measurable. We will typically only consider
discrete spaces of countable carriers, as these are precisely the
discrete standard Borel spaces. Let $I$ be a set indexing some
measurable spaces $X_i$. The measurable subsets of the disjoint union
$\sum_{i\in I} X_i$ are the disjoint unions $\sum_{i\in I} A_i$ of
measurable subsets $A_i$ in $X_i$. When $I$ is countable, the
measurable subsets of the cartesian product $\prod_{i\in I} X_i$ are
given by taking the boxes $\prod_{i\in I} A_i$, with each $A_i$
measurable in $X_i$, and closing under countable unions and
complements. These three constructions are the free measurable space
over a set, categorical coproduct, and categorical countable product
in the category of measurable spaces.

\paragraph{Constructing kernels}
The \emph{Dirac} kernel $\dirac- : X \leadsto X$ assigns to each
$x \in \carrier X$ its \emph{Dirac distribution} defined by $\dirac x
(A) = 1$ for $x \in A$ and $\dirac x(A) = 0$ otherwise.  Every
measurable function $f : X \to Y$ inudces a kernel $f : X \leadsto Y$
by setting $f(x, A) \definedby \dirac {f(x)}(A)$.  Each kernel $k :
X \leadsto Y$ acts on measures $\mu$ over $X$ yielding the
unique measure $\mu \bind k$ over $Y$ satisfying, for all measurable
$\phi : Y \to [0, \infty]$:
$$
\int_Y (\mu \bind k)(\dif y)\phi(y) = \int_X\mu(\dif x)\int_Yk(x,\dif y)\phi(y)
$$
Moreover, given $k : X \leadsto Y$ and $l : Y \leadsto Z$, the mapping
$x\mapsto k(x) \bind l$ is a kernel $k \fbind l :
X \leadsto Z$. The s-finite kernels are closed under this
composition~\cite{sfinite,
kallenberg:random-measures-theory-and-applications}.
 
\subsection{Syntax Spaces}\subseclabel{syntax-spaces}
Consider the set of terms of type $\ty$ in variable context $\ctx$,
$\Expr\ctx\ty \definedby \set{\trm \suchthat \Ginf \trm \ty}$. The set
of \emph{values} of type $\ty$ in context $\ctx$ consists of the
subset $\Val\ctx\ty \subset \Expr\ctx\ty$ given inductively by:
\[
  \val,\val[2],\val[3] \gdefinedby \var
  \gor \cnst{r}
  \gor \tUnit
  \gor \tpair{\val}{\val[2]}
  \gor \tInj\ty\Cns\val
  \gor \troll\ty(\val)
  \gor \expfun \var\ty \trm\qquad\qquad \synname{values}.
\]
We want to equip $\Expr\ctx\ty$ and $\Val\ctx\ty$ with a measurable
space structure.  Let $\lctx$ range over \emph{non-symmetric linear}
variable contexts of type $\Real$, namely finite sequences of
variables, and let \emph{mixed} variable contexts $\ctx;\lctx$ be
pairs of variable context $\ctx$ and linear contexts $\lctx$ with
disjoint sets of variables. Each such mixed context can be turned into
an ordinary variable context $\ctx,\lctx:\Real$ by treating the
identifiers in $\lctx$ as variables of type $\Real$. We will use the
identifiers $\rbox, \rbox[1], \rbox[2]$ for the variables in
$\lctx$. A term \emph{template} is a term that does not involve
nullary primitives $\cnst r$. We refine the SFPC type-system to
judgements of the form $\Ginf[;\lctx]{\trm}{\ty}$ by treating $\lctx$
non-symmetric linearly, e.g.:
\[
  \inferrule{
    ~
  }{
    \Ginf[; \rbox] \rbox\Real
  }
  \quad
    \inferrule{
    \Ginf[;\lctx_1]{\trm}{\ty[2]\To\ty}
    \\
    \Ginf[;\lctx_2]{\trm[2]}{\ty[2]}
  }{
    \Ginf[;\lctx_1,\lctx_2]{\trm\, \trm[2]}{\ty}
  }
\]
Let $\LExpr{\ctx;\lctx}\ty \subset \Expr{\ctx,\lctx:\Real}\ty$ be the
set of well-formed templates $\Ginf[;\lctx]\trm\ty$ and
${\LVal{\ctx;\lctx}\ty\! \subset \LVal{\ctx;\lctx}\ty}$ be the set of
well-formed template values. Given any
$\lctx = \rbox[1], \ldots, \rbox[k]$, and
$r_1, \ldots, r_k \in \reals$, we define the following function
\[
-[r_1, \ldots, r_n] : \LExpr{\ctx;\lctx}\ty \to \Expr\ctx\ty
\qquad
\trm{}[r_1, \ldots, r_n] \defeq
\trm{}[\rbox[1] \mapsto \cnst{r_1}, \ldots, \rbox[k] \mapsto \cnst{r_k}]
\]
Combining these functions we have a bijection
$\sum_{n \in \naturals, \trm \in \LExpr{\ctx; \rbox[1], \ldots, \rbox[n]}\ty}
\reals^n \isomorphic \Expr\ctx\ty$.
\begin{example}
The Bayesian linear regression program from \figref{regression}(a) is represented by:
\[
\parent{
11,
\begin{array}{@{}lr@{}l@{}}
\letin{a}{\cnst{\mathrm{normal\textit-rng}\,}(\rbox[1], \rbox[2])}{}
&(&{0},{2},
\\\quad{\tscore[](\cnst{\textit{normal-pdf}\,} (\rbox[3]\mid a* \rbox[4],\rbox[5] ))};
&&{1.1},{1},{0.25},
\\\quad{\tscore[](\cnst{\textit{normal-pdf}\,}(\rbox[6]\mid a* \rbox[7], \rbox[8]))};
&,\quad\vphantom{,}&{1.9},{2},{0.25},
\\\quad{\tscore[](\cnst{\textit{normal-pdf}\,}(\rbox[9]\mid a* \rbox[10], \rbox[11] ))};a
&&{2.7},{3},{0.25})
\end{array}
}
\]
\end{example}
The representation $\sum_{n \in \naturals, \trm \in \LExpr{\ctx;
    \rbox[1], \ldots, \rbox[n]}\ty} \reals^n$ has a canonical
measurable space structure, combining the Borel $\sigma$-algebra on
$\reals^n$ and the coproduct $\sigma$-algebra. As a consequence, we
equip $\Expr\ctx\ty$ with the measurable space structure making the
representation a measurable isomorphism. Similarly, we equip
$\Val\ctx\ty$ with a measurable space structure. Summing over all
types, we get the measurable spaces of closed terms, $\Expr[]{}{}$,
and closed values, $\Val[]{}{}$.

\subsection{Operational Semantics}\subseclabel{opsem}
\ConferenceArxiv{}{
  \begin{figure}
  \[
\begin{array}{@{}c@{}}
  \trm\evalto[0]A \defeq 0
  \qquad
  \inferrule{
    \trm_1 \evalto[n] \cnst {r_1}
    \quad\ldots\quad
    \trm_k \evalto[n] \cnst {r_k}
  }{
    \cnst f(\trm_1, \ldots, \trm_k) \evalto[n] \cnst {f(r_1, \ldots, r_n)}
  }
  \qquad
  \begin{array}{@{}l@{}}
    \parent{\tsample[]\evalto[n]} \defeq
    \\
    \qquad
     \uniform{[0,1]} \bind \lambda r. \dirac{\cnst r}
  \end{array}
  \qquad
  \begin{array}{@{}l@{}}
    \parent{\tscore[](\trm)\evalto[n]} \defeq
    \\
    \quad
    \parent{\trm\evalto[n]} \bind \lambda \cnst r . \abs r \cdot \dirac \tUnit
  \end{array}
  \\
  \inferrule{
    \trm\evalto[n] \cnst 0
    \quad
    \trm[2]_1\evalto[n] \val
  }{
    \ifz{\trm}{\trm[2]_1}{\trm[2]_2})\evalto[n]\val
  }
  \quad
  \inferrule{
    \trm\evalto[n] \cnst r
    \quad
    \trm[2]_2\evalto[n] \val
  }{
    \ifz{\trm}{\trm[2]_1}{\trm[2]_2})\evalto[n]\val
  }(r \neq 0)
  \\
  \inferrule{
    ~
  }{
    \val \evalto[n] \val
  }(\val = \tUnit, \expfun\var\ty\trm)
  \quad
  \inferrule{
    \trm\evalto[n] \tUnit
    \quad
    \trm[2]\evalto[n] \val
  }{
    \umatch
           \trm
           {\trm[2]}
    \evalto[n]
    \val
  }
  \quad
  \inferrule{
    \trm[1]\evalto\val[1]
    \quad
    \trm[2]\evalto\val[2]
  }{
    \tpair{\trm[1]}{\trm[2]}
    \evalto[n]
    \tpair{\val[1]}{\val[2]}
  }
  \quad
    \inferrule{
    \trm\evalto[n]\val
  }{
    \tInj\ty\Cns\trm
    \evalto[n]
    \tInj\ty\Cns\val
  }
  \\
  \inferrule{
    \trm\evalto[n]\tpair{\val[2]}{\val[3]}
    \quad
    \trm[2][{\var[1] \mapsto \val[2], \var[2] \mapsto \val[3]}]
    \evalto[n-1] \val
  }{
    \pmatch
           \trm
           {\var[1]}{\var[2]}
           {\trm[2]}
    \evalto[n]
    \val
  }
  \
  \inferrule{
    \trm \evalto[n] \Inj{\Cns_{\mathnormal i}}{\val[2]}
  \quad
  \trm[2]_i[\var_i \mapsto \val[2]] \evalto[n-1] \val
  }{
    \begin{array}[t]{@{}r@{\,}l@{}l@{}}
    \vmatch \trm{
                    &\Inj[]{\Cns_{\mathnormal 1}\ }{\var_1}&\To{\trm[2]_1}
                    \vor\cdots\\
                    \vor&\Inj[]{\Cns_{\mathnormal m}}{\var_m}&\To{\trm[2]_m}
    }
    \evalto[n]
    \val
      \end{array}
  }
  \
  \inferrule{
    \trm\evalto[n] \val
  }{
    \troll\ty(\trm) \evalto[n] \troll\ty\val
  }
  \\
  \inferrule{
    \trm \evalto[n] \troll\ty\val[2]
    \ \ \
    \trm[2][\var \mapsto \val[2]] \evalto[n-1] \val
  }{
    \rmatch
           \trm
           {\var}
           {\trm[2]}
    \evalto[n]
    \val
  }
  \
  \inferrule{
    \trm\evalto[n] \expfun{\var}{\ty}{\trm[3]}
    \ \ \
    \trm[2]\evalto[n] \val[2]
    \ \ \
    \trm[3][\var \mapsto \val[2]]
    \evalto[n-1]
    \val
  }{
    \trm\, \trm[2]
    \evalto[n] \val
  }
  \end{array}
\]
   \ConferenceArxiv
      {\caption*{\figref{operational-semantics}.
          \,SFPC big-step operational semantics ($n > 0$)
      }}
      {\caption{
          SFPC big-step operational semantics ($n > 0$)
      }}
    \ConferenceArxiv{}{\figlabel{operational-semantics}}
\end{figure}
 }
We define a structural operational semantics as a kernel ${\evalto} :
\Expr[]{}{} \leadsto \Val[]{}{}$. As usual, the operational semantics
is \emph{not} compositional. E.g., the semantics of function
application is not defined in terms of that of a subterm, but of a
substituted subterm. As a consequence, we construct an increasing
sequence of kernels ${\evalto[n]} : \Expr[]{}{}\leadsto\Val[]{}{}$
indexed by the natural numbers and set ${\evalto} \definedby \lub_n
{\evalto[n]}$. At level $0$, the evaluation kernel is the least
kernel, namely $\trm\evalto[0]A \defeq 0$ for every closed term $\trm$ and
measurable $A \subset \Val[]{}{}$. For positive levels $n > 0$, we use
the following three notational conventions that highlight that the
semantics is a standard adaptation of the more familiar operational
semantics. First:
\[
\begin{array}{@{}l@{}}
\inferrule{
  k_1(\trm[1])\,\val[2]_1
  \quad
  k_2(\trm[1], \val[2]_1)\,\val[2]_2
  \quad
  \ldots
  \quad
  k_n(\trm[1], \val[2]_1, \ldots, \val[2]_n)\,\val
}{
  l(\trm[1])\, f(\trm[1], \val[2]_1, \ldots, \val[2]_n, \val)
}
\quad
\text{ means }
\\~\\\text{Second: }\hspace{.3cm}
l(t) \definedby k_1(t) \ \
l(t) \definedby k_2(t)
\text{ means }
l(t) \definedby k_1(t) + k_2(t)
\end{array}
\begin{array}{@{}l@{}}
l(\trm[1]) \definedby
\begin{array}[t]{@{}l@{}l@{}l@{}}
k_1&(\trm[1])                               &\bind \lambda \val[2]_1.\\
k_2&(\trm[1], \val[2]_1)                    &\bind \lambda \val[2]_2.\ \ldots\\
k_n&(\trm[1], \val[2]_1, \ldots, \val[2]_n) &\bind
\lambda\val.\\
\multicolumn{3}{@{}l@{}}{\dirac{f(\trm[1], \val[2]_1, \ldots, \val[2]_n, \val)}}
\end{array}
\end{array}
\]
i.e., we sum overlapping definitions. Finally, if one of the
intermediate kernels $k_i$ is undefined (on a measurable subset), we
define $l(t)$ to have measure $0$ on this subset.

\ConferenceArxiv
    {\refstepcounter{figure}
      \figlabel{operational-semantics}
      The crucial definitions involved in the indexed evaluation
      kernels are sampling and conditioning, which evaluate using the
      interpreter's corresponding primitives (see
      \figref{operational-semantics} in
      Appx.~\ref{appx:semantic-figures}):
     \[
    \parent{\tsample[]\evalto[n]} \defeq
     \uniform{[0,1]} \bind \lambda r. \dirac{\cnst r}
  \qquad
    \parent{\tscore[](\trm)\evalto[n]} \defeq
    \parent{\trm\evalto[n]} \bind \lambda \cnst r . \abs r \cdot \dirac \tUnit
    \]
    The
    }
    {
          \figref{operational-semantics}
presents the indexed evaluation
kernels. Primitives evaluate their arguments left-to-right.
 Sampling
and conditioning evaluate using the interpreter's statistical sampling
and conditioning primitives. We could equivalently use a lower-level
interpreter supporting only (sub)-probabilistic sampling and
maintaining an aggregate weight, i.e., a kernel
$\evalto[n]' : \Expr[]{}{} \leadsto \reals\times\Val[]{}{}$ and the rule:
\[
  \inferrule{
    t \evalto[n]' {\tpair {r_1}{\cnst{r_2}}}
  }{
    \tscore(t) \evalto[n]' \tpair{r_1\cdot \abs{r_2}}{\tUnit}
  }
\]
and the other rules changing analogously. The remaining rules are
standard. The index decreases only when we apply the kernel
non-compositionally in the elimination forms for binary products,
variants, recursive types, and function application.
 The evaluation
    }
kernel ${\evalto}$ is s-finite, as the s-finite kernels are
    closed under composition and lubs~\cite{sfinite}.

\subsection{Contextual Equivalence}
The operational semantics lets us compare the meaning of terms using
the following standard notions. Given variable context $\ctx_1,
\ctx_2$, we say that $\ctx_2$ \emph{extends} $\ctx_1$, and write
$\ctx_2 \geq \ctx_1$ when, for all $(\var : \ty)\in \ctx_1$, we have
$(\var : \ty) \in \ctx_2$.  \emph{Program contexts of type $\ty[2]$
  with a hole $-$ of type $\ctx\vdash\ty$} are terms
$C[\ctx\vdash-:\ty]$ of type $\ty[2]$ with a single variable of type
$\ty$, where this variable $-$ always occurs inside the term in
contexts $\ctx' \geq \ctx$.  Write $C[\trm]$ for the \emph{capturing}
substitution $\subst{C[\ctx \vdash -:{\ty}]}{\sfor{-}{\trm}}$.

Two terms $\trm,\trm[2]\in\Expr{\ctx}{\ty}$ are in the
\emph{contextual preorder} $\trm\precsim\trm[2]$ when for all program
contexts $C[\ctx \vdash -: {\ty}]$ of type $\Real$, we have that $
{C[\trm]\evalto} \leq {C[\trm[2]]\evalto}$ in the usual (pointwise)
order of measures on $\Val{}{\Real}$.  We say that $\trm$ and
$\trm[2]$ are \emph{contextually equivalent}, writing $\trm\approx
\trm[2]$, when $\trm\precsim \trm[2]$ and $\trm[2]\precsim \trm$

The observational preorder and equivalence are the same even if we
vary the definitions:
\begin{enumerate}
\item using contexts of type $\Unit$ and observing only the weight of
  convergence;
\item using contexts of ground type (i.e. iterated sums and products
  of primitive types) and observing the distribution over
  $\Val{}{\ty}$;
\item using contexts of arbitrary type $\ty$ and observing the induced
distribution over $\Val{}{\ty}/\sim$, where $\sim$ is the smallest congruence
identifying all $\lambda$-abstractions (as done in \cite{pitts1996relational}).
\end{enumerate}
Indeed, for the equivalence of 1., 2. and 3., use characteristic
functions $\reals^n\to \reals$.  For the equivalence with our notion,
observe that $\Unit$ embeds into $\Real$ and, conversely, distinguish
any distribution on $\Val{}{\Real}$ with contexts of type $\Unit$ by
using $ \ifz{\cnst{[-\in A]}(-_{\Real})}{\bot}{\tUnit} $, for
Borel $A\subseteq \reals$.

\subsection{Idealised Church and SPCF}

Since both Idealised Church (\subsecref{idealised-church}) and SPCF
(\subsecref{spcf}) are fragments of SFPC, they inherit an operational
semantics and a notion of contextual equivalence w.r.t.~contexts in
the fragment.  For Idealised Church, these contexts translate into
certain SFPC program contexts of type $\UntypedL\Real$.  The induced
notion of contextual preorder is now that for two Idealised Church
terms we set $\trm\precsim \trm[2]$ when ${C[\trm]^\dagger\evalto}\leq
C[\trm[2]]^\dagger\evalto$ as distributions on
$\Val{}{\UntypedL\Real}/\sim$ for all Idealised Church program
contexts $C[-]$, where $\sim$ identifies all values of the form
$\Inj{Fun}{(\lambda x.\trm)}$. For CBV SPCF program contexts of type
$\Real$ induce an analogous notion of contextual preorder and
equivalence.

\begin{lemma}\lemmalabel{ICadequacy}\lemmalabel{PCFadequacy}
  The translations $\text{Idealised Church} \xto{(-)^\dagger}
  \text{SFPC} \xfrom{(-)^\ddagger} \text{CBV SPCF}$ are adequate:
  \[
  \trm_1^\dagger\precsim_{\text{SFPC}} \trm_2^\dagger
  \implies
  \trm_1 \precsim_{\text{Idealised Church}} \trm_2
  \qquad
  \trm[2]_1^\ddagger\precsim_{\text{SFPC}} \trm[2]_2^\ddagger
  \implies
  \trm[2]_1 \precsim_{\text{CBV SPCF}} \trm[2]_2
  \]
\end{lemma}
\begin{proof}
Both languages are fragments of SFPC with the induced operational
semantics.  Moreover, the source contexts are a subset of those of
SFPC.  Therefore, the SFPC contextual preorder is as fine-grained
as each fragment's contextual preorder.
\end{proof}

\subsection{Denotational Semantics}
Recall the type semantics of \subsecref{bilimit-compact} for SFPC:
closed types $\ty$ denote \wqbs{}es $\sem \ty$. We extend this
assignment to contexts by: $\sem{\ctx}\defeq \prod_{(\var : \ty) \in
  \ctx}\sem\ty$.
\ConferenceArxiv{
  We define
}{
  \figref{interpretation-terms} defines
} semantics for
values and terms:
\[
\sem{-}^v:\Val{\Gamma}{\ty}\to \wQbs(\sem{\Gamma},\sem{\ty})\qquad\qquad
\sem{-}:\Expr{\Gamma}{\ty}\to \wQbs(\sem{\Gamma},T\sem{\ty})
\]
such that $\return^T_{\sem{\ty}}\parent{\sem{\val}^v\env}
=\sem{\val}\env$ for every $\val\in\Val{\Gamma}{\ty}$ and $\env \in
\sem\ctx$.  This interpretation is the standard semantics of a
call-by-value calculus using a monad over a bi-cartesian closed
category, where we interpret the operations $\sem{\tsample[]}\env
\defeq \ssample()$ and
$\sem{\tscore{\trm}}\gamma \defeq
\sem{\trm}\gamma\bind^T \sscore$.
\ConferenceArxiv{
  The full details are in \figref{interpretation-terms} in
  Appx.~\ref{appx:semantic-figures}.
}{}

\ConferenceArxiv{
  \refstepcounter{figure}
      \figlabel{interpretation-terms}
}{
  \begin{figure}
  \[
\begin{array}{@{}c@{}}
  \textbf{Values:\hspace{1.25cm}}
  \quad
  \sem{\var}^v\env \defeq \projection_{\var}\env
  \quad
  \sem{\cnst{r}}^v\env \defeq r
  \quad
  \sem{\tUnit}^v\env \defeq \sUnit
  \quad
  \sem{\tpair{\val}{\val[2]}}^v\env \defeq \spair{\sem{\val}^v\env}{\sem{\val[2]}^v\env}
  \\
  \sem{\tInj\ty{\Cns_i}\val}^v\env \defeq \injection_i(\sem{\val}^v\env)
  \qquad
  \sem{\troll\ty(\val) }^v \defeq \semroll(\sem{\val}^v\env)
  \qquad
  \sem{\expfun{\var}{\ty[2]}{\trm}}^v\env \defeq \fun a\sem{\trm}\env[\var \mapsto a]
\end{array}
\]
   \[
\begin{array}{@{}c@{}}
  \begin{array}{@{}r@{}}
      \textbf{Terms:\quad }
  \sem{\val}\env \defeq  \return(\sem{\val}^v\env)\\
  \quad
  \text{ for }\val = \var, \tUnit, \expfun\var\ty\trm
  \\
    \begin{array}[t]{@{}l@{}}
    \sem{\tsample[]}\env
    \\\quad
    \defeq \ssample()
  \end{array}
  \
  \begin{array}[t]{@{}l@{}}
    \sem{\tscore{\trm}}\gamma \defeq
    \\\quad
    \sem{\trm}\gamma\bind^T \sscore
  \end{array}
  \end{array}
  \quad
  \begin{array}{@{}l@{}}
    \sem{\cnst{f}(\trm_1,\ldots,\trm_n)}(\gamma) \defeq\\
    \quad
    \begin{array}[t]{@{}l@{}}
      \sem{\trm_1}\gamma \bind \fun{r_1} \ldots\\
    \sem{\trm_n}\gamma\bind \fun{r_n} \\
    \return(f(r_1,\ldots,r_n))
  \end{array}
  \end{array}
  \quad
  \begin{array}[t]{@{}l@{}}
  \sem{
    \begin{array}{@{}l@{}}
      \ifz[\\]{\trm}{\trm[2]\\}{\trm[3]}
    \end{array}
  }\env
  \end{array}
  \defeq
  \begin{array}{@{}l@{}}
    \sem{\trm}\env\bind\\{}
          [\set 0. \sem{\trm[2]}\env,\\
            \hphantom[ \set0^{\complement}.\sem{\trm[3]}\env]
  \end{array}
  \\
  \begin{array}[t]{@{}l@{}}
    \sem{\tpair{\trm}{\trm[2]}}\env \defeq\\
    \quad
    \begin{array}[t]{@{}l@{}}
    \sem{\trm}\env \bind \fun{a}\\
    \sem{\trm[2]}\env\bind\fun b\\
    \return(\spair ab))
    \end{array}
  \end{array}
    \qquad
  \begin{array}[t]{@{}l@{}}
    \sem{\tInj\ty{\Cns_i}\trm}\env \defeq
    \\\quad
    T\injection_i(\sem{\trm}\env)
    \\
    \sem{\troll\ty\trm }\env \defeq
    \\\quad
    T\semroll (\sem{\trm}\env)
  \end{array}
  \quad
  \begin{array}[t]{@{}l@{}}
  \sem{\trm\ \trm[2]}\env \defeq\\
  \quad\begin{array}[t]{@{}l@{}}
      \sem{\trm}\env \bind \fun f\\
      \sem{\trm[2]}\env \bind \fun a \\
      f\,a
  \end{array}
  \end{array}
  \quad
    \begin{array}[t]{@{}l@{}}
      \sem{
        \begin{array}{@{}l@{}}
        \umatch[\\]
           {\trm}
           {\trm[2]}
        \end{array}
      }\env \defeq
      \begin{array}{@{}l@{}}
        \sem{\trm}\env \bind \fun\_\\
        \sem{\trm[2]}\env
      \end{array}
    \\[10pt]
      \sem{
        \begin{array}{@{}l@{}}
        \pmatch[\\]
           {\trm}
           {\var[1]}{\var[2]}
           {\trm[2]}
        \end{array}
      }\env \defeq
      \begin{array}{@{}l@{}}
        \sem{\trm}\env \bind\fun{\spair ab}\\
        \sem{\trm[2]}\env[\var[1] \mapsto a, \var[2] \mapsto b]
      \end{array}
    \end{array}
    \\
\sem{
  \begin{array}{@{}c@{}}
    \begin{array}[t]{@{}l@{}}
    \vmatch[\\] {\trm }{
                      \Inj{\Cns_1}{\var_1}\To{\trm[2]_1}
                \vor \cdots\\
                 \ \vor\Inj{\Cns_n}{\var_n}\To{\trm[2]_n}
                }\end{array}
          \end{array}
}\env \defeq
\begin{array}{@{}l@{}}
  \sem{\trm}\env\bind \\{}
    [\fun {a_1\,} \sem{\trm[2]_1}\env[\var_1 \mapsto a_1],\ldots,\\
    \hphantom[ \fun {a_n} \sem{\trm[2]_n}\env[\var_n \mapsto a_n]]
\end{array}
\quad
\begin{array}[t]{@{}l@{}}
  \sem{
    \begin{array}{@{}l@{}}
    \rmatch[\\]
           {\trm}
           \var{\trm[2]}
    \end{array}
  }\env
  \defeq
  \begin{array}{@{}l@{}}
    \sem{\trm}\env\bind \fun a\\
    \sem{\trm[2]}\ctx[\var \mapsto\semunroll\,a]
  \end{array}
\end{array}
\end{array}
\]
     \ConferenceArxiv
      {\caption*{\figref{interpretation-terms}.
          \,The denotational interpretation of values (top), and
          terms (bottom).}
      }
      {\caption{The denotational interpretation of values (top), and
          terms (bottom).}
        \figlabel{interpretation-types}
        \figlabel{interpretation-values}
        \figlabel{interpretation-terms}
      }
\end{figure}
 }

Induction on terms and values proves that the semantics has the
following standard properties:
\begin{lemma}[Substitution] \lemmalabel{valsubst}
Let $\vdash \val_{\var}:\ty[2]$, $(\var : \ty[2]) \in \ctx$ be closed
SFPC values and $\ctx \vdash \trm:\ty$ be an SFPC term.  Then $
\sem{\subst{\trm}{\sfor{\var}{\val_{\var}}}_{(\var : \ty[2]) \in
    \ctx}} = \sem{\trm}\coseq[{(\var : \ty[2]) \in \ctx}]{\var \mapsto
  \sem{\val_{\var}}^v()}$.
\end{lemma}

\begin{theorem}[Compositionality] \theoremlabel{monsemctxt}
Let $C[\ctx \vdash -:{\ty}]$ be an SFPC program context and let $\ctx
\vdash\trm,\trm[2]:\ty$ be SFPC terms. If $\sem{\trm}\leq
\sem{\trm[2]}$ then $\sem{C[\trm]} \leq \sem{C[\trm[2]]}$. As a
consequence, the meaning of a term depends only on the meaning of its
sub-terms: if $\sem\trm = \sem{\trm[2]}$ then $\sem{C[\trm]} =
\sem{C[\trm[2]]}$.
\end{theorem}

\subsection{Enriched Semantics}
These semantic definitions can be phrased inside the category $\wQbs$.
Recall from \subsecref{qbs} the adjunction $\Sigma_{-}\dashv M_{-} :
\Qbs \to \Meas$. The coproduct representation of $\Expr{\ctx}{\ty}$ as
$\sum_{n\in\naturals,
  \trm\in\LExpr{\Gamma;\rbox[1],\ldots,\rbox[n]}{\ty}} \reals^n$ is
meaningful in $\wQbs$ too. Because left adjoints preserve co-products,
its underlying measurable space is the measurable space structure of
$\Expr{\ctx}{\ty}$ from \subsecref{syntax-spaces}.  Similarly,
$\Val{\Gamma}{\ty}$ is the underlying measurable space of a qbs.  Both
become \wqbs es via the discrete order.  The denotational
interpretation functions are \wqbs-morphisms
$\sem{-}^v:\Val{\Gamma}{\ty}\to \sem{\ty}^{\sem{\Gamma}}$ and
$\sem{-}:\Expr{\Gamma}{\ty}\to (T \sem{\ty})^{\sem{\Gamma}}$.
\begin{lemma}\lemmalabel{faithobs}
\lemmalabel{faithful} The denotational semantics $ \sem{-}^v:
\Val{~}{\Real}\to \reals $ is a an isomorphism. Therefore, $\sem-^v_T
\defeq T\sem-^v : T\Val{~}{\Real}\to T\reals $ is an isomorphism.
\end{lemma}
The notational conventions defining the big-step operational semantics
kernel internalise in $\wQbs$, if we read $\return^T$ for $\dirac {}$
and $\ssample$ for $\uniform{[0,1]}$.  We can therefore define an
abstract interpreter as an $\omega$-qbs map ${\evalto} :
\Expr{}{\tau}\to T\Val{}{\ty}$. As in \subsecref{valuations},
because we equipped $\Val{}\ty$ with the discrete order, the Kleisli
arrow ${\evalto}: \Expr{}\tau \to T\Val{}\ty$ induces at most one
s-finite kernel ${\evalto}: \Expr{}\tau \leadsto \Val{}\ty$ by
restriction to characteristic functions. This kernel is in fact the
big-step semantics from \subsecref{opsem}.

\subsection{Adequacy}
The key to relating the denotational and operational semantics is to
establish that $\sem{\trm} = \sem{\trm\evalto}^v_T$ as expectation
operators in $T\sem\ty$. As usual, we prove one inequality by
induction on the syntax.
\begin{lemma}[Computational Soundness] \lemmalabel{sound}
For every closed term $\vdash \trm : \ty$, we have $ \sem{\trm} \geq
\sem{\trm\evalto}^v_T$.
\end{lemma}

\begin{figure}
  \[
\begin{array}{@{}c@{}}
  r \trianglelefteq^{\vdash\Real}_v \cnst{r}   \shortiff \top
  \qquad
  \sUnit\trianglelefteq^{\vdash\Unit}_v \tUnit   \shortiff \top
  \qquad
  \spair{a_1}{a_2} \trianglelefteq^{\vdash\ty_1\t* \ty_2}_v \tpair{\val_1}{\val_2}
  \shortiff a_1\trianglelefteq^{\vdash\ty_1}_v \val_1
  \wedge a_2\trianglelefteq^{\vdash\ty_2}_v \val_2
  \\
  \begin{array}[t]{@{}r@{}}
  \injection_i(a)\trianglelefteq^{\vdash\ty}_v\tInj\ty\Cns_j\val
  \shortiff
  i = j \wedge
  a\trianglelefteq^{\vdash\ty_i}_v \val
  \\
  (\ty=\Variant{
                \Inj{\Cns_1}{\ty_1}
                \vor \ldots \vor
                \Inj{\Cns_n}{\ty_n}})
  \end{array}
  \qquad
  \begin{array}[t]{@{}l@{}}
\semroll\,a\trianglelefteq^{\vdash\ty}_v\troll\ty\val
\shortiff\\\quad
a\trianglelefteq^{\vdash\subst{\ty[2]}{\sfor{\tvar}{\ty}}}_v \val
\quad
(\ty=\rec{\tvar}{\ty[2]})
  \end{array}
  \qquad
  \begin{array}[t]{@{}l@{}}
\Lambda(k)\trianglelefteq^{\vdash\ty\To\ty[2]}_v\expfun \var\ty \trm
\shortiff
\\\quad
k\leq \sem{\trm}\wedge k\trianglelefteq^{\var:\ty\vdash \ty[2]}_c \trm
  \end{array}
  \\
  \begin{array}[t]{@{}l@{}}
    \mu\trianglelefteq^{\vdash \ty}_c\trm   \shortiff
    \\\quad
    \mu \leq \sem{t \evalto}^v_T
  \end{array}
  \qquad
  \begin{array}[t]{@{}*3{l@{}}}
b\trianglelefteq_v^{\Gamma\vdash\ty}\val[2] &\shortiff \forall
\seq[{(\var : \ty[2]) \in \ctx}]{a_{\var}\trianglelefteq^{\vdash
    \ty[2]}_v \val_{\var}}.
b\parent{a_{\var}}_{(\var : \ty[2]) \in \ctx}\trianglelefteq^{\vdash\ty}_v
\subst{\val[2]&}{\sfor{\var}{\val_{\var}}}_{(\var : \ty[2]) \in \ctx}\\
k\trianglelefteq_c^{\Gamma\vdash\ty}\trm
&\shortiff \forall
\seq[{(\var : \ty[2]) \in \ctx}]{a_{\var}\trianglelefteq^{\vdash
    \ty[2]}_v \val_{\var}}. k\parent{a_{\var}}_{(\var : \ty[2]) \in \ctx}
\trianglelefteq^{\vdash\ty}_c
\subst{\trm&}{\sfor{\var}{\val_{\var}}}_{(\var : \ty[2]) \in \ctx}\\
  \end{array}
\end{array}
\]
   \caption{The properties of the relational interpretation (logical
    relation) $\trianglelefteq_v$ and
    $\trianglelefteq_c$.}\figlabel{logical-relation}
\end{figure}
For the other direction, we construct a logical relation,
using a variation of Pitts' minimal invariant relations method
in bilimit compact categories \cite{pitts1996relational,levy2012call}.
\begin{lemma}[Relational Interpretation] \lemmalabel{logrelconstr}
  For every context $\Gamma$ and type $\ty$ of SFPC, there are
  relations $\trianglelefteq_v^{\Gamma\vdash \ty} \subseteq
  \wQbs(\sem{\Gamma},\sem{\ty})\times \Val{\Gamma}{\ty}$ and
  $\trianglelefteq_c^{\Gamma\vdash\ty}\subseteq \wQbs(\sem{\Gamma},
  T\sem{\ty}) \times \Expr{\Gamma}{\ty}$ satisfying
  \figref{logical-relation}.
\end{lemma}

We have that $a\trianglelefteq_v^{\vdash\ty} \val$ implies that $a\leq
\sem{\val}^v$.  Therefore, using the Substitution \lemmaref{valsubst}
and the definitions of $\evalto$, $\sem{-}^v$ and $\sem{-}$, we
establish the following fundamental lemma, by a lengthy mutual
induction on the structure of $\val[2]$ and $\trm[2]$.
\begin{lemma}[Fundamental] \lemmalabel{fundamental}
For every $\val[2]\in\Val{\Gamma}{\ty}$ and $\trm[2]\in\Expr{\Gamma}{\ty}$
$
\sem{\val[2]}^v\trianglelefteq^{\Gamma\vdash\ty}_v \val[2]
$ and $
\sem{\trm[2]}\trianglelefteq^{\Gamma\vdash\ty}_c \trm[2]
$.
\end{lemma}
In particular, for closed terms $\vdash \trm : \ty$ we have $\sem{\trm}\leq
\sem{\trm\evalto}^v_T$. Adequacy now follows:
\begin{theorem}[Adequacy]\theoremlabel{adequacy}
For all types $\ty$ of SFPC and all closed terms
$\trm,\trm[2]\in\Expr{~}{\ty}$,
we have that $\sem{\trm}\leq\sem{\trm[2]}$ implies that $\trm\precsim\trm[2]$.
In particular, $\sem{\trm}=\sem{\trm[2]}$ implies that $\trm\approx \trm[2]$.
\end{theorem}
\begin{proof}
Assume
$\sem{\trm}\leq\sem{\trm[2]}$ and consider any any context $C[\ctx \vdash -: {\ty}]$
of type $\Real$.
By the Compositionality \theoremref{monsemctxt},
we have:
$
\sem{C[\trm]}
\leq
\sem{C[\trm[2]]}.
$
Therefore:
$
\sem{C[\trm]\evalto}^v_T
=\sem{C[\trm]}
\leq\sem{C[\trm[2]]}
=\sem{C[\trm[2]]\evalto}^v_T.
$
By \lemmaref{faithful}, we deduce
$
C[\trm]{\evalto}\leq C[\trm[2]]{\evalto}.
$
So $\trm\precsim \trm[2]$.
\end{proof}

By \lemmaref{ICadequacy}, it now follows
that the induced denotational semantics of Idealised Church at the
\wqbs{} $\sem{\UntypedL \Real}$ and the induced denotational semantics
of CBV PPCF are adequate.
\begin{corollary}
For Idealised Church or CBV PPCF terms
$\trm,\trm[2]$,
 $\sem{\trm}\leq\sem{\trm[2]}$ implies that $\trm\precsim\trm[2]$.
\end{corollary}

We conclude with two applications of the Adequacy \theoremref{adequacy}.
First, evaluation order in SFPC and its sub-fragments does not matter, as
a consequence of the monad commutativity (\theoremref{commutative}):
\begin{corollary}\corollarylabel{comm-sem}
  For every $\ctx \vdash \trm[1] : \ty[1]$, $\ctx \vdash \trm[2] :
  \ty[2]$, and $\ctx, \var[1] : \ty[1], \var[2] : \ty[2] \vdash
  \trm[3] : \ty[3]$ we have:
  \[
  \begin{array}{@{}l@{}}
    \expletin {\var[1]}{\ty[1]}{\trm[1]}\\
    \expletin {\var[2]}{\ty[2]}{\trm[2]}
    \trm[3]
  \end{array}
\quad  \approx\quad
  \begin{array}{@{}l@{}}
    \expletin {\var[2]}{\ty[2]}{\trm[2]}\\
    \expletin {\var[1]}{\ty[1]}{\trm[1]}
    \trm[3]
  \end{array}
  \]
\end{corollary}

Second, we show that our definable term-level recursion operator is
indeed a fixed-point operator:
\begin{corollary}
  Let $\ty = \ty_1 \To \ty_2$ be a function type. For every $\ctx,
  \var : \ty \vdash \trm : \ty$ we have $\rec{\var:\ty}{\trm} \approx
  \trm{}[\var \mapsto \rec{\var:\ty}{\trm}]$. Therefore, the following
  derivation rule is admissible:
  \[
  \inferrule{
      \trm{}[\var \mapsto \rec{\var:\ty}{\trm}]
      \evalto \val
    }{
      \rec{\var:\ty}{\trm}
      \evalto
      \val
    }
  \]
\end{corollary}
 \ConferenceArxiv{}{\section{Characterising Quasi-Borel Pre-domains}\seclabel{characterising-wqbs}
\label{characterising-wqbs}
We describe three additional categories equivalent to $\wQbs$.
Each approaches the question of how to combine qbs and \wcpo{}
structures in a different way. We summarise the characterisations:
\ConferenceArxiv{\begin{reptheorem}{7.1}}{\begin{theorem}}
\theoremlabel{characterisation}
  We have equivalences of categories:
  \(
  \wQbs \equivalent \wE \equivalent \IntwCpo\Qbs \equivalent \Mod(\wqbspres, \Set)
  \).
\ConferenceArxiv{\end{reptheorem}}{\end{theorem}}
We describe these equivalent categories. For our semantics, we only
need these consequences:
\ConferenceArxiv{\begin{repcorollary}{7.2}}{\begin{corollary}}
\corollarylabel{cocomplete}
  The category of \wqbs{}es, $\wQbs$, is locally
  $\suc\continuum$-presentable, where $\suc\continuum$ is the
  successor cardinal of the continuum. In particular, it has all small
  limits and colimits.
\ConferenceArxiv{\end{repcorollary}}{\end{corollary}}

\subsection{$\wE$: a Domain-Theoretic Completion of Standard Borel Spaces}

Quasi-Borel spaces are a quasi-topos completion of the category $\Sbs$
of standard Borel spaces. \citet{HeunenKSY-lics17} establish an
equivalence $\QBS- : \E \equivalent \Qbs$, where $\E$ is the full
sub-category of presheaves $[\opposite\Sbs, \Set]$ consisting of those
functors $F : \opposite\Sbs \to \Set$ that are:
\begin{itemize}
\item countable coproduct preserving, a sheaf condition: for every
  countable $I$ and coproduct cocone
  $\pair{\sum_{i \in I}S_i}{\seq[i \in I]{\injection_i}}$, the pair
  $\pair{F\sum_{i \in I}S_i}{\seq[i \in I]{F\injection_i}}$ forms a
  product cone of $\prod_{i \in I}FS_i$; and
  \item separated: the function
    $m_F \defeq \seq[r \in \RR]{F\constantly r} : F\RR \to \prod_{r \in
      \RR}F\terminal$ is injective.
\end{itemize}
This equivalence is given on objects by mapping $F$ to the qbs
$\QBS F$ whose carrier is $F\terminal$ and whose random elements are
the image $m_F[F\RR]$ under the injection from the separatedness
condition. Therefore, $m_F : F\RR \to \prod_{r \in \RR}F\terminal$ restricts to
a bijection $\xi_F : F\RR \isomorphic M_{\QBS F}$.
The equivalence is given on morphisms $\alpha : F \to G$
by $\QBS\alpha \defeq \alpha_{\terminal}$. Conversely, given any $\Qbs$-morphism
$f : \QBS F \to \QBS G$, by setting:
\[\textstyle
  \alpha_{[n]} : F[n] \xto{\isomorphic}
             \prod_{i \in [n]} F\terminal
             \xto{f^n}
             \prod_{i \in [n]} G\terminal
             \xto{\isomorphic}
             G[n]
  \quad
  \alpha_{\RR} :
                 F\RR
                 \xto{\xi_F}
                 M_{\QBS F}
                 \xto{f\compose (-)}
                 M_{\QBS G}
                 \xto{\xi_G}
                 G\RR
\]
we obtain the natural transformation $\alpha : F \to G$ for which
$\QBS\alpha = f$.
Here, we write $[n]$ for a standard Borel space with $n$ elements.
As an adjoint equivalent to $\QBS-$, we can choose,
for every qbs $X$, the separated sheaf $\invQBS Xa \definedby X^{a}$
where $a$ is either an object or a morphism, and the (co)unit of this
adjoint equivalence is given by
$\eta_X : \QBS{\invQBS X{}} = \invQBS X\terminal = X^{\terminal}
\xto{\isomorphic} X$.

We define a similar category $\wE$ to be the full subcategory of
$[\opposite\Sbs, \wCpo]$ consisting of those functors
$F : \opposite\Sbs \to \wCpo$ that are:
\begin{itemize}
\item countable coproduct preserving, i.e., the same sheaf condition; and
\item $\wCpo$-separated: the function
  $m_F \defeq \seq[r \in \RR]{F\constantly r} : F\RR \to \prod_{r \in
    \RR}F\terminal$ is a full mono.
\end{itemize}
Post-composing with the forgetful functor $\carrier - : \wCpo \to \Set$
yields a forgetful functor $\carrier - : \wE \to \E$, as the
underlying function of a full mono is injective.

We equip $\wE$ with an $\wCpo$-category structure by setting the order
componentwise, defining $ \alpha \leq \beta$ when $ \forall S \in
\Sbs. \alpha_S \leq \beta_S$.  Recall that an
\emph{$\wCpo$-equivalence} is an adjoint pair of locally-continuous
functors whose unit and counit are isomorphisms. To specify an
$\wCpo$-equivalence, it suffices to give a fully-faithful
locally-continuous essentially surjective functor in either way,
together with a choice of sources and isomorphisms for the essential
surjectivity.

\ConferenceArxiv{\begin{repproposition}{7.3}}{\begin{proposition}}
  The equivalence ${\wQBS-} : \wE \equivalent\wQbs$ of
  \theoremref{characterisation} is $\wCpo$-enriched, and given by:
  \[
    \wQBS F \definedby
    \triple{\carrier{F\terminal}}{m_F[F\RR]}{\leq_{F\terminal}}
    \qquad\qquad
    \wQBS {\parent{\alpha : F \to G}} \definedby \alpha_{\terminal}
  \]
  Moreover, the two forgetful functors $\carrier- : \wE \to \E$ and
  $\carrier- : \wQbs \to \Qbs$ form a map of adjoints.
\ConferenceArxiv{\end{repproposition}}{\end{proposition}}

\subsection{$\IntwCpo\Qbs$: \wcpo s Internal to $\Qbs$}
The category $\Qbs$ is a Grothendieck quasi-topos, which means it has
a canonical notion of a sub-space: \emph{strong monos} (see after
\exampleref{scott open}). Sub-spaces let us define
relations/predicates, and interpret a fragment of higher-order logic
formulae as subspaces. In particular, we can interpret \wcpo{}'s
definition \emph{internally} to $\Qbs$ and Scott-continuous morphisms
between such \wcpo{}s as a subspace of the $\Qbs$-function space, to
form the category $\IntwCpo\Qbs$.  We interpret functional operations,
like the $\sup$-operation of a \wcpo{} and the homorphisms of \wcpo s,
as internal functions, rather than as internal functional relations,
as the two notions differ in a non-topos quasi-topos such as $\Qbs$.

\subsection{$\Mod(\wqbspres, \Set)$: Models of an Essentially Algebraic Theory}

Both $\wCpo$ and $\Qbs$ are locally presentable categories. Therefore,
there are essentially algebraic theories $\wcpopres$ and $\qbspres$
and equivalences $\Mod(\wcpopres, \Set) \equivalent \wCpo$ and
$\Mod(\qbspres, \Set)\equivalent \Qbs$ of their categories of
set-theoretic algebras.  We combine these two presentations into a
presentation $\wqbspres$ for \wqbs{}, given in
Appx.~\ref{appendix:essentially-algebraic}, by taking their union,
identifying the element sorts and adding a sup operation for \wchain s
of the random elements, and a single axiom stating this sup is
computed pointwise. Local presentability, for example, implies the
existence of all small limits and colimits.
 }

\section{Related work and Concluding Remarks}\seclabel{conclusion}
There is an extensive body of literature on probabilistic power-domain
constructions. We highlight the first denotational models of
higher-order probabilistic languages with
recursion~\cite{saheb1980cpo} and the work of
\citet{jones1989probabilistic} who give an adequate model
of FPC with \emph{discrete} probabilistic choice based on a category
of pre-domains and a probabilistic power-domain construction using
valuations.  Crucially, in this approach, {commutativity of the
  probabilistic power-domain may fail} unless one restricts to certain
subcategories of continuous domains which are known to not be
Cartesian closed.  \citet{JUNG199870} survey the challenges in the
search for such a sub-category of continuous domains that admits
function spaces and probabilistic power-domains. One ingredient of
this problem is that the measurable space structure used is generated
from the Scott topology of the domain at hand. Subsequently
\citet{goubault2011continuous, barker2016monad, mislove2016domains};
and \citet{mardare-scott}
proposed variations on this in which random variables play a crucial
role, as they do in qbses. We extend their reach by establishing
commutativity.

\citet{ehrhard2017measurable} gave an adequate semantics for a
\emph{call-by-name} PCF with continuous probabilistic choice.  The
semantics is based on a variation of \emph{stable} domain theory,
using cones and stable functions, equipped with a notion of random
elements analogous to a qbs-structure which they call
\emph{measurability tests}.  We extend their reach by interpreting
soft constraints, recursive types, and commutativity.  A large
technical difference with our work arises from their choice to work
with cones rather than \wcpo s, and with stable functions, rather than mere
continuous functions, and from our choice to use monadic semantics.

\citet{BorgstromLGS-corr15} conducted a thorough {operational}
treatment of an untyped probabilistic calculus, \`a la Idealised
Church.  Our work complements this analysis with a denotational
counterpart.

\emph{Topological domain theory (TDT)}~\cite{battenfeld2007convenient}
accommodates most of the features that we have considered in this
paper. In particular, it provides a cartesian closed category with an
algebraically compact expansion and a commutative probabilistic
powerdomain. Indeed, \citet{huang-morrisett-spitters} have already
proposed it as a basis for statistical probabilistic programming.  We
leave the formal connections between topological domains and \wqbs es
for future investigation, but we compare briefly, following
\S\ref{sec:summary-of-wqbs}.  In traditional domain theory, the Borel
structure is derived from the order; in TDT, both order and Borel
structure are derived from the topology; and in \wqbs es, both order
and Borel structure are independent (see \subsecref{simple
  types}). Concretely, TDT disallows topological discontinuities such
as our zero-testing conditionals;
moreover it makes a direct connection to computability~\cite{afr-computability}.

To conclude, we developed pre-domains (\S\ref{wqbs}) for statistical
probabilistic recursive programs via:
\begin{itemize}
\item a \emph{convenient} category $\wQbs$, being Cartesian closed and
  (co)complete (\corollaryref{wqbs-structure});
\item a commutative probabilistic power-domain modelling synthetic measure
theory~(\S\ref{monad});
\item an interpretation of recursive types, through axiomatic domain theory (\S\ref{axiomatic-domain-theory});
\item adequate models of recursive languages with continuous probabilistic
choice and soft constraints, of recursively typed, untyped and simply types varieties (\S\ref{semantics});
\item canonicity through four independent
characterisations of $\wQbs$~(\S\ref{characterising-wqbs}).
\end{itemize}
This semantics gives sound reasoning principles about recursive
probabilistic programs.

\appendix
\begin{acks}

  Research supported by Balliol College Oxford (Career Development
  Fellowship), EPSRC (grants EP/N007387/1, EP/M023974/1), and the
  Royal Society.
  We are grateful to the reviewers for their suggestions.
  It has been helpful to discuss this work with the Oxford PL and Foundations groups,
  at the Domains 2018 and HOPE 2018 workshops, and with Marcelo Fiore, Chris Heunen, Paul Levy, Carol Mak, Gordon Plotkin, Alex Simpson, Hongseok Yang, amongst others.
\hide{  We thank A.~Bauer, S.~Castellan, S.~Dash, M.~Escard\'o, M.~Fiore,
  C.~Heunen, M.~Huot, A.~Jung, Y.~Kaddar, P.~Levy, C.~Mak, C.~Matache,
  M.~Mislove, S.~Moss, M.~New, M.~Pagani, H.~Paquet, P.~Panangaden,
  G.~Plotkin, M.~Pretnar, A.~\'Scibior, D.~Scott, A.~Simpson,
  D.~Stein, I.~Stark, C.~Tasson, A.~Vandenbroucke, M.~Wand, H.~Yang,
  and the anonymous reviewers for fruitful discussions and
  suggestions.}
\end{acks}
 \clearpage
\bibliography{references}

\clearpage
\ConferenceArxiv{\section{Explicit semantics}\label{appx:semantic-figures}
We give explicit description of the operational and denotational
semantics. Thanks to our conventions and set-up, they are standard.

\begin{figure}
  \[
\begin{array}{@{}c@{}}
  \trm\evalto[0]A \defeq 0
  \qquad
  \inferrule{
    \trm_1 \evalto[n] \cnst {r_1}
    \quad\ldots\quad
    \trm_k \evalto[n] \cnst {r_k}
  }{
    \cnst f(\trm_1, \ldots, \trm_k) \evalto[n] \cnst {f(r_1, \ldots, r_n)}
  }
  \qquad
  \begin{array}{@{}l@{}}
    \parent{\tsample[]\evalto[n]} \defeq
    \\
    \qquad
     \uniform{[0,1]} \bind \lambda r. \dirac{\cnst r}
  \end{array}
  \qquad
  \begin{array}{@{}l@{}}
    \parent{\tscore[](\trm)\evalto[n]} \defeq
    \\
    \quad
    \parent{\trm\evalto[n]} \bind \lambda \cnst r . \abs r \cdot \dirac \tUnit
  \end{array}
  \\
  \inferrule{
    \trm\evalto[n] \cnst 0
    \quad
    \trm[2]_1\evalto[n] \val
  }{
    \ifz{\trm}{\trm[2]_1}{\trm[2]_2})\evalto[n]\val
  }
  \quad
  \inferrule{
    \trm\evalto[n] \cnst r
    \quad
    \trm[2]_2\evalto[n] \val
  }{
    \ifz{\trm}{\trm[2]_1}{\trm[2]_2})\evalto[n]\val
  }(r \neq 0)
  \\
  \inferrule{
    ~
  }{
    \val \evalto[n] \val
  }(\val = \tUnit, \expfun\var\ty\trm)
  \quad
  \inferrule{
    \trm\evalto[n] \tUnit
    \quad
    \trm[2]\evalto[n] \val
  }{
    \umatch
           \trm
           {\trm[2]}
    \evalto[n]
    \val
  }
  \quad
  \inferrule{
    \trm[1]\evalto\val[1]
    \quad
    \trm[2]\evalto\val[2]
  }{
    \tpair{\trm[1]}{\trm[2]}
    \evalto[n]
    \tpair{\val[1]}{\val[2]}
  }
  \quad
    \inferrule{
    \trm\evalto[n]\val
  }{
    \tInj\ty\Cns\trm
    \evalto[n]
    \tInj\ty\Cns\val
  }
  \\
  \inferrule{
    \trm\evalto[n]\tpair{\val[2]}{\val[3]}
    \quad
    \trm[2][{\var[1] \mapsto \val[2], \var[2] \mapsto \val[3]}]
    \evalto[n-1] \val
  }{
    \pmatch
           \trm
           {\var[1]}{\var[2]}
           {\trm[2]}
    \evalto[n]
    \val
  }
  \
  \inferrule{
    \trm \evalto[n] \Inj{\Cns_{\mathnormal i}}{\val[2]}
  \quad
  \trm[2]_i[\var_i \mapsto \val[2]] \evalto[n-1] \val
  }{
    \begin{array}[t]{@{}r@{\,}l@{}l@{}}
    \vmatch \trm{
                    &\Inj[]{\Cns_{\mathnormal 1}\ }{\var_1}&\To{\trm[2]_1}
                    \vor\cdots\\
                    \vor&\Inj[]{\Cns_{\mathnormal m}}{\var_m}&\To{\trm[2]_m}
    }
    \evalto[n]
    \val
      \end{array}
  }
  \
  \inferrule{
    \trm\evalto[n] \val
  }{
    \troll\ty(\trm) \evalto[n] \troll\ty\val
  }
  \\
  \inferrule{
    \trm \evalto[n] \troll\ty\val[2]
    \ \ \
    \trm[2][\var \mapsto \val[2]] \evalto[n-1] \val
  }{
    \rmatch
           \trm
           {\var}
           {\trm[2]}
    \evalto[n]
    \val
  }
  \
  \inferrule{
    \trm\evalto[n] \expfun{\var}{\ty}{\trm[3]}
    \ \ \
    \trm[2]\evalto[n] \val[2]
    \ \ \
    \trm[3][\var \mapsto \val[2]]
    \evalto[n-1]
    \val
  }{
    \trm\, \trm[2]
    \evalto[n] \val
  }
  \end{array}
\]
   \ConferenceArxiv
      {\caption*{\figref{operational-semantics}.
          \,SFPC big-step operational semantics ($n > 0$)
      }}
      {\caption{
          SFPC big-step operational semantics ($n > 0$)
      }}
    \ConferenceArxiv{}{\figlabel{operational-semantics}}
\end{figure}
     \figref{operational-semantics}
presents the indexed evaluation
kernels. Primitives evaluate their arguments left-to-right.
 Sampling
and conditioning evaluate using the interpreter's statistical sampling
and conditioning primitives. We could equivalently use a lower-level
interpreter supporting only (sub)-probabilistic sampling and
maintaining an aggregate weight, i.e., a kernel
$\evalto[n]' : \Expr[]{}{} \leadsto \reals\times\Val[]{}{}$ and the rule:
\[
  \inferrule{
    t \evalto[n]' {\tpair {r_1}{\cnst{r_2}}}
  }{
    \tscore(t) \evalto[n]' \tpair{r_1\cdot \abs{r_2}}{\tUnit}
  }
\]
and the other rules changing analogously. The remaining rules are
standard. The index decreases only when we apply the kernel
non-compositionally in the elimination forms for binary products,
variants, recursive types, and function application.

\figref{interpretation-terms} presents the denotational semantics for
SFPC using the semantic structures we developed.

\begin{figure}
  \[
\begin{array}{@{}c@{}}
  \textbf{Values:\hspace{1.25cm}}
  \quad
  \sem{\var}^v\env \defeq \projection_{\var}\env
  \quad
  \sem{\cnst{r}}^v\env \defeq r
  \quad
  \sem{\tUnit}^v\env \defeq \sUnit
  \quad
  \sem{\tpair{\val}{\val[2]}}^v\env \defeq \spair{\sem{\val}^v\env}{\sem{\val[2]}^v\env}
  \\
  \sem{\tInj\ty{\Cns_i}\val}^v\env \defeq \injection_i(\sem{\val}^v\env)
  \qquad
  \sem{\troll\ty(\val) }^v \defeq \semroll(\sem{\val}^v\env)
  \qquad
  \sem{\expfun{\var}{\ty[2]}{\trm}}^v\env \defeq \fun a\sem{\trm}\env[\var \mapsto a]
\end{array}
\]
   \[
\begin{array}{@{}c@{}}
  \begin{array}{@{}r@{}}
      \textbf{Terms:\quad }
  \sem{\val}\env \defeq  \return(\sem{\val}^v\env)\\
  \quad
  \text{ for }\val = \var, \tUnit, \expfun\var\ty\trm
  \\
    \begin{array}[t]{@{}l@{}}
    \sem{\tsample[]}\env
    \\\quad
    \defeq \ssample()
  \end{array}
  \
  \begin{array}[t]{@{}l@{}}
    \sem{\tscore{\trm}}\gamma \defeq
    \\\quad
    \sem{\trm}\gamma\bind^T \sscore
  \end{array}
  \end{array}
  \quad
  \begin{array}{@{}l@{}}
    \sem{\cnst{f}(\trm_1,\ldots,\trm_n)}(\gamma) \defeq\\
    \quad
    \begin{array}[t]{@{}l@{}}
      \sem{\trm_1}\gamma \bind \fun{r_1} \ldots\\
    \sem{\trm_n}\gamma\bind \fun{r_n} \\
    \return(f(r_1,\ldots,r_n))
  \end{array}
  \end{array}
  \quad
  \begin{array}[t]{@{}l@{}}
  \sem{
    \begin{array}{@{}l@{}}
      \ifz[\\]{\trm}{\trm[2]\\}{\trm[3]}
    \end{array}
  }\env
  \end{array}
  \defeq
  \begin{array}{@{}l@{}}
    \sem{\trm}\env\bind\\{}
          [\set 0. \sem{\trm[2]}\env,\\
            \hphantom[ \set0^{\complement}.\sem{\trm[3]}\env]
  \end{array}
  \\
  \begin{array}[t]{@{}l@{}}
    \sem{\tpair{\trm}{\trm[2]}}\env \defeq\\
    \quad
    \begin{array}[t]{@{}l@{}}
    \sem{\trm}\env \bind \fun{a}\\
    \sem{\trm[2]}\env\bind\fun b\\
    \return(\spair ab))
    \end{array}
  \end{array}
    \qquad
  \begin{array}[t]{@{}l@{}}
    \sem{\tInj\ty{\Cns_i}\trm}\env \defeq
    \\\quad
    T\injection_i(\sem{\trm}\env)
    \\
    \sem{\troll\ty\trm }\env \defeq
    \\\quad
    T\semroll (\sem{\trm}\env)
  \end{array}
  \quad
  \begin{array}[t]{@{}l@{}}
  \sem{\trm\ \trm[2]}\env \defeq\\
  \quad\begin{array}[t]{@{}l@{}}
      \sem{\trm}\env \bind \fun f\\
      \sem{\trm[2]}\env \bind \fun a \\
      f\,a
  \end{array}
  \end{array}
  \quad
    \begin{array}[t]{@{}l@{}}
      \sem{
        \begin{array}{@{}l@{}}
        \umatch[\\]
           {\trm}
           {\trm[2]}
        \end{array}
      }\env \defeq
      \begin{array}{@{}l@{}}
        \sem{\trm}\env \bind \fun\_\\
        \sem{\trm[2]}\env
      \end{array}
    \\[10pt]
      \sem{
        \begin{array}{@{}l@{}}
        \pmatch[\\]
           {\trm}
           {\var[1]}{\var[2]}
           {\trm[2]}
        \end{array}
      }\env \defeq
      \begin{array}{@{}l@{}}
        \sem{\trm}\env \bind\fun{\spair ab}\\
        \sem{\trm[2]}\env[\var[1] \mapsto a, \var[2] \mapsto b]
      \end{array}
    \end{array}
    \\
\sem{
  \begin{array}{@{}c@{}}
    \begin{array}[t]{@{}l@{}}
    \vmatch[\\] {\trm }{
                      \Inj{\Cns_1}{\var_1}\To{\trm[2]_1}
                \vor \cdots\\
                 \ \vor\Inj{\Cns_n}{\var_n}\To{\trm[2]_n}
                }\end{array}
          \end{array}
}\env \defeq
\begin{array}{@{}l@{}}
  \sem{\trm}\env\bind \\{}
    [\fun {a_1\,} \sem{\trm[2]_1}\env[\var_1 \mapsto a_1],\ldots,\\
    \hphantom[ \fun {a_n} \sem{\trm[2]_n}\env[\var_n \mapsto a_n]]
\end{array}
\quad
\begin{array}[t]{@{}l@{}}
  \sem{
    \begin{array}{@{}l@{}}
    \rmatch[\\]
           {\trm}
           \var{\trm[2]}
    \end{array}
  }\env
  \defeq
  \begin{array}{@{}l@{}}
    \sem{\trm}\env\bind \fun a\\
    \sem{\trm[2]}\ctx[\var \mapsto\semunroll\,a]
  \end{array}
\end{array}
\end{array}
\]
     \ConferenceArxiv
      {\caption*{\figref{interpretation-terms}.
          \,The denotational interpretation of values (top), and
          terms (bottom).}
      }
      {\caption{The denotational interpretation of values (top), and
          terms (bottom).}
        \figlabel{interpretation-types}
        \figlabel{interpretation-values}
        \figlabel{interpretation-terms}
      }
\end{figure}
  }{}
\ConferenceArxiv{\section{Characterising Quasi-Borel Pre-domains}\seclabel{characterising-wqbs}
\label{characterising-wqbs}
We describe three additional categories equivalent to $\wQbs$.
Each approaches the question of how to combine qbs and \wcpo{}
structures in a different way. We summarise the characterisations:
\ConferenceArxiv{\begin{reptheorem}{7.1}}{\begin{theorem}}
\theoremlabel{characterisation}
  We have equivalences of categories:
  \(
  \wQbs \equivalent \wE \equivalent \IntwCpo\Qbs \equivalent \Mod(\wqbspres, \Set)
  \).
\ConferenceArxiv{\end{reptheorem}}{\end{theorem}}
We describe these equivalent categories. For our semantics, we only
need these consequences:
\ConferenceArxiv{\begin{repcorollary}{7.2}}{\begin{corollary}}
\corollarylabel{cocomplete}
  The category of \wqbs{}es, $\wQbs$, is locally
  $\suc\continuum$-presentable, where $\suc\continuum$ is the
  successor cardinal of the continuum. In particular, it has all small
  limits and colimits.
\ConferenceArxiv{\end{repcorollary}}{\end{corollary}}

\subsection{$\wE$: a Domain-Theoretic Completion of Standard Borel Spaces}

Quasi-Borel spaces are a quasi-topos completion of the category $\Sbs$
of standard Borel spaces. \citet{HeunenKSY-lics17} establish an
equivalence $\QBS- : \E \equivalent \Qbs$, where $\E$ is the full
sub-category of presheaves $[\opposite\Sbs, \Set]$ consisting of those
functors $F : \opposite\Sbs \to \Set$ that are:
\begin{itemize}
\item countable coproduct preserving, a sheaf condition: for every
  countable $I$ and coproduct cocone
  $\pair{\sum_{i \in I}S_i}{\seq[i \in I]{\injection_i}}$, the pair
  $\pair{F\sum_{i \in I}S_i}{\seq[i \in I]{F\injection_i}}$ forms a
  product cone of $\prod_{i \in I}FS_i$; and
  \item separated: the function
    $m_F \defeq \seq[r \in \RR]{F\constantly r} : F\RR \to \prod_{r \in
      \RR}F\terminal$ is injective.
\end{itemize}
This equivalence is given on objects by mapping $F$ to the qbs
$\QBS F$ whose carrier is $F\terminal$ and whose random elements are
the image $m_F[F\RR]$ under the injection from the separatedness
condition. Therefore, $m_F : F\RR \to \prod_{r \in \RR}F\terminal$ restricts to
a bijection $\xi_F : F\RR \isomorphic M_{\QBS F}$.
The equivalence is given on morphisms $\alpha : F \to G$
by $\QBS\alpha \defeq \alpha_{\terminal}$. Conversely, given any $\Qbs$-morphism
$f : \QBS F \to \QBS G$, by setting:
\[\textstyle
  \alpha_{[n]} : F[n] \xto{\isomorphic}
             \prod_{i \in [n]} F\terminal
             \xto{f^n}
             \prod_{i \in [n]} G\terminal
             \xto{\isomorphic}
             G[n]
  \quad
  \alpha_{\RR} :
                 F\RR
                 \xto{\xi_F}
                 M_{\QBS F}
                 \xto{f\compose (-)}
                 M_{\QBS G}
                 \xto{\xi_G}
                 G\RR
\]
we obtain the natural transformation $\alpha : F \to G$ for which
$\QBS\alpha = f$.
Here, we write $[n]$ for a standard Borel space with $n$ elements.
As an adjoint equivalent to $\QBS-$, we can choose,
for every qbs $X$, the separated sheaf $\invQBS Xa \definedby X^{a}$
where $a$ is either an object or a morphism, and the (co)unit of this
adjoint equivalence is given by
$\eta_X : \QBS{\invQBS X{}} = \invQBS X\terminal = X^{\terminal}
\xto{\isomorphic} X$.

We define a similar category $\wE$ to be the full subcategory of
$[\opposite\Sbs, \wCpo]$ consisting of those functors
$F : \opposite\Sbs \to \wCpo$ that are:
\begin{itemize}
\item countable coproduct preserving, i.e., the same sheaf condition; and
\item $\wCpo$-separated: the function
  $m_F \defeq \seq[r \in \RR]{F\constantly r} : F\RR \to \prod_{r \in
    \RR}F\terminal$ is a full mono.
\end{itemize}
Post-composing with the forgetful functor $\carrier - : \wCpo \to \Set$
yields a forgetful functor $\carrier - : \wE \to \E$, as the
underlying function of a full mono is injective.

We equip $\wE$ with an $\wCpo$-category structure by setting the order
componentwise, defining $ \alpha \leq \beta$ when $ \forall S \in
\Sbs. \alpha_S \leq \beta_S$.  Recall that an
\emph{$\wCpo$-equivalence} is an adjoint pair of locally-continuous
functors whose unit and counit are isomorphisms. To specify an
$\wCpo$-equivalence, it suffices to give a fully-faithful
locally-continuous essentially surjective functor in either way,
together with a choice of sources and isomorphisms for the essential
surjectivity.

\ConferenceArxiv{\begin{repproposition}{7.3}}{\begin{proposition}}
  The equivalence ${\wQBS-} : \wE \equivalent\wQbs$ of
  \theoremref{characterisation} is $\wCpo$-enriched, and given by:
  \[
    \wQBS F \definedby
    \triple{\carrier{F\terminal}}{m_F[F\RR]}{\leq_{F\terminal}}
    \qquad\qquad
    \wQBS {\parent{\alpha : F \to G}} \definedby \alpha_{\terminal}
  \]
  Moreover, the two forgetful functors $\carrier- : \wE \to \E$ and
  $\carrier- : \wQbs \to \Qbs$ form a map of adjoints.
\ConferenceArxiv{\end{repproposition}}{\end{proposition}}

\subsection{$\IntwCpo\Qbs$: \wcpo s Internal to $\Qbs$}
The category $\Qbs$ is a Grothendieck quasi-topos, which means it has
a canonical notion of a sub-space: \emph{strong monos} (see after
\exampleref{scott open}). Sub-spaces let us define
relations/predicates, and interpret a fragment of higher-order logic
formulae as subspaces. In particular, we can interpret \wcpo{}'s
definition \emph{internally} to $\Qbs$ and Scott-continuous morphisms
between such \wcpo{}s as a subspace of the $\Qbs$-function space, to
form the category $\IntwCpo\Qbs$.  We interpret functional operations,
like the $\sup$-operation of a \wcpo{} and the homorphisms of \wcpo s,
as internal functions, rather than as internal functional relations,
as the two notions differ in a non-topos quasi-topos such as $\Qbs$.

\subsection{$\Mod(\wqbspres, \Set)$: Models of an Essentially Algebraic Theory}

Both $\wCpo$ and $\Qbs$ are locally presentable categories. Therefore,
there are essentially algebraic theories $\wcpopres$ and $\qbspres$
and equivalences $\Mod(\wcpopres, \Set) \equivalent \wCpo$ and
$\Mod(\qbspres, \Set)\equivalent \Qbs$ of their categories of
set-theoretic algebras.  We combine these two presentations into a
presentation $\wqbspres$ for \wqbs{}, given in
Appx.~\ref{appendix:essentially-algebraic}, by taking their union,
identifying the element sorts and adding a sup operation for \wchain s
of the random elements, and a single axiom stating this sup is
computed pointwise. Local presentability, for example, implies the
existence of all small limits and colimits.
 }{}
\section{An Essentially Algebraic Theory for $\wQbs$}
\label{appendix:essentially-algebraic}
An \wcpo{} or a qbs are essentially algebraic, in a precise
sense~\cite[Chapter~3.D]{adamek-rosicky}. We will use this algebraic
nature to analyse the quasi-Borel pre-domains and see how the quasi-Borel
structure interacts with the \wcpo{} structure.

\subsection{Presentations}
For the following, fix a
regular cardinal $\kappa$. Given a set $\Sorts$ with cardinality
$\cardinality \Sorts < \kappa$, whose elements we call \emph{sorts}, an
\emph{$\Sorts$-sorted ($\kappa$-ary) signature} $\Sig$ is a pair $\Sig
= \pair\Ops\arity$ consisting of a set of \emph{operations} and
$\arity : \Ops \to \Sorts^{<\kappa}\times\Sorts$ assigns to each
operation $\op \in \Ops$ a sequence $\seq[i \in I]{\s_i}$, indexed by
some set $I$ of cardinality $\cardinality I < \kappa$, assigning to
each index $i \in I$ its \emph{argument sort}, together with
another \emph{result sort} $\s$. We write $(\op : \prod_{i \in
  I}{\s_i} \to \s) \in \Sig$ for $\arity(\op) = \pair{\seq[i \in
    I]{\s_i}}\s$.

Given an $\Sorts$-sorted signature $\Sig$, and a $\Sorts$-indexed
sequence of sets $\Var = \seq[\s \in \Sorts]{\Var_s}$ of
\emph{variables} we define the collections of $\Sorts$-sorted terms
$\Term^{\Sig}\Var = \seq[\s \in \Sorts]{\Term^{\Sig}_{\s}\Var}$ over $\Var$
inductively as follows:
\[
  \inferrule{
    ~
  }{
    x \in \Term_{\s}\Var
  }(x \in \Var_{\s})
  \qquad
  \inferrule{
    \text{for all $i \in I$, } t_i \in \Term_{s_i}\Var
  }{
    \op\seq[n \in A]{t_i} \in \Term_{\s}\Var
  }((\op : \prod_{i \in I}{\s_i} \to \s) \in \Sig)
\]
Given a signature $\Sig$ and a sort $\s$, an \emph{equation of sort
  $\s$} is a pair of terms $\pair{t_1}{t_2} \in \Term_{\s}\Var$ over
some set of variables. As each term must involve less than
$\kappa$-many variables, due to $\kappa$'s regularity, and so we may
fix the indexed set of variables $\Var$ to be any specified collection
of sets of cardinality $\kappa$.

\begin{definition}
  An \emph{essentially algebraic presentation} $\Pres$ is a tuple
  $\seq{\Sorts, \tSig, \pSig, \Def, \Eq}$ containing:
  \begin{itemize}
  \item a set $\Sorts$ of \emph{sorts};
  \item two $\Sorts$-sorted signatures with disjoint sets of operations:
    \begin{itemize}
    \item a signature $\tSig$ of \emph{total} operations;
    \item a signature $\pSig$ of \emph{partial} operations;
    \end{itemize}
    we denote their combined signature by $\uSig \defeq
    \pair{\Ops_{\tSig}\union \Ops_{\pSig}}{[\arity_{\tSig},
        \arity_{\pSig}]}$;
  \item for each $(\op : \prod_{i \in I}\s_i \to \s)\in \pSig$, a set
    $\Def(\op)$ of $\tSig$-equations over the variables $\set {x_i :
    \s_i \suchthat i \in I}$ which we call the \emph{assumptions of
    $\op\seq[i \in I]{x_i}$}; and
  \item a set $\Eq$ of $\uSig$-equations which we call the \emph{axioms}.
  \end{itemize}
\end{definition}

The point of this definition is just to introduce the relevant
vocabulary. We will only be considering the following presentation for
posets, then \wcpo s, then qbses, then \wqbs {es}:

\begin{example}[{poset presentation cf.~\cite[Examples~3.35(1),(4)]{adamek-rosicky}}]
  The presentation of \emph{posets}, $\pospres$, has two sorts:
  \begin{itemize}
  \item $\elem$, which will be the carrier of the poset; and
  \item $\ineq$, which will describe the poset structure.
  \end{itemize}
  The total operations are:
  \begin{itemize}
  \item $\source : \ineq \to \elem$, assigning to each inequation its lower element;
  \item $\target : \ineq \to \elem$, assigning to each inequation its upper element;
  \item $\refl : \elem \to \ineq$, used to impose reflexivity;
  \end{itemize}
  The partial operations are:
  \begin{itemize}
  \item $\irrelevance : \ineq \times \ineq \to \ineq$, used to impose
    proof-irrelevance on inequations, with $\Def(\irrelevance(e_1, e_2))$:
    \begin{mathpar}
      \source(e_1) = \source (e_2)

      \target(e_1) = \target(e_2)
    \end{mathpar}
  \item $\antisym : \ineq \times \ineq \to \elem$, used to impose anti-symmetry on inequations, with $\Def(\antisym(e, \opposite e))$:
    \begin{mathpar}
      \source(e) = \target(\opposite e)

      \target(e) = \source(\opposite e)
    \end{mathpar}
  \item $\trans   : \ineq \times \ineq \to \ineq$, used to impose transitivity on inequations, with $\Def(\trans(e_1, e_2))$:
    \begin{mathpar}
      \target(e_1) = \source(e_2)
    \end{mathpar}
  \end{itemize}

    The axioms are:
  \begin{gather*}
    \tag{proof irrelevance}\label{proof irrelevance}
    e_1 = \irrelevance(e_1, e_2) = e_2
    \\
    \tag{reflexivity}\label{reflexivity}
    \source(\refl(x)) = x = \target(\refl(x))
    \\
    \tag{anti-symmetry}\label{anti-symmetry}
    \source(e_1) = \antisym(e_1, e_2) = \source(e_2)
    \\
    \tag{transitivity}\label{transitivity}
    \source(\trans(e_1, e_2)) = \source(e_1)
    \qquad
    \target(\trans(e_1, e_2)) = \target(e_2)
  \end{gather*}
\end{example}

\begin{example}[\wcpo{} presentation]
  In addition to the operations and axioms for posets, the
  presentation $\wcpopres$ of \wcpo{}s includes the following partial
  operations:
  \begin{itemize}
  \item $\lub     : \prod_{n \in \NN}\ineq \to \elem$, used to express lubs of \wchain s, with
    $\Def(\lub_{n \in \NN}e_n)$:
    \begin{mathpar}
      \target(e_n) = \source(e_{n+1}), \text{for each $n \in \NN$}
    \end{mathpar}
  \item for each $k \in \NN$, $\ub_k : \prod_{n \in \NN}\ineq \to
    \ineq$, collectively used to impose the lub being an upper-bound, with $\Def(\ub_k\seq[n \in \NN]{e_n})$:
    \begin{mathpar}
      \target(e_n) = \source(e_{n+1}), \text{for each $n \in \NN$}
    \end{mathpar}
  \item $\least : \elem\times\prod_{n \in \NN} \ineq\times \prod_{n \in \NN}\ineq
    \to \ineq$, used to express that the lub is the least bound, with
    $\Def(\least(x, \seq[n\in\NN]{e_n}, \seq[n \in \NN]{b_n}))$:
    \begin{mathpar}
      \target(e_n) = \source(e_{n+1})

      \target(b_n) = x

      \source(e_n) = \source(b_n), \text{for each $n \in \NN$}
    \end{mathpar}
  \end{itemize}

  The axioms are:
  \begin{gather*}
    \tag{upper bound}\label{upper bound}
    \source(\ub_k\seq[n]{e_n}) = \source(e_k)
    \qquad
    \target(\ub_k\seq[n]{e_n}) = \lub\seq[n]{e_n}
    \\
    \tag{least upper bound}\label{least upper bound}
    \source(\least(x, \seq[n]{e_n}, \seq[n]{b_n})) = \lub\seq[n]{e_n}
    \qquad
    \target(\least(x, \seq[n]{e_n}, \seq[n]{b_n})) = x
  \end{gather*}
\end{example}

\begin{example}[qbs presentation]
  The presentation of \emph{qbses}, $\qbspres$, has two sorts:
  \begin{itemize}
  \item $\elem$, which will be the carrier of the qbs; and
  \item $\rand$, which will be the random elements.
  \end{itemize}
  The total operations are:
  \begin{itemize}
  \item $\ev_r : \rand \to \elem$, for each $r \in \RR$, evaluating a
    random element at $r \in \RR$;
  \item $\const : \elem \to \rand$ assigning to each element $x$ the
    constantly-$x$ random element;
  \item $\rearrange_{\phi} : \rand \to \rand$, for each $\phi : \RR \to
    \RR$ Borel-measurable, precomposing random elements with $\phi$; and
  \item $\match_{\seq[i \in I]{S_i}} : \prod_{i \in I}\rand \to
    \rand$, for each $I \subset \NN$ and measurable partition $\RR =
    \sum_{i \in I}S_i$, pasting together a countable collection
    of random elements into a countable-case split.
  \end{itemize}
  There is one partial operation:
  \begin{itemize}
  \item $\extensionality : \rand\times\rand \to \rand$, used for
    establishing that a random element is uniquely determined
    extensionally, with $\Def(\extensionality(\alpha, \beta))$ given by
    \[
    \set{\ev_r(\alpha) = \ev_r(\beta) : \elem \suchthat r \in \RR}
    \]
  \end{itemize}

  The axioms are:
  \begin{gather*}
  \tag{extensionality}\label{extensionality}
  \alpha = \extensionality(\alpha, \beta) = \beta
  \\\tag{constantly}\label{const}
  \set{ \ev_r(\const{(x)}) = x \suchthat r \in \RR}
  \\\tag{rearrange}\label{rearrange}
  \set{ \ev_r(\rearrange_{\phi}\alpha) = \ev_{\phi(r)}\alpha
    \suchthat \text{$\phi : \RR \to \RR$ Borel-measurable}, r \in \RR}
  \\\tag{match}\label{match}
  \set{
    \ev_r\parent{\match_{\seq[i \in I]{S_i}}\seq[i \in I]{\alpha_i}}
    =
    \ev_r(\alpha_i)
    \suchthat I \subset \NN,
    \text{$\RR = \sum_{i \in I}S_i$ Borel-measurable}, i \in I, r \in S_i
    }
  \end{gather*}
\end{example}

We can now present the \wqbs{}es:

\begin{example}[\wqbs{} presentation]
  The presentation $\wqbspres$ of \wqbs{}es extends the presentations
  $\wcpopres$ and $\qbspres$, identifying the $\elem$ sorts, with the
  following additional partial operation:
  \begin{itemize}
  \item
    $\rlub : \prod_{n \in \NN}\rand \times \prod_{n \in \NN, r \in
      \RR}\ineq \to \rand$, used for establishing that the
    random-elements are closed under lubs w.r.t.~the pointwise order,
    with
    $\Def{\rlub(\seq[n \in \NN]{\alpha_n}, \seq[n \in \NN, r \in
      \RR]{e_n^r})}$ given by:
    \[
      \set{
        \source(e^r_n) = \ev_r(\alpha_n),
        \target(e^r_n) = \ev_r(\alpha_{n+1}),
        \suchthat n \in \NN, r \in \RR}
    \]
  \end{itemize}
  The additional axioms are:
  \[
    \tag{pointwise lubs}\label{pointwise lubs}
    \set{
      \ev_r\parent{\rlub\parent{
          \seq[n \in \NN]{\alpha_n},
          \seq[n \in \NN, r \in \RR]{e_n^r}
        }
      }
      =
      \lub\seq[n \in \NN]{e_n^r}
      \suchthat
      r \in \RR
    }
  \]
\end{example}

\subsection{Algebras}

Every essentially algebraic presentation induces a category of
set-theoretic models, and this category for the \wcpo{} presentation
is equivalent to $\wCpo$. Moreover, we can interpret such
presentations in any category with sufficient structure, namely
countable products and equalisers (i.e., countable limits). We briefly
recount how to do this.

Let $\cat$ be a category with $\lambda$-small limits, with $\lambda$
regular.  As usual, if $\Sig$ is any $\Sorts$-sorted $\lambda$-ary
signature, we define a (multi-sorted) \emph{$\Sig$-algebra} $\Alg =
\pair{\seq[\s \in \Sorts]{\sem{\s}}}{\sem-}$ to consist of an
$\Sorts$-indexed family of objects $\seq[\s \in \Sorts]{\sem\s}$, the
\emph{carrier} of the algebra, and an assignment, to each $\op :
\prod_{i \in I}\s_i \to \s$ in $\Sig$, of a morphism:
\[
\sem\op : \prod_{i \in I}\sem{\s_i} \to \sem\s
\]
Given such an algebra $\Alg$, and an $\Sorts$-indexed set $\Var$ of
variables with $\cardinality{\Var_{\s}} < \lambda$ for each $\s \in
\Sorts$, each term $t$ in $\Term_{\s}\Var$ denotes a morphism:
\[
\sem t_{\s} : \prod_{\s \in \Sorts}\sem{\s}^{\Var_{\s}} \to \sem\s
\]
as follows:
\begin{mathpar}
  \sem{x}_{\s} : \prod_{\s \in \Sorts}\sem{\s}^{\Var_{\s}} \xto{\projection_{\s}}
  \sem{\s}^{\Var_{\s}}
  \xto{\projection_{x}}
  \sem\s

  \sem{f\seq[i \in I]{t_i}} :
  \prod_{\s \in \Sorts}\sem{\s}^{\Var_{\s}}
  \xto{\seq[i \in I]{\sem{t_i}}}
  \prod_{i \in I}\sem{\s_i}
  \xto{\sem f}
  \sem{\s}
\end{mathpar}
(When $\cardinality {\Var_{\s}} \geq \lambda$, there are less than
$\lambda$ different variables that actually appear in $t$, and so we
can find a smaller set of sorts and variables for which to define as
above.)

A \emph{$\Sig$-homomorphism} $h : \Alg \to \Alg[2]$ between
$\Sig$-algebras $\Alg$, $\Alg[2]$ is an $\Sorts$-indexed family of
functions $h_{\s} : \Alg\sem{\s} \to \Alg[2]\sem\s$ such that, for
every operation symbol $\op : \prod_{i \in I}\s_i \to \s$ in $\Sig$:
\insertdiagram{01}

We denote the category of $\Sig$-algebras in $\cat$ and their
homomorphisms by $\Mod(\Sig, \cat)$.

\end{document}